%% file: main_all.tex
\documentclass[twoside,journey]{IEEEtran}

\usepackage{makecell}
\usepackage{array}
\usepackage[nointegrals]{wasysym} 
\usepackage{float}
\usepackage{pdfpages}
\usepackage{amsmath}
\usepackage{graphicx,amssymb}
\usepackage{multicol}
\usepackage[noadjust]{cite}
\usepackage{setspace}
\usepackage{subfig}
\usepackage{graphicx}
\usepackage{url}
\usepackage{stfloats}
\usepackage{amsthm,pifont}
\usepackage{flushend}
\usepackage{cases,subeqnarray}
\usepackage{bm,multirow,bigstrut}
\usepackage{textcomp}
\usepackage{latexsym,bm}
\usepackage{booktabs}
\usepackage{xcolor}
\usepackage{mathtools}
\usepackage{dsfont}
\usepackage{extarrows}
\usepackage{epsfig}
\usepackage{epsfig}
\usepackage{epstopdf}
\usepackage[noend]{algpseudocode}
\usepackage{algorithmicx,algorithm}
\usepackage{catchfilebetweentags}
\usepackage{xr-hyper}
\usepackage[pagebackref=false,breaklinks=true,colorlinks,citecolor=blue,linkcolor=blue,bookmarks=true]{hyperref}

\usepackage{tikz}
\usetikzlibrary{positioning}
\usetikzlibrary{mindmap}
\definecolor{lime}{HTML}{A6CE39}
\DeclareRobustCommand{\orcidicon}{%
\begin{tikzpicture}
\draw[lime, fill=lime] (0,0) 
circle [radius=0.16] 
node[white] {{\fontfamily{qag}\selectfont \tiny ID}};\draw[white, fill=white] (-0.0625,0.095) 
circle [radius=0.007];\end{tikzpicture}
\hspace{-2mm}}
\foreach \x in {A, ..., Z}{%
\expandafter\xdef\csname orcid\x\endcsname{\noexpand\href{https://orcid.org/\csname orcidauthor\x\endcsname}{\noexpand\orcidicon}}
}

\hypersetup{
 colorlinks=true,
 linkcolor=black,
 filecolor=black,
 urlcolor=black,
}
\theoremstyle{plain}

\theoremstyle{plain}

\newcommand{\greencheck}{\textcolor{green!70!black}{\ding{51}}}
\newcommand{\redcross}{\textcolor{red!70!black}{\ding{55}}}

\IEEEoverridecommandlockouts
\begin{document}
\title{Edge-Cloud Collaborative Computing on Distributed\\Intelligence and Model Optimization: A Survey}

\author{
Jing Liu\orcidA{},~\IEEEmembership{Member,~IEEE},
 Yao Du\orcidB{},~\IEEEmembership{Graduate Student Member,~IEEE},
 Kun Yang\orcidC{},
 Jiaqi Wu\orcidK{},
 Yan Wang\orcidD{},
 Xiping Hu\orcidE{},~\IEEEmembership{Senior Member,~IEEE},
 Zehua Wang\orcidF{},~\IEEEmembership{Member,~IEEE},
 Yang Liu\orcidG{},~\IEEEmembership{Member,~IEEE},
 Peng Sun\orcidH{},~\IEEEmembership{Senior Member,~IEEE},
 Azzedine Boukerche\orcidI{},~\IEEEmembership{Fellow,~IEEE},
 Victor C.M. Leung\orcidJ{},~\IEEEmembership{Life Fellow,~IEEE}
\IEEEcompsocitemizethanks{
This work is supported in part by the Professional Discretionary Fund under Grant Nos. 26AKUG0088 and 00AKUG0343, the Guangdong Pearl River Talents Recruitment Program under Grant No. 2019ZT08X603, the Guangdong Pearl Rivers Talent Plan under Grant No. 2019JC01X235, the National Natural Science Foundation of China under Grant Nos. 62576213 and 62406075, the Natural Sciences and Engineering Research Council (NSERC) of Canada under Grants Nos. RGPIN-2021-02970 and DGECR-202100187, the National Key Research and Development Program of China (2023YFC3604802), the Shanghai Key Technology R\&D Program under Grant No. 25511107200, and the Mitacs Project under Grant Nos. IT44479 and QJLI GR037230. (Corresponding authors: Xiping Hu, Zehua Wang, Yang Liu, Peng Sun).}
\IEEEcompsocitemizethanks{
\IEEEcompsocthanksitem Jing Liu is with the Division of Natural and Applied Sciences, Duke Kunshan University, Kunshan 215316, China, and with the Department of Electrical and Computer Engineering, The University of British Columbia, Vancouver, BC V6T 1Z4, Canada, and also with the College of Future Information Technology, Fudan University, Shanghai 200433, China (e-mail: jing.liu@ieee.org).
\IEEEcompsocthanksitem Yao Du and Zehua Wang are with the Department of Electrical and Computer Engineering, The University of British Columbia, Vancouver, BC V6T 1Z4, Canada (e-mail: \{yaodu, zwang\}@ece.ubc.ca).
\IEEEcompsocthanksitem Kun Yang is with the Ant Group, also with the College of Information Science and Electronic Engineering, Zhejiang University, Hangzhou 310013, China (e-mail: kunyang20@zju.edu.cn).
\IEEEcompsocthanksitem Jiaqi Wu is with the Department of Automation, Tsinghua University, Beijing 100084, China, and with the School of Artificial Intelligence, China University of Mining and Technology (Beijing), Beijing 100083, China (e-mail: wujiaqi@mail.tsinghua.edu.cn).
\IEEEcompsocthanksitem Yan Wang is with the School of Data Science and Engineering, East China Normal University, Shanghai 200062, China (e-mail: yanwang@dase.ecnu.edu.cn).
\IEEEcompsocthanksitem Xiping Hu is with the Academy of Artificial Intelligence, SMBU, Shenzhen 518115, Guangdong, China (e-mail: huxp@smbu.edu.cn).
\IEEEcompsocthanksitem Yang Liu is with the College of Electronic and Information Engineering, Tongji University, Shanghai 201804, China (e-mail: yang\_liu@ieee.org)
\IEEEcompsocthanksitem Peng Sun is with the Division of Natural and Applied Sciences, Duke Kunshan University, Kunshan 215316, China (e-mail: peng.sun568@duke.edu).
\IEEEcompsocthanksitem Azzedine Boukerche is with the Paradise Research Laboratory, EECS, University of Ottawa, Ottawa, ON K1N 6N5, Canada (e-mail: boukerche@site.uottwa.ca).
\IEEEcompsocthanksitem Victor C. M. Leung is with the Academy of Artificial Intelligence, SMBU, Shenzhen 518115, Guangdong, China, and with the College of Computer Science and Software Engineering, Shenzhen University, Shenzhen 518060, China, and also with the Department of Electrical and Computer Engineering, The University of British Columbia, Vancouver, BC V6T 1Z4, Canada (e-mail: vleung@ece.ubc.ca).
}
}

\markboth{IEEE COMMUNICATIONS SURVEYS \& TUTORIALS}%
{Liu \MakeLowercase{\textit{et al}}: ECCC on Distribute Intelligence and Model Optimization: A Survey}

\IEEEtitleabstractindextext{
\begin{abstract}
 Edge-cloud collaborative computing (ECCC) has emerged as a pivotal paradigm for addressing the computational demands of modern intelligent applications, integrating cloud resources with edge devices to enable efficient, low-latency processing across distributed communication networks. Recent advancements in AI, particularly deep learning and large language models (LLMs), have dramatically enhanced the capabilities of these networked systems, yet introduce significant challenges in model deployment, network resource management, and cross-layer optimization. In this survey, we comprehensively examine the intersection of distributed intelligence and model optimization within edge-cloud environments, providing a structured tutorial on fundamental architectures, communication protocols, and network-aware computing frameworks. Additionally, we systematically analyze model optimization approaches, including compression, adaptation, and neural architecture search, alongside AI-driven resource management strategies that balance performance, energy efficiency, and communication overhead across heterogeneous networks. We further explore critical aspects of privacy protection and security enhancement within ECCC systems and examine practical deployments through diverse networked applications, spanning autonomous driving, healthcare, and industrial automation. Performance analysis and benchmarking techniques are also thoroughly explored to establish evaluation standards for these complex distributed systems. Furthermore, the review identifies critical research directions including LLMs deployment, 6G integration, neuromorphic computing, and quantum computing, offering a roadmap for addressing persistent challenges in heterogeneity management, real-time processing, and scalability. By bridging theoretical advancements in communications with practical deployments, this survey offers researchers and practitioners a holistic perspective on leveraging AI to optimize distributed computing environments over next-generation communication networks, fostering innovation in intelligent networked systems.

\end{abstract}
}

\maketitle
\IEEEdisplaynontitleabstractindextext
\IEEEpeerreviewmaketitle

\section{Introduction}\label{sec1}
\subsection{Background}\label{sec1.1}

\IEEEPARstart{T}{he} increasing prevalence of Internet of things (IoT) devices and the demand for real-time applications have propelled a shift from centralized cloud computing to a distributed edge-cloud paradigm \cite{masaracchia20256genabled,wang2024endedgecloud,vanhuynh2022edge}.  While cloud computing initially offers scalability and cost-effectiveness, limitations such as latency and bandwidth constraints spur the emergence of edge computing, bringing computation closer to data sources \cite{yang2025edge,moreschini2024edge,yang2025growthadaptive}. Consequently, edge-cloud collaborative computing (ECCC) arises, integrating the strengths of both paradigms to facilitate efficient resource utilization and enable complex AI tasks across the edge-cloud continuum \cite{khalyeyev2023characterization,sun2025latencyaware}.  AI integration within ECCC systems addresses the inherent complexity of distributed environments, enabling optimized resource allocation and strategic model deployment for enhanced system performance \cite{wang2025efficient}.  In addition, AI empowers intelligent decision-making at the edge, enabling real-time responses.  However, realizing distributed intelligence in ECCC systems introduces challenges, including efficient model partitioning and resource-aware model optimization, along with ensuring seamless edge-cloud collaboration and addressing security and privacy concerns \cite{liu2023distributed,yao2023edgecloud}.

Technical foundations underlying ECCC span multiple critical domains that require comprehensive understanding. Model optimization represents a fundamental challenge in deploying sophisticated AI models across resource-constrained edge devices while maintaining computational efficiency and performance \cite{sharma2024deep}. Model compression techniques including pruning, quantization, and knowledge distillation, alongside adaptive learning approaches such as transfer learning, federated learning, and continual learning enable models to evolve with changing data distributions \cite{yang2025growthadaptive}. Resource management in distributed edge-cloud environments involves complex orchestration of computational, communication, and storage resources across heterogeneous infrastructure, requiring intelligent task offloading strategies, dynamic resource allocation algorithms, and energy-efficient optimization techniques to balance performance, cost, and sustainability \cite{lu2024a2cdrl}. Security and privacy protection mechanisms are paramount in ECCC systems due to the distributed nature of data processing and the involvement of multiple stakeholders, necessitating advanced cryptographic protocols, privacy-preserving machine learning techniques, and secure multi-party computation approaches to protect sensitive information while enabling collaborative intelligence \cite{seifelnasr2024privacypreserving}.

ECCC has emerged as a crucial paradigm to support the growing demands of real-time, intelligent applications, particularly with the proliferation of artificial intelligence of things (AIoT) devices \cite{tuli2023ai}.  However, inherent limitations in edge devices, such as constrained computational power and limited battery life, hinder their ability to independently handle computationally intensive AI workloads \cite{zhou2024enhancing}.  Additionally, the dynamic nature of workloads in these environments requires sophisticated resource allocation and task offloading strategies to maintain performance and prevent bottlenecks \cite{paul2025quantumenhanced}.  Applications like autonomous driving \cite{poularakis2019joint} and industrial automation \cite{ren2021synergy} further necessitate low latency and high reliability, demanding intelligent resource management and optimization \cite{jansen2023specrg}. Consequently, AI offers a powerful set of tools, including deep reinforcement learning (DRL) for dynamic resource allocation \cite{xu2024fusion} and multi-objective optimization for task scheduling \cite{zhang2024manyobjective}, to address these challenges and improve the efficiency and performance of ECCC systems.  For instance, federated learning enables collaborative model training across distributed devices while preserving privacy \cite{zhou2024enhancing}, ultimately contributing to enhanced quality of service (QoS) by optimizing resource utilization and minimizing latency.

Communication network evolution plays a pivotal role in enabling effective edge-cloud collaboration by providing the foundational infrastructure that connects distributed computational resources. Recent advancements in network technologies, particularly software-defined networking (SDN) \cite{guo2024madrlom} and network function virtualization (NFV) \cite{bensalah2024vnf}, have significantly enhanced the flexibility and programmability of communication infrastructures supporting ECCC. Advanced technologies facilitate dynamic network resource allocation and traffic engineering, which are essential for optimizing data flows between edge devices and cloud servers \cite{bensalem2023scaling}. Furthermore, emerging wireless communication paradigms, including millimeter-wave communications, massive MIMO, and non-orthogonal multiple access (NOMA) \cite{letaief2022edge}, are revolutionizing the capacity, reliability, and efficiency of edge-cloud connectivity. Moreover, the integration of network slicing techniques \cite{li2021slicingbased} enables the creation of virtualized, isolated network segments with tailored quality-of-service guarantees for diverse edge-cloud applications. Moreover, cross-layer optimization approaches that jointly consider communication, computation, and caching resources \cite{kanduri2024edgecentric} are becoming increasingly important for minimizing end-to-end latency and maximizing system throughput. As communication networks evolve toward 6G, the synergistic optimization of networking and computing resources will be instrumental in addressing the growing demands of distributed intelligence across the edge-cloud continuum \cite{kokkonen2023autonomy}.

The selection and optimization of communication protocols significantly impact the performance, security, and reliability of edge-cloud collaborative systems \cite{liu2021mpcenabled}. Protocol optimization encompasses multiple layers with distinct considerations: (1) \textit{Application-layer protocols} such as HTTP/HTTPS remain fundamental for web-based edge services, while MQTT and CoAP provide lightweight alternatives specifically designed for resource-constrained IoT devices with intermittent connectivity~\cite{phung2020enhancing}. MQTT's publish-subscribe model offers excellent scalability for sensor data aggregation scenarios, whereas CoAP's REST-based approach provides seamless integration with existing web infrastructures, though MQTT introduces broker dependency that can create single points of failure, while CoAP's UDP-based nature may require additional mechanisms for reliable delivery in lossy networks~\cite{qu2024mobile}; (2) \textit{Network-layer protocols} including IPv6 and emerging protocols like named data networking (NDN) address the scalability challenges of massive IoT deployments. IPv6 provides abundant address space essential for large-scale edge device connectivity, while NDN's content-centric approach enables efficient data distribution and caching across edge networks, nevertheless IPv6 adoption remains inconsistent across heterogeneous edge environments, and NDN requires significant infrastructure changes that may limit near-term deployment~\cite{gundogan2020iot}; and (3) \textit{Transport-layer innovations} such as QUIC and specialized protocols like BBR congestion control offer improved performance over traditional TCP in high-latency, variable-bandwidth edge-cloud connections~\cite{joarder2024exploring}. Specifically, QUIC's built-in encryption and reduced connection establishment overhead particularly benefit mobile edge scenarios, whereas BBR's model-based congestion control adapts better to the diverse network conditions encountered in edge deployments. Consequently, the choice of appropriate protocols depends on specific application requirements, device capabilities, and network characteristics, with optimal edge-cloud systems often employing protocol stacks that dynamically adapt to changing conditions and requirements~\cite{kawana2024communication}.

The imperative for advanced model optimization and resource management in ECCC is driven by the demanding requirements of emerging application scenarios. For instance, autonomous driving requires ultra-low latency for real-time decision-making and must process vast streams of sensor data while preserving privacy~\cite{poularakis2019joint}. Augmented and virtual reality (AR/VR) applications demand high-bandwidth, low-latency rendering to deliver immersive user experiences, which is computationally intensive and energy-draining for mobile devices~\cite{vanhuynh2022edge}. Similarly, smart manufacturing relies on real-time fault diagnosis and predictive maintenance, necessitating reliable and continuous model adaptation in complex industrial environments~\cite{ren2021synergy}. Collectively, these applications create a tension between computational complexity, resource constraints, and performance requirements, motivating the development of the enabling technologies surveyed in this survey.
A systematic mapping of these application scenarios to their technical challenges and enabling solutions is provided in the Supplementary Material (see \autoref{tab:app_summary} in \autoref{sec:supp-application-scenarios}), where we detail how different domains leverage the technologies surveyed in this paper.

\subsection{Related Works}\label{sec1.3}

Over the past few years, several surveys have explored various aspects of edge computing and cloud collaboration, each with distinct scopes and technical emphases. Early works like Abbas et al.~\cite{abbas2018mobile} and Ferrer et al.~\cite{ferrer2019decentralised} establish a foundational understanding of edge computing and decentralized cloud architectures respectively, with their primary focus on architectural foundations rather than AI-driven model optimization. Similarly, Liu et al.~\cite{liu2019survey} provide valuable insights into edge computing systems and tools, while Wang et al.~\cite{wang2019edge} contribute significantly to offloading algorithms, with their emphasis primarily on different aspects of edge-cloud integration rather than comprehensive AI technique integration. 
More recent surveys have begun exploring particular dimensions of this space. Cao et al.~\cite{cao2021survey} examine edge-cloud assisted cyber-physical systems with strong coverage of resource management, with their primary emphasis on system-level considerations rather than model optimization techniques. Domain-specific surveys like Arthurs et al.~\cite{arthurs2022taxonomy} focus on intelligent transportation systems with specialized depth in this domain. Despite promising advancements, emerging paradigms such as large language models (LLMs) at the network edge remain an active area of exploration, with Qu et al.~\cite{qu2025mobile} providing a dedicated investigation into mobile edge intelligence for LLMs, focusing primarily on resource-efficient techniques within their specialized scope. Additionally, specialized applications of edge intelligence for digital twin-enabled metaverse environments, as examined by Van Huynh et al.~\cite{vanhuynh2022edge}, offer valuable insights into ultra-reliable and low-latency communications with their emphasis on domain-specific considerations. 

The most relevant contemporary works include Yang et al.~\cite{yang2025edge}, who extensively analyze reinforcement learning methodologies for mobile edge computing through an advanced exploration of optimization techniques, with their work emphasizing network dynamics as their primary contribution. Similarly, the comprehensive works by Yao et al.~\cite{yao2023edgecloud}, Gu et al.~\cite{gu2024aienhanced}, and Wang et al.~\cite{wang2024endedgecloud} each make valuable contributions with distinct emphases. Specifically, \cite{yao2023edgecloud} provides excellent coverage of edge-cloud collaboration and AI techniques with particular strength in these core areas. \cite{gu2024aienhanced} delivers strong analysis of AI-enhanced cloud-edge-terminal networks with good application coverage, emphasizing network architecture aspects. \cite{wang2024endedgecloud} offers comprehensive model optimization coverage across the end-edge-cloud spectrum with particular depth in optimization techniques.

Two surveys are particularly relevant to our work and warrant detailed comparison. Yang et al.~\cite{yang2025edge} provides a comprehensive survey of reinforcement learning approaches for mobile edge computing with excellent coverage of resource management and autonomous driving applications. Their work has a specific focus that differs from our survey in several key aspects: (1) \textit{Specialized computing paradigm emphasis}, while they focus extensively on edge computing and resource management, our work provides comprehensive coverage of cloud integration and edge-cloud collaboration frameworks; (2) \textit{Domain-specific technical focus}, their emphasis on reinforcement learning, while thorough, represents a specialized approach compared to our comprehensive treatment of model optimization techniques such as compression, adaptation, and neural architecture search that are central to modern ECCC systems; (3) \textit{Different scope for privacy and security}, their work emphasizes other technical aspects, while our survey provides systematic attention to privacy-preserving mechanisms and security enhancement for real-world ECCC deployments; (4) \textit{Specialized application coverage}, while strong in autonomous driving, our work provides comprehensive coverage of manufacturing and healthcare applications that increasingly rely on edge-cloud collaboration. Qu et al.~\cite{qu2025mobile} presents a focused investigation into mobile edge intelligence for large language models with strong coverage of LLMs, privacy considerations, and healthcare applications. Their work complements our survey with different emphases: (1) \textit{Specialized technical scope}, while excellent in LLM-specific optimizations, our survey provides systematic coverage of general model optimization techniques applicable across diverse AI models and architectures; (2) \textit{Mobile edge intelligence focus}, their focus on mobile edge intelligence represents a specialized area, while our work emphasizes comprehensive edge-cloud collaboration frameworks and distributed intelligence paradigms; (3) \textit{Specialized application domain}, while strong in healthcare applications, our survey provides detailed analysis of manufacturing and industrial IoT scenarios that represent major application domains for ECCC; (4) \textit{Domain-specific security analysis}, their privacy coverage addresses specific requirements, while our work provides comprehensive treatment of security mechanisms and adversarial considerations essential for robust ECCC systems.

In contrast to these works, our survey uniquely provides: (1) \textit{Comprehensive computing continuum coverage} spanning edge, cloud, edge-cloud computing, and collaborative frameworks; (2) \textit{Holistic technical integration} systematically covering model optimization (compression, adaptation, NAS), resource management, privacy, security, and emerging LLM paradigms; (3) \textit{Diverse application analysis} with detailed coverage across manufacturing, autonomous driving, and healthcare domains; (4) \textit{Unified framework approach} that synthesizes distributed intelligence principles with practical implementation considerations.
As summarized in \autoref{tab:1}, these surveys collectively address different aspects of ECCC, with each making valuable contributions to specific areas of the field. Our survey builds upon these foundational works by providing integrated coverage of model optimization, privacy, security, and diverse applications within a unified framework. Most importantly, existing works have established strong foundations in specific technical aspects (e.g., model optimization or resource management) or particular application domains (e.g., transportation or healthcare), while our work provides a holistic synthesis that spans the entire computing continuum. Consequently, our survey contributes by providing comprehensive analysis that encompasses advanced technical focuses including LLMs, and practical applications across manufacturing, autonomous driving, and healthcare, thereby complementing and extending the existing literature.

While the recent survey by~\cite{qu2025mobile} provides a valuable overview of AI models for edge computing, our work offers several distinct contributions. First, we present a more fine-grained taxonomy for model optimization techniques, particularly in the context of co-design with resource management (as detailed in \autoref{sec3} and \autoref{sec4}). Second, our survey uniquely focuses on the interplay between task characteristics and adaptive AI strategies, a dimension not explicitly covered in~\cite{qu2025mobile}. Finally, we provide a more forward-looking perspective on open challenges, integrating insights from both system-level and algorithm-level research.

\input{t1_survey_1.tex}

\subsection{Scope and Contributions}\label{sec1.4}

\begin{figure*}[th!]
  \centering
\includegraphics[width=1\textwidth]{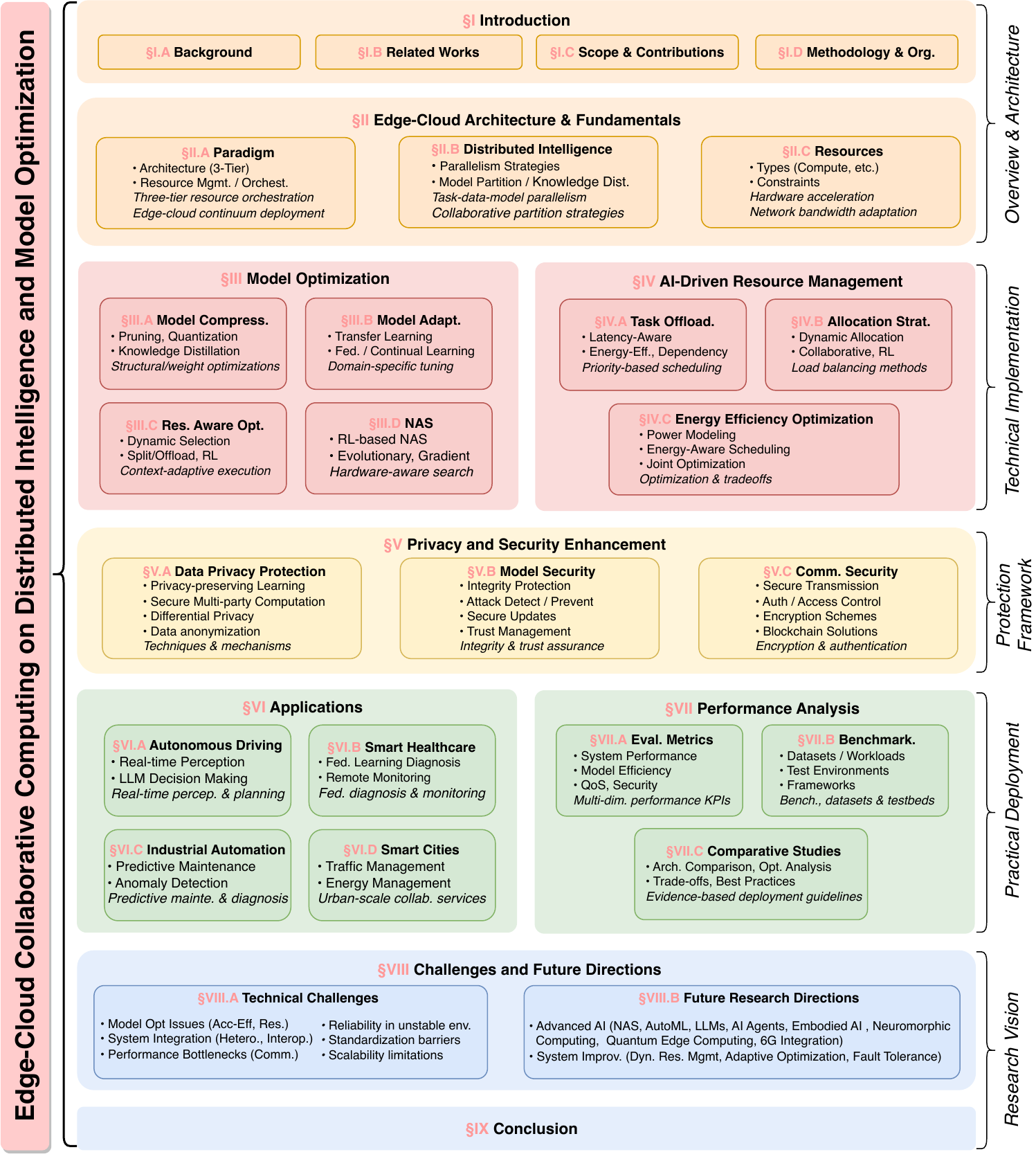}
\caption{Overall structure of this survey, wherein the framework is organized into five layers: (1) Overview \& architecture covering introduction and edge-cloud fundamentals, (2) technical implementation focusing on model optimization and resource management techniques, (3) protection framework addressing privacy and security enhancement, (4) practical deployment discussing applications and performance analysis, and (5) research vision outlining challenges and future directions.}
\label{fig:2}
\vspace{-5px}
\end{figure*}

As illustrated in \autoref{fig:2}, we provide a systematic and multi-layered examination of ECCC, addressing the critical intersection of distributed intelligence and model optimization. Edge-cloud systems face unprecedented challenges due to heterogeneous infrastructures, resource limitations, and growing computational demands from AI applications, making a comprehensive review of this rapidly evolving field both timely and necessary. 
To structure our comprehensive review, this survey is organized according to the five-layer framework depicted in \autoref{fig:2}, which reflects a top-down approach from foundational concepts to future research. Specifically, our organizational design provides a clear and logical progression through the complex ECCC landscape, starting with the \textit{Overview \& Architecture layer} that establishes the groundwork by covering introductory concepts and fundamental architectures. Building upon this foundation, the \textit{Technical Implementation layer} delves into the core enabling technologies, including the model optimization and resource management strategies detailed in \autoref{sec3} and \autoref{sec4}. Subsequently, the framework addresses security, practical deployment, and forward-looking research through dedicated layers, thereby ensuring a holistic examination of the field.

By systematically analyzing ECCC architectures, resource management strategies, and performance metrics, the survey bridges existing gaps in understanding how AI can enhance decentralized computing environments. Moreover, it investigates critical enabling technologies such as model compression, adaptation, resource-aware optimization, and security mechanisms, which are pivotal for overcoming current limitations, spanning from architectural foundations and optimization techniques to practical applications and future research directions, thus offering researchers and practitioners a holistic perspective of this transformative paradigm. 
Regarding the three primary technical categories of ECCC, each addresses distinct challenges and exhibits specific applicability constraints. \textit{Model optimization techniques} (\autoref{sec3}) primarily target the fundamental challenge of deploying sophisticated AI models across resource-constrained edge devices while maintaining computational efficiency. Model optimization approaches are particularly suitable for scenarios involving deep learning model deployment, privacy-sensitive distributed training, and real-time inference applications. However, their effectiveness is constrained by the inherent trade-offs between model accuracy and compression ratios, the heterogeneity of edge device capabilities, and the complexity of managing distributed learning processes. Although model compression methods excel in reducing computational overhead, they may compromise accuracy, whereas federated learning approaches provide privacy preservation at the cost of increased communication complexity and convergence challenges with non-IID settings.

\textit{AI-driven resource management strategies} (\autoref{sec4}) focus on optimizing resource utilization and task orchestration across the heterogeneous edge-cloud continuum, proving most effective in scenarios with dynamic workloads, multi-tenant environments, and applications requiring intelligent load balancing across diverse computational resources. Nevertheless, primary limitations include the computational complexity of multi-objective optimization problems, the difficulty in predicting system behavior under varying conditions, and scalability challenges in large-scale deployments. Although task offloading mechanisms can significantly reduce latency and energy consumption, they require sophisticated algorithms to handle dependency relationships and may introduce communication overhead that negates performance benefits. Furthermore, \textit{privacy and security enhancement frameworks} (\autoref{sec5}) address the critical requirements for protecting sensitive data and ensuring system trustworthiness in distributed environments, particularly essential for healthcare, financial, and government applications where regulatory compliance and data protection are paramount. Nevertheless, privacy-preserving mechanisms often introduce computational overhead and complexity in key management, while security measures must continuously evolve to address emerging threats in heterogeneous edge environments.
To provide a high-level overview of the core technical domains, \autoref{t-tech-comparison} presents a comparative analysis of model optimization (\autoref{sec3}), AI-driven resource management (\autoref{sec4}), and privacy and security enhancement (\autoref{sec5}). Specifically, this summary table evaluates each area across its core techniques, advantages, limitations, and applicable scenarios, offering a structured guide to the key trade-offs and design considerations discussed in the subsequent sections.
The contributions of this survey are summarized as follows:
\input{t1.1_tech_comparison.tex}
\begin{itemize}
    \item \textbf{First Comprehensive Survey.} To the best of our knowledge, this is the first comprehensive survey that specifically targets the intersection of distributed intelligence and model optimization within the broader field of edge-cloud collaborative computing, reviewing recent advances across architectures, techniques, security, and privacy.
    \item \textbf{Unified Architectural Framework.} Our survey establishes a cohesive framework for understanding edge-cloud collaborative computing architectures, resource constraints, and distributed intelligence models. Through systematic categorization, researchers gain a structured understanding of how computational resources can be effectively distributed across the edge-cloud continuum.
    \item \textbf{Multi-domain Application Analysis.} Through detailed applications spanning autonomous driving, smart healthcare, industrial automation, and smart cities, our work demonstrates the practical impact of AI-driven edge-cloud computing across diverse domains and identifies domain-specific requirements and challenges.
    \item \textbf{Future Research Roadmap.} By identifying technical challenges and promising research directions, including advanced AI techniques, system improvements, and emerging applications, we provide a comprehensive roadmap for researchers and practitioners to advance the field of AI-driven edge-cloud collaborative computing.
\end{itemize}

\subsection{Research Methodology and Organization}\label{sec1.5}

\begin{figure}[t!]
  \centering
\includegraphics[width=0.49\textwidth]{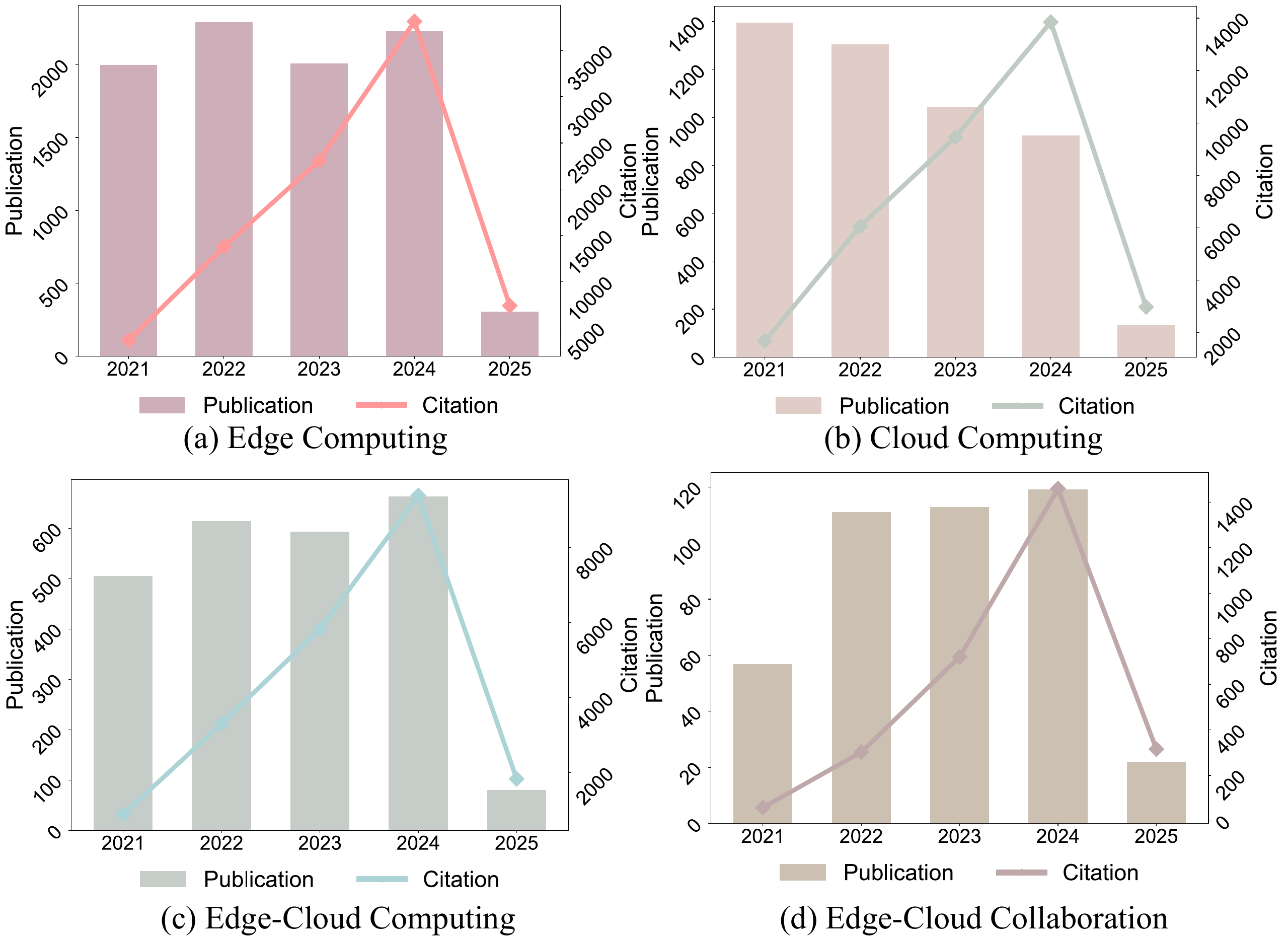}
\caption{Publication trends and citation analysis of research in distributed computing paradigms from 2021 to 2025.}
\label{fig:1}
\vspace{-15px}
\end{figure}

Recent analysis reveals significant research momentum in ECCC, as illustrated in \autoref{fig:1}, where publication trends and citation patterns demonstrate the evolving landscape across four interrelated domains (The statistics data is collected as of 2025/03/05). While traditional edge computing maintains consistent publication volumes with steadily increasing citation impact, cloud computing exhibits characteristics of a mature research area with decreasing publication numbers yet sustained high citation rates. Most notably, edge-cloud computing shows clear upward trajectories in both publications and citations, signaling growing research interest, whereas the specialized subfield of edge-cloud collaboration displays exponential citation growth peaking in 2024, indicating its emerging importance. Our systematic review methodology employed comprehensive literature searches across prominent scholarly databases including IEEE Xplore, ACM Digital Library, ScienceDirect, and Web of Science, utilizing carefully selected keywords such as ``edge-cloud computing'', ``distributed intelligence'', ``model optimization'', ``AI'', and ``machine learning''. Following retrieval, selected papers pertinent to ECCC were carefully reviewed and categorized based on their primary contributions aligned with the scope of this work.

The remainder of this survey provides a comprehensive overview of AI-driven ECCC. We begin with edge-cloud computing fundamentals including architecture and distributed intelligence in \autoref{sec2}. \autoref{sec3} examines model optimization techniques including compression and neural architecture search. \autoref{sec4} explores AI-driven resource management covering task offloading and energy efficiency, while \autoref{sec5} addresses privacy and security enhancement. \autoref{sec6} presents practical applications spanning autonomous driving, healthcare, and smart cities. \autoref{sec7} delivers performance analysis through benchmarking and evaluation, followed by challenges and future directions in \autoref{sec8}, with conclusions in \autoref{sec9}.

\section{Architecture and Fundamentals}\label{sec2}
\begin{figure}[t!]
  \centering
\includegraphics[width=0.49\textwidth]{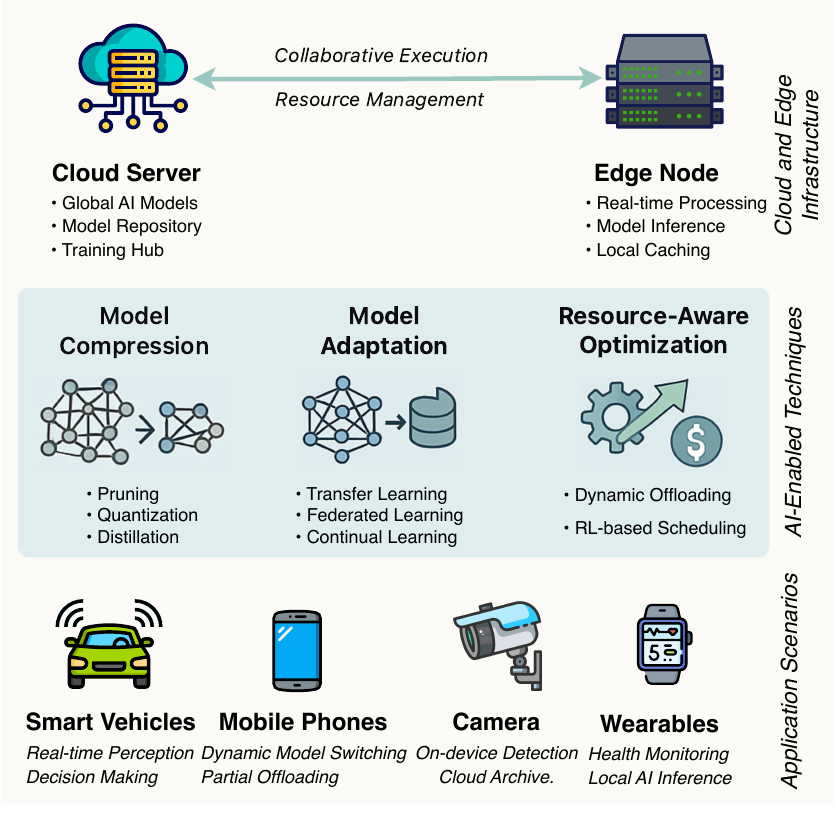}
  \caption{Concepts of architecture, wherein this diagram provides a holistic view of the technical scope covered in this survey, where the architecture integrates centralized cloud resources, decentralized edge nodes, and diverse mobile/edge devices while AI-enabled technologies including model compression, adaptation, and resource-aware optimization jointly support distributed intelligence and efficient model execution across multiple application scenarios.}
  \label{fig:3}
\vspace{-15px}
\end{figure}

Modern intelligent applications demand architectures capable of processing vast data volumes with minimal latency, driving innovation toward distributed computing paradigms that bridge resource-rich cloud infrastructure with proximity-based edge processing. \autoref{fig:3} provides a conceptual illustration of AI-driven edge-cloud collaborative systems, depicting the core infrastructure interplay between cloud servers and edge nodes, alongside essential AI-enabled techniques (e.g., model adaptation, resource-aware optimization) that facilitate distributed intelligence across various application scenarios. Building upon this visual overview, the subsequent subsections delve into the foundational aspects of this architecture. We begin by outlining the core computing paradigm in \autoref{sec2.1}, subsequently examining the distributed intelligence framework enabling collaboration in \autoref{sec2.2}. Following this, an analysis of crucial resource types and their inherent constraints is presented in \autoref{sec2.3}.
\subsection{Edge-Cloud Computing Paradigm}\label{sec2.1}
Distributed computing architectures optimize performance by positioning computational and storage resources closer to data sources, implementing a three-tier hierarchy encompassing cloud infrastructure, edge nodes, and end devices \cite{vo2022edge}.  Within this hierarchy, the cloud tier provides extensive computational and storage resources for large-scale data processing and model training, while the edge tier, comprising edge servers and gateways, offers localized processing to reduce latency \cite{tocze2018taxonomy}. At the periphery, end devices such as smartphones and IoT sensors generate and consume data at the network edge, necessitating effective resource management and orchestration for efficient resource allocation across these tiers.  Notably, distributing intelligence across the network, a key characteristic of this paradigm, can be represented by a function $D(R_i, L_j)$, quantifying the proportion of resource $R_i$ allocated to layer $L_j$ \cite{hamm2019edge}.  Edge-cloud systems aim to optimize resource allocation, considering factors like application requirements and network conditions, to maximize a utility function $U(P, C, L)$ encompassing performance $P$, cost $C$, and latency $L$, subject to resource constraints \cite{tosic2019blockchainbased}. Consequently, processing data closer to its source reduces latency and bandwidth consumption.  Additionally, distributing workloads enhances scalability and fault tolerance, improving the system's ability to handle increasing data volumes and withstand node failures \cite{cao2020edge}.  Edge federation models further enable collaborative resource sharing, optimizing service delivery across a federated network \cite{ramu2023edge}, especially for mission-critical applications \cite{skarin2018missioncritical}.

\subsection{Distributed Intelligence Framework}\label{sec2.2}

Distributed intelligence frameworks optimize performance by strategically coordinating computational resources across heterogeneous infrastructure, leveraging both proximity-based edge processing and scalable cloud capabilities to address unique constraints and opportunities inherent in collaborative environments. Within this framework, the distributed intelligence model in ECCC adheres to three key principles specifically adapted for edge-cloud deployment: task parallelism, data parallelism, and model parallelism. 

Task parallelism in edge-cloud contexts executes different parts of complex AI tasks concurrently on separate devices, considering edge device limitations and cloud processing capabilities \cite{wang2022ace}, represented as $T = \{t_1^{edge}, t_2^{cloud}, ..., t_n^{hybrid}\}$, where tasks are strategically assigned based on latency requirements, computational complexity, and data locality. Data parallelism trains identical models on different data subsets distributed across multiple edge devices while maintaining privacy \cite{yao2023edgecloud}, expressed as $D = \{d_1^{local}, d_2^{local}, ..., d_m^{local}\}$, where data remains decentralized to address privacy concerns and network bandwidth limitations. Model parallelism deploys different parts of large models on separate devices within the edge-cloud continuum \cite{kim2024dnn}, formulated as $M = \{m_1^{edge}, m_2^{fog}, ..., m_p^{cloud}\}$, where model partitioning considers device capabilities, network latency, and energy constraints.

Key design challenges include managing heterogeneous device capabilities across the edge-cloud spectrum, where edge devices have limited computational power but offer low latency, while cloud resources provide extensive computation but introduce communication delays. Network reliability and bandwidth variability require adaptive model partitioning strategies that can dynamically adjust based on real-time network conditions. Energy efficiency becomes paramount for battery-powered edge devices, necessitating intelligent workload distribution that balances computational load with energy consumption. Privacy and data locality requirements demand sophisticated federated learning approaches that enable collaborative intelligence without centralizing sensitive data. 

Edge intelligence focuses on real-time processing with sub-10ms latency requirements \cite{asim2020review}, making it suitable for time-critical applications like autonomous driving and industrial control. Cloud intelligence handles complex computations and large-scale training using virtually unlimited resources \cite{mungoli2023scalable}, enabling sophisticated model development and global knowledge aggregation. Model partitioning strategies must balance workloads by considering model complexity, data size, network topology, and resource constraints specific to edge-cloud deployments \cite{banitalebi-dehkordi2021autosplit}. Resource allocation algorithms dynamically assign computational, storage, and communication resources based on real-time demands, device capabilities, and edge-cloud topology \cite{alvar2021paretooptimal}. Performance optimization techniques like model compression enhance efficiency across the distributed architecture by reducing communication overhead and enabling deployment on resource-constrained devices \cite{wang2024socialized}. Additionally, fault tolerance mechanisms ensure system reliability despite potential edge device failures, network partitions, or cloud service interruptions, requiring redundancy strategies and graceful degradation protocols.

\subsection{Resource Types and Constraints}\label{sec2.3}
Fundamental system resources, including computation, communication bandwidth, and storage capacity, present inherent constraints that significantly impact distributed architecture design and performance optimization across heterogeneous infrastructure \cite{tocze2018taxonomy}.  Specifically, computation resources, denoted as $C_k$ for node $k$, represent processing power; communication bandwidth, $B_{ij}$, limits data transfer rates between nodes $i$ and $j$, directly affecting latency; and storage capacity, $S_k$, dictates the data volume node $k$ can store, influencing data-intensive applications \cite{luo2021resource}.  Managing these constrained and heterogeneous resources across diverse devices, under dynamic workloads and with limited edge resources, especially energy for battery-powered devices \cite{hamm2019edge}.

Dynamic resource allocation addresses these challenges by adapting to workload fluctuations, aiming to find an optimal mapping of tasks $\mathcal{T}$ to nodes $\mathcal{N}$ that minimizes a cost function (e.g., latency) subject to resource constraints \cite{rublein2024improved}.  Mathematically, this involves minimizing $\sum_{t \in \mathcal{T}} Cost(M(t), t)$ subject to $\forall n \in \mathcal{N}: \sum_{t: M(t) = n} R(t) \leq C_n$, where $Cost(M(t), t)$ is the cost of executing task $t$ on node $M(t)$, and $R(t)$ is the resource requirement of task $t$.  In addition, energy-efficient offloading minimizes energy consumption, $E_{ij}(t) = E_{comm}(t, B_{ij}) + E_{comp}(t, C_j)$, by strategically offloading tasks, where $E_{comm}$ and $E_{comp}$ represent communication and computation energy, respectively \cite{wang2019edge}.  Efficient edge storage management, leveraging techniques like data caching and prefetching \cite{sonbol2020edgekv}, is also crucial, especially for data-intensive applications, alongside integrating QoS considerations \cite{farhoudi2023qosaware}.

\section{Model Optimization for Edge-Cloud Systems}\label{sec3}

\input{taxo-sec3-model-opt.tex}
To systematically address these challenges, this section presents a comprehensive taxonomy of model optimization techniques tailored for ECCC, as visually summarized in \autoref{taxo:sec3}. Our classification is structured around four fundamental pillars that represent distinct yet complementary approaches to enhancing model efficiency and performance in distributed settings. \textit{Model compression} (\autoref{sec3.1}) focuses on reducing model size and computational complexity. \textit{Model adaptation} (\autoref{sec3.2}) enables models to adjust to new data or environments. \textit{Resource-aware model optimization} (\autoref{sec3.3}) dynamically tailors models to hardware and network constraints. Finally, \textit{neural architecture search} (\autoref{sec3.4}) automates the discovery of optimal model designs. Collectively, this taxonomy provides a structured framework for navigating the diverse landscape of optimization strategies.

\subsection{Model Compression}\label{sec3.1}
Model compression techniques are essential for deploying complex AI models in resource-constrained environments by reducing their size and computational demands~\cite{tang2024anomaly,vuruma2024cloud}. In line with established literature, this survey categorizes these techniques into three widely recognized pillars: pruning, quantization, and knowledge distillation~\cite{wang2024endedgecloud}. Notably, this classification is justified by the distinct mechanism each approach employs. \textit{Pruning} involves structural modification to eliminate redundant model parameters or components. \textit{Quantization} focuses on numerical precision optimization by reducing the bit-width of parameters. \textit{Knowledge distillation} facilitates knowledge transfer from a large ``teacher'' model to a smaller ``student'' model. As summarized in \autoref{tab:2}, each category offers unique trade-offs between compression ratio, accuracy, and computational overhead, making them suitable for different ECCC scenarios.

Each compression category exhibits distinct advantages and limitations for edge-cloud deployment scenarios. \textit{Pruning techniques} excel in reducing model size and computational complexity by eliminating redundant parameters, making them particularly effective for convolutional neural networks where structured pruning can remove entire channels or layers. However, pruning may require fine-tuning to recover accuracy losses and can be sensitive to the pruning ratio selection~\cite{ko2024twophase}. \textit{Quantization methods} provide significant memory reduction and accelerated inference by reducing parameter precision, with minimal accuracy degradation when carefully implemented. They are especially suitable for deployment on specialized hardware with quantization support, though they may face challenges with extremely low-bit quantization or complex model architectures~\cite{liu2024quasyncfl}. \textit{Knowledge distillation} offers the unique advantage of training compact models from scratch while leveraging teacher model knowledge, enabling flexible student model architectures. However, it requires access to both teacher models and training data, making it computationally intensive during the training phase~\cite{wang2024clouddevice}.

The choice of compression technique should consider several factors: (1) \textit{Resource constraints}, pruning is preferred for memory-limited devices, quantization for computational acceleration, and knowledge distillation for balanced performance-efficiency trade-offs; (2) \textit{Model complexity}, structured pruning works well for over-parameterized models, quantization suits models with redundant precision, and knowledge distillation excels for complex teacher models; (3) \textit{Deployment requirements}, pruning enables immediate deployment, quantization requires hardware compatibility, and knowledge distillation demands training infrastructure; (4) \textit{Performance sensitivity}, knowledge distillation typically achieves the best accuracy preservation, followed by quantization and pruning respectively.

Notably, pruning techniques demonstrate significant improvements in inference latency and energy efficiency by strategically eliminating less important parameters~\cite{wu2022serving}, as further detailed in \autoref{sec3.1.1}. Meanwhile, quantization methods, explored in \autoref{sec3.1.2}, reduce parameter precision requirements, with implementations achieving remarkable energy efficiency in healthcare applications. Knowledge distillation approaches, discussed in \autoref{sec3.1.3}, enable smaller models to effectively mimic larger counterparts, with recent innovations specifically optimized for efficient LLM inference across edge-cloud boundaries. Across these diverse techniques, a common pattern emerges wherein most successful implementations leverage both edge and cloud resources collaboratively, highlighting the synergistic nature of modern compression strategies in distributed computing environments.
\input{t2_mod_compre_3.1.tex}

\subsubsection{Pruning}\label{sec3.1.1}
Parameter elimination strategies reduce model complexity by removing redundant components, enabling efficient AI deployment on resource-constrained devices through targeted removal of less critical weights or architectural elements \cite{yan2024hybrid}. We explore various strategies targeting different architectural aspects, such as pruning entire neurons or filters based on performance contribution, as in the two-phase split computing framework (TSCF) \cite{ko2024twophase}, or focusing on individual weights. Applied during or after training, dynamic and static pruning balance accuracy, and computational cost. For instance, Hybrid SD \cite{yan2024hybrid} employs structural pruning of the Stable Diffusion Model's U-Net to maintain image quality on edge devices, while the adaptive pruning-split federated learning (PSFed) method \cite{zheng2024semantic} optimizes semantic coder updates in a SemCom-assisted SEC framework, reducing training delay and energy consumption.

\subsubsection{Quantization}\label{sec3.1.2}
Precision reduction techniques effectively minimize memory footprint and computational cost by employing lower-precision representations of model parameters \cite{dong2020cdc}, making quantized models particularly suitable for resource-constrained devices.  Instead of 32-bit floating-point representations, quantization employs lower precision formats such as 8-bit integers, consequently diminishing memory requirements and accelerating computations on these lower-precision numbers.  For example, \cite{chen2024implementing} explored model quantization for edge implementation in arrhythmia monitoring to reduce energy consumption.  Additionally, QuAsyncFL \cite{liu2024quasyncfl} integrates quantization with asynchronous federated learning to enhance communication efficiency in cloud-edge-terminal collaborations within the AIoT domain.

\subsubsection{Knowledge Distillation}\label{sec3.1.3}

Knowledge transfer approaches enable compact ``student'' models to achieve comparable performance with reduced complexity by learning from larger ``teacher'' models \cite{yao2024gkt}, facilitating efficient deployment in resource-constrained environments \cite{tang2024anomaly}.  Instead of learning from ground truth labels, the student model learns the teacher model's output distribution, providing richer information than hard labels and enabling more nuanced representations and better generalization \cite{wang2024clouddevice}.  KD benefits ECCC by placing the teacher model in the resource-rich cloud and the distilled student model on edge devices \cite{yang2025growthadaptive}, optimizing overall system performance. Agglomerative federated learning (FedAgg) \cite{wu2024agglomerative} uses a bridge sample based online distillation protocol (BSBODP) for knowledge transfer between nodes across different tiers, enabling larger model training while maintaining privacy and flexibility.  Guidance-based knowledge transfer (GKT) \cite{yao2024gkt} employs a larger LLM as a ``teacher'' to generate guidance prompts for a smaller ``student'' model, enhancing accuracy and speed without fine-tuning.  
For example, in autonomous driving, a powerful cloud-based LLM can distill its complex decision-making logic into a smaller, faster model suitable for real-time navigation on an in-vehicle computer.
Growth-adaptive distillation \cite{yang2025growthadaptive} automatically adjusts the student model structure for efficient, low-power edge deployment.  In cloud-edge collaborative reinforcement learning, algorithm-independent knowledge distillation \cite{zeng2024offlinetransferonline} accelerates online RL agent convergence by leveraging pre-trained cloud models and offline agent interactions.  A federated bi-directional knowledge distillation strategy \cite{zhang2024cloud} enhances fault detection in manufacturing by integrating quality knowledge from both cloud and edge, strengthening interaction between sub-processes and hierarchical levels.

\subsection{Model Adaptation}\label{sec3.2}
\begin{figure}[t!]
  \centering
\includegraphics[width=0.49\textwidth]{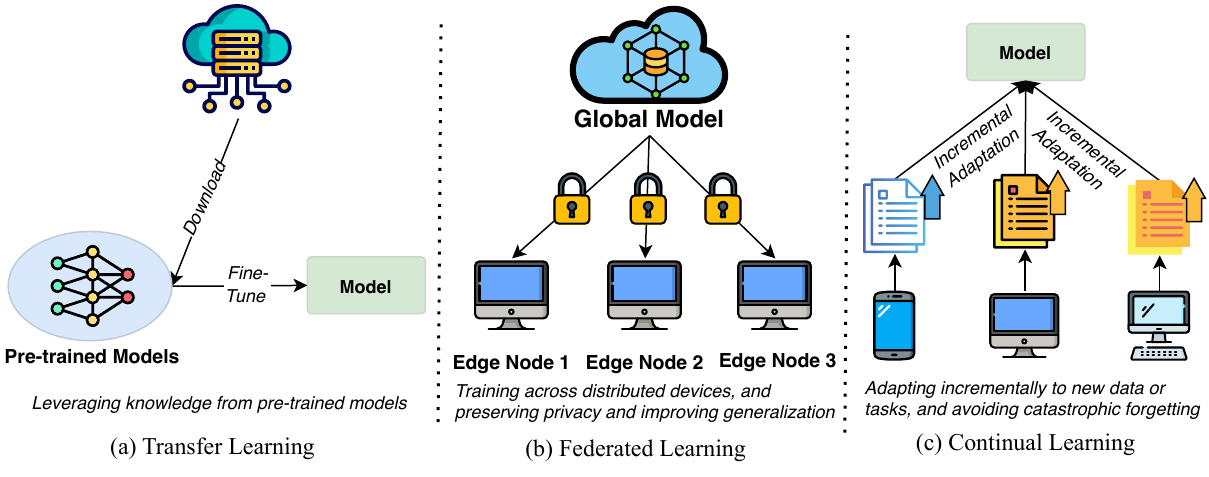}
  \caption{Illustration of model adaptation, detailing (a) transfer learning, (b) federated learning, and (c) continual learning.}
  \label{fig:4}
\vspace{-15px}
\end{figure}

Model adaptation enables AI models to respond to new tasks, datasets, or evolving data distributions in dynamic ECCC environments. Our taxonomy for model adaptation is organized based on the nature of the adaptation challenge being addressed. Accordingly, this leads to a classification into four key paradigms, as depicted in \autoref{fig:4}: (1) \textit{Transfer learning} (\autoref{sec3.2.1}) focuses on adapting knowledge from a source domain to a target domain, (2) \textit{Federated learning} (\autoref{sec3.2.2}) addresses adaptation across distributed, private datasets, (3) \textit{Continual learning} (\autoref{sec3.2.3}) handles adaptation to sequential data streams over time, and (4) \textit{Split learning} (\autoref{sec3.2.4}) facilitates adaptation by partitioning models across devices. Each paradigm offers a distinct approach to achieving model adaptability under the unique constraints of ECCC.

While some taxonomies treat federated learning as a distinct resource management or privacy technique, our classification positions FL within model adaptation because it fundamentally addresses how models adapt to distributed data patterns and heterogeneous environments, which is a core adaptation challenge in ECCC systems. More specifically, we distinguish three adaptation mechanisms: (1) \textit{Domain adaptation techniques} (transfer learning) enable models to adapt across different domains or tasks by leveraging pre-existing knowledge; (2) \textit{Distributed adaptation methods} (federated learning) enable models to adapt collaboratively across multiple data sources while preserving privacy and handling non-IID distributions; (3) \textit{Temporal adaptation approaches} (continual learning) enable models to adapt incrementally to evolving data and new tasks over time. Notably, this classification emphasizes the adaptation aspect of each technique while acknowledging that FL also serves privacy and resource distribution functions, making it inherently multi-faceted in ECCC systems. 

Each learning approach offers unique advantages for different edge-cloud deployment scenarios and data distribution characteristics. \textit{Transfer learning} excels in scenarios with limited target domain data by leveraging pre-trained models from related domains, achieving faster convergence and reduced computational requirements during training~\cite{li2024transfer}. However, transfer learning performance heavily depends on the similarity between source and target domains, potentially struggling with domain shifts or highly specialized tasks. \textit{Federated learning} provides exceptional privacy preservation by keeping data distributed across edge devices while enabling collaborative model training, making it ideal for scenarios with sensitive data or regulatory constraints~\cite{wu2024agglomerative}. Nevertheless, FL faces challenges with non-IID data distributions, communication overhead, and convergence difficulties in heterogeneous environments~\cite{qu2024blockchained}. \textit{Continual learning} offers superior adaptability to evolving data distributions and new tasks without requiring access to historical data, enabling lifelong learning capabilities crucial for dynamic edge environments~\cite{yang2024continual}. However, CL techniques often struggle with catastrophic forgetting and require sophisticated regularization mechanisms to maintain performance on previous tasks.

The choice of learning paradigm should align with specific data distribution characteristics: (1) \textit{Homogeneous data distributions}, transfer learning works well when source and target domains share similar characteristics, while federated learning benefits from consistent data patterns across participants; (2) \textit{Non-IID data distributions}, specialized FL techniques (e.g., FedProx, FedNova) address statistical heterogeneity, while transfer learning may require domain adaptation methods; (3) \textit{Temporal data shifts}, continual learning excels in adapting to evolving data patterns, while transfer learning may need periodic retraining; (4) \textit{Limited local data}, transfer learning provides quick adaptation with few samples, while FL can leverage collective knowledge across devices even with sparse local datasets. Notably, the optimal selection often involves hybrid approaches that combine multiple paradigms to address specific edge-cloud requirements and constraints.
Furthermore, \autoref{tab:3} presents a comparative summary of recent research efforts utilizing TL, FL, and CL, highlighting their specific approaches, application domains, and system characteristics within ECCC. Following this, the subsequent sections delve into the transfer learning in \autoref{sec3.2.1}, federated learning in \autoref{sec3.2.2}, and continual learning in \autoref{sec3.2.3}.
\input{t3_mod_ada_3.2.tex}

\subsubsection{Transfer Learning}\label{sec3.2.1}

Pre-trained model adaptation to new tasks or datasets leverages transfer learning principles to reduce training time and data requirements, thereby improving deployment efficiency in resource-constrained environments~\cite{li2024transfer}.  For example, in surface defect detection, a model pre-trained on a general image dataset can be fine-tuned with a smaller, specialized dataset to identify specific defects~\cite{li2024transfer}. Similarly, in cloud-device collaborative learning for multimodal large language models (MLLMs), techniques like adapter-based knowledge distillation (AKD) facilitate knowledge transfer from larger cloud models to smaller, compressed device models~\cite{wang2024clouddevice}, maintaining performance while minimizing computational burden.  In cloud-edge collaborative fault diagnosis, transfer learning addresses data heterogeneity and distribution shifts by fine-tuning source models trained on diverse user data to bridge distribution gaps and enable effective diagnosis on target datasets~\cite{wang2024cloudedgea}.

\subsubsection{Model Adaptation in Federated Learning}\label{sec3.2.2}
The explosive growth of FL research has made it an indispensable paradigm in distributed AI systems, driving critical need to explore how model adaptation mechanisms can be effectively implemented within federated frameworks \cite{zhou2024accelerating}. As one of the most active research domains in distributed machine learning, FL cannot be overlooked when examining model adaptation techniques, particularly given its unique challenges in handling heterogeneous data distributions and diverse edge environments \cite{khuencheng2024integration}. We focus specifically on model adaptation strategies within FL contexts, exploring how adaptive mechanisms enable models to effectively respond to distributed, non-IID data patterns while maintaining collaborative training capabilities \cite{loconte2024expanding}. Recent frameworks demonstrate sophisticated adaptation approaches, such as FedCAE \cite{yu2024fedcae} for machine fault diagnosis that adapts models to diverse industrial environments, and QuAsyncFL \cite{liu2024quasyncfl} for AIoT applications that enables adaptive synchronization strategies across heterogeneous devices. Advanced adaptation mechanisms like SFETEC \cite{le2024sfetec} optimize model adaptation for thing-edge-cloud environments by implementing adaptive model splitting strategies that dynamically adjust to varying communication constraints and device capabilities, addressing non-IID data challenges through intelligent partitioning. Security-aware adaptation frameworks such as FlexibleFL \cite{zhao2024flexiblefl} integrate adaptive evaluation mechanisms that assess uploaded parameters and participant contributions, enabling models to adapt against evolving poisoning attacks while maintaining collaborative learning effectiveness. Despite its privacy-preserving architecture, FL remains vulnerable to model privacy attacks such as membership inference, which are discussed in detail in \autoref{sec:supp-model-privacy-attacks}. Furthermore, adaptive participant selection and topology optimization strategies \cite{wei2024joint} enhance FL efficiency by dynamically adjusting collaboration patterns, while blockchain integration approaches \cite{qu2024blockchained} provide adaptive security mechanisms that evolve with emerging threats, demonstrating how model adaptation principles are fundamental to advancing FL systems in distributed edge-cloud environments.

\subsubsection{Continual Learning}\label{sec3.2.3}

Incremental adaptation strategies enable models to evolve with dynamic data and emerging tasks while addressing catastrophic forgetting challenges through lifelong learning capabilities \cite{yang2024continual}. For example, in a smart factory, a continual learning model for predictive maintenance can adapt to new machine behaviors and fault patterns over time without being completely retrained, ensuring the system remains accurate as equipment ages.  Several strategies, including regularization, replay, and dynamic architectures, mitigate this issue, ensuring robust and adaptable systems.  Specifically, continual learning enables digital twins to synchronize with their evolving physical counterparts and sensor data in edge computing, providing accurate predictions \cite{li2024digital,li2024conditionadaptive}.  For instance, in intelligent transportation systems (ITS), cloud-edge collaborative continual adaptation frameworks, leveraging visual prompts and knowledge distillation, enhance the generalization of lightweight edge models for object detection in dynamic environments \cite{lian2024cloudedge}.  Similarly, online management frameworks, such as EdgeC3, optimize model aggregation and dynamic offloading for continuous data streams in edge-cloud collaborative continual learning \cite{lin2024online}. Cross-edge federated continual learning (Cross-FCL) frameworks, employing parameter decomposition and various cross-edge strategies, address knowledge retention across devices in multiple FL systems, balancing memory and adaptation for new tasks \cite{zhang2024crossfcl}.

\subsubsection{Split Learning}\label{sec3.2.4}

Distributed learning paradigms partition neural network models across multiple computational nodes to enable collaborative training without sharing raw data, addressing computational limitations while preserving privacy where early layers typically run on edge devices for low-latency feature extraction and deeper layers are offloaded to servers \cite{gao2020endend,lin2024split}. Strategic architectural division effectively addresses device limitations and privacy, as seen in federated split learning (FSL) for smart cities \cite{ahmed2024heterogeneous} and MADRL-based partitioning systems that optimize cloud-edge collaboration \cite{fan2025madrlbased}. Furthermore, selecting optimal split points minimizes latency and communication overhead, with frameworks like EPSL \cite{lin2024efficient} and SFETEC \cite{le2024sfetec} demonstrating efficiency in wireless and thing-edge-cloud environments respectively. Moreover, specialized architectures address incremental learning challenges \cite{kim2021splitbridge}, while implementations like CSE-FSL \cite{mu2025federated} and FedSL \cite{ni2024fedsl} balance communication efficiency with model performance, enabling secure participation of low-end devices in diverse 6G networks \cite{lin2024split}.

\subsection{Resource-Aware Model Optimization}\label{sec3.3}

Resource-aware model optimization focuses on dynamically adapting AI models to the heterogeneous and constrained computational environments found in ECCC systems. Building upon foundational techniques discussed in prior surveys~\cite{qu2025mobile}, we classify advanced resource-aware strategies into three primary approaches based on their operational focus. \textit{Dynamic model selection} adapts to resource availability by choosing the most appropriate model from a pool of candidates at runtime. \textit{Model splitting and offloading} partitions model components across the edge-cloud infrastructure to balance latency and computational load. \textit{Reinforcement learning-based optimization} uses learning-based methods for adaptive resource allocation and decision-making. Working together, these strategies collectively address the challenges of resource heterogeneity and dynamic operational constraints. For instance, in AR applications, model splitting can offload scene recognition to an edge server while keeping real-time pose rendering on the local device to ensure a fluid user experience~\cite{hua2024energyefficient}. Similarly, reinforcement learning can optimize resource allocation in IoT monitoring applications to improve stability and reduce failure rates~\cite{xu2024fusion,goudarzi2024mmddrl}. For comprehensive technical details on each approach, please see \autoref{sec:supp-resource-aware} of the Supplementary Material.

\subsection{Neural Architecture Search}\label{sec3.4}
NAS automates the design of efficient neural networks, a critical task for creating models tailored to the specific constraints of ECCC. In line with established surveys, we categorize NAS techniques based on their underlying search strategy. Building upon this foundation, the three dominant paradigms are: reinforcement learning-based search, where an agent learns optimal architectures through environmental feedback; evolutionary algorithm-based search, which mimics natural selection to evolve populations of architectures; and gradient-based search, which uses gradient descent for efficient optimization of a continuous architectural representation~\cite{goudarzi2023distributed,he2024efficient,zhang2024manyobjective}. Each of these paradigms offers distinct advantages for discovering architectures that balance performance with the resource limitations of edge devices. For a detailed technical analysis of each search strategy, please see \autoref{sec:supp-resource-nas} of the Supplementary Material.

\subsection{Lessons Learned on Model Optimization}\label{sec3.5}
Our analysis of model optimization techniques for ECCC systems reveals several critical insights. A recurring theme is the inherent trade-off between performance and efficiency, which manifests differently across various optimization strategies. Key lessons learned include: (1) Model compression techniques, such as pruning, quantization, and knowledge distillation, effectively reduce model complexity but often incur accuracy degradation, with the optimal balance being highly model- and task-dependent. (2) Model adaptation methods like federated and continual learning address data privacy and model evolution but introduce significant communication bottlenecks and are sensitive to non-IID data distributions, which remains a primary research obstacle. (3) Neural architecture search automates the design of efficient models, yet its substantial computational overhead limits direct application in resource-constrained edge scenarios, thereby motivating the development of lightweight and one-shot NAS methods. (4) A holistic optimization strategy must dynamically consider device capabilities, network conditions, and application requirements, where dynamic model selection and partitioning are crucial for achieving resource-aware optimization.
Notably, the model optimization techniques detailed in this section primarily address ``design-time'' challenges, creating AI models that are inherently smaller, faster, and more adaptable. Accordingly, an efficiently optimized model fundamentally simplifies the subsequent ``run-time'' challenges of task offloading and resource allocation, whereby the AI-driven resource management strategies discussed in \autoref{sec4} are specifically designed to effectively deploy and run the AI models optimized in this section, thus serving as a critical foundation for translating model-level improvements into system-wide performance gains.

\section{AI-Driven Resource Management}\label{sec4}
\input{taxo-sec4-resource-mgmt.tex}
Building upon the foundation of optimized AI models from \autoref{sec3}, this section investigates the critical challenge of ``run-time" resource management. Specifically, while model optimization prepares AI models for efficient execution, resource management strategies govern their actual deployment, scheduling, and orchestration within the dynamic and heterogeneous ECCC environment. To provide a structured overview of this domain, \autoref{taxo:sec4} presents a hierarchical taxonomy of AI-driven resource management, organized into three primary branches. First, task offloading mechanisms (\autoref{sec4.1}) determine optimal task placement. Second, resource allocation strategies (\autoref{sec4.2}) manage the distribution of computational resources. Third, energy efficiency optimization (\autoref{sec4.3}) focuses on minimizing power consumption. Accordingly, the following subsections are structured according to this taxonomy, exploring how approaches within each category enable the full potential of optimized models to be realized, thereby ensuring robust and efficient system performance.

\subsection{Task Offloading Mechanisms}\label{sec4.1}
\begin{figure}[t!]
  \centering
\includegraphics[width=0.49\textwidth]{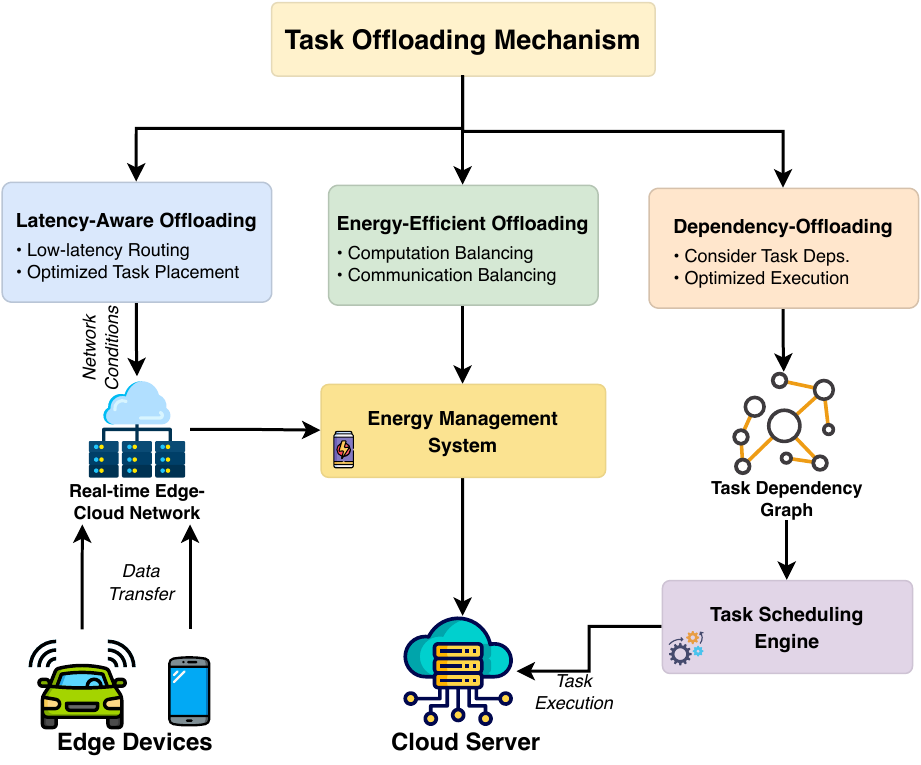}
\caption{Task offloading mechanisms, including latency-aware, energy-efficient, and dependency-aware offloading strategies.}
  \label{fig:6}
\vspace{-15px}
\end{figure}
\input{t4_task_4.1.tex}

Effective resource utilization and performance optimization in ECCC depend critically on sophisticated task offloading mechanisms. As depicted in \autoref{fig:6}, a framework outlines three core strategies: latency-aware, energy-efficient, and dependency-aware offloading, detailing their interactions with edge devices, cloud servers, and supporting systems to improve task placement and scheduling. Specifically, these mechanisms address distinct objectives such as real-time performance, energy consumption, and task dependencies, thereby boosting overall system efficiency. To further illustrate their practical impact, \autoref{tab:4} summarizes recent task offloading methods in edge-cloud environments. 
Notably, the table reveals significant performance gains, such as up to a 33.42× reduction in delay for sensitive workloads \cite{zhai2024edgecloud}, alongside improvements in energy consumption and system reliability, highlighting the versatility of these strategies across diverse application domains.

\subsubsection{Latency-Aware Offloading}\label{sec4.1.1}
Minimizing response time through latency-aware offloading represents a crucial strategy for time-sensitive applications in edge-cloud systems, with numerous approaches demonstrating significant performance gains. According to the comparative analysis in \autoref{tab:4}, methods like iGATS~\cite{sun2025latencyaware} and DRLIS~\cite{wang2024deep} have achieved substantial reductions in processing delays through intelligent placement decisions. For instance, Lyapunov optimization techniques employed by Abouaomar et al.~\cite{abouaomar2021resource} enable dynamic resource allocation at the edge, while Poularakis et al.~\cite{poularakis2019joint} address multidimensional constraints through joint optimization of service placement and request routing. Similarly, Fan et al.~\cite{fan2024collaborative} propose a collaborative framework that minimizes total processing delay while ensuring queuing stability in multi-task scenarios. When considering delay-sensitive workloads specifically, Zhai et al.~\cite{zhai2024edgecloud} demonstrate a remarkable 33.42× reduction in latency compared to traditional cloud deployments by intelligently distributing tasks based on their characteristics. Additionally, the double deep Q-network (DDQN) approach introduced by Pang et al.~\cite{pang2024multimobile} effectively manages mobility while optimizing offloading decisions, while Qi et al.~\cite{qi2024service} present a two-timescale scheduling algorithm that balances service provider costs with stringent latency requirements.
Furthermore, the local computation delay of a task is not static; it can be significantly reduced through the model compression techniques detailed in \autoref{sec3.1}, such as pruning and quantization, thereby making local execution more feasible.

\subsubsection{Energy-Efficient Offloading}\label{sec4.1.2} 

Strategic energy management across edge-cloud systems requires balancing computation and communication costs while optimizing resource allocation through specialized techniques. For instance, in mobile AR, offloading computationally intensive rendering tasks from a headset to a nearby edge server can significantly conserve battery life, extending the device's operational time while maintaining an interactive experience.
In ECCC, prolonging the battery life of edge devices necessitates balancing computation and communication energy costs, especially for battery-powered devices \cite{hu2022energy}. As shown in \autoref{tab:4}, numerous approaches including PDCO~\cite{su2024primaldualbased}, Hua et al.~\cite{hua2024energyefficient}, and ETCRA~\cite{li2024energyaware} have demonstrated significant improvements in energy efficiency across edge-cloud systems. Particularly for battery-powered devices, strategic decision-making has to consider both computational and communication energy costs, weighing factors like task complexity, network conditions, and hardware energy profiles~\cite{hu2022energy,he2019collaborative}. Notably, the choice of model architecture has a decisive impact on computational energy consumption, a key optimization target for the neural architecture search (NAS) techniques discussed in \autoref{sec3.4}. While offloading computationally intensive tasks to cloud resources can reduce device energy consumption, communication costs under limited bandwidth or unstable networks often counterbalance these benefits, necessitating sophisticated optimization approaches. Beyond basic consumption metrics, advanced energy-efficient mechanisms consider additional dimensions such as power control, transmission scheduling~\cite{hua2024energyefficient}, and user mobility patterns~\cite{arroba2024sustainable}. Furthermore, in heterogeneous IoT deployments with diverse device capabilities and constraints, efficient resource allocation becomes increasingly crucial for sustainable delivery~\cite{chen2024fast}, with some approaches even incorporating renewable energy sources to create more environmentally responsible edge-cloud ecosystems~\cite{zhai2024edgecloud}.

Energy-efficient offloading mechanisms strive to minimize overall energy consumption by jointly considering computation and communication costs.  Additionally, optimizing power control and transmission scheduling can further enhance energy efficiency \cite{hua2024energyefficient}.  User mobility introduces another layer of complexity to energy optimization, as dynamic network conditions and varying proximity to edge servers impact energy trade-offs \cite{arroba2024sustainable}.  In heterogeneous IoT environments with diverse device capabilities and energy constraints, efficient resource allocation becomes even more crucial for energy-efficient service delivery \cite{chen2024fast}.  Prioritizing workloads based on their energy requirements and leveraging renewable energy sources can contribute to a more sustainable edge-cloud ecosystem \cite{zhai2024edgecloud}.

\subsubsection{Dependency-Aware Offloading}\label{sec4.1.3}
Complex applications often consist of multiple interconnected subtasks with dependencies, forming a dependency graph where some subtasks must finish before others can start \cite{feng2024dependencyaware}.  Complex interdependence introduces complexities in task scheduling and resource allocation, potentially leading to suboptimal performance due to idle resources and increased latency if ignored.  Therefore, efficient offloading mechanisms must consider these precedence constraints to optimize overall application execution. For example, the PRAISE algorithm in \cite{feng2024dependencyaware} addresses triangular dependencies by reconfiguring the task call graph into sequential layers for efficient resource allocation.  Similarly, modeling task dependencies as a directed acyclic graph (DAG) optimizes offloading decisions, particularly in vehicle-edge-cloud environments \cite{pang2024multimobile}.  Additionally, deep Q-networks (DQN) for real-time offloading in cloud-edge computing benefit from incorporating dependency awareness \cite{chen2024realtime}, enabling dynamic adaptation to changing environments without pre-set task priorities.

\subsubsection{Multi-Objective Optimization Strategies}\label{sec4.1.4}
Contemporary edge-cloud systems increasingly demand sophisticated offloading strategies that simultaneously optimize conflicting performance metrics like latency, energy consumption, and reliability through Pareto optimization methods \cite{zhang2024manyobjective,jain2022latencymemory}. Approaches like the many-objective resource scheduling model proposed by Zhang et al. \cite{zhang2024manyobjective} effectively navigate complex solution spaces considering time, cost, and load balance, whilst other frameworks frame service deployment as high-dimensional problems incorporating network reliability \cite{guo2023edge}. Additionally, joint optimization techniques combining energy efficiency with latency minimization prove effective in battery-powered environments \cite{hua2024energyefficient}, and integrating security considerations addresses trust-sensitive application demands \cite{li2024energyaware}, highlighting the critical need for efficient algorithms capable of handling multi-dimensional constraints in real-time.

\subsection{Resource Allocation Strategies}\label{sec4.2}
\input{t5_res_alloc_4.2.tex}

Strategic allocation of computational resources across the edge-cloud continuum remains fundamental for maximizing performance in AI-driven collaborative environments. Accordingly, we explore three pivotal strategies: dynamic resource allocation, collaborative resource management, and reinforcement learning-based optimization, each designed to enhance task execution and system efficiency. By effectively distributing resources, advanced allocation approaches address the challenges of varying workloads and resource constraints in distributed environments. To illustrate their practical impact, \autoref{tab:5} provides a comparative analysis of recent dynamic resource allocation methods in ECCC systems. 
Notably, the table reveals a wide range of application domains, from IoT networks to vehicular edge clouds, alongside key performance metrics like latency, energy consumption, and quality of service, highlighting the adaptability and effectiveness of these strategies in diverse scenarios.

Resource allocation strategies in ECCC systems can be systematically organized according to their adaptation mechanisms and coordination approaches: (1) \textit{Temporal adaptation strategies} (dynamic resource allocation) that respond to time-varying workloads and resource availability through real-time monitoring and reactive adjustment mechanisms; (2) \textit{Spatial coordination strategies} (collaborative resource management) that optimize resource distribution across geographically distributed edge nodes and cloud infrastructure through inter-node cooperation and hierarchical coordination; (3) \textit{Learning-based optimization strategies} (reinforcement learning approaches) that continuously improve resource allocation decisions through environmental feedback and experience accumulation. Accordingly, the classification framework emphasizes the complementary nature of these approaches, where temporal adaptation handles immediate demand fluctuations, spatial coordination optimizes cross-infrastructure efficiency, and learning-based methods provide long-term performance improvement through adaptive policy refinement.

\subsubsection{Dynamic Resource Allocation}\label{sec4.2.1}
Adaptive resource management addresses fluctuating workloads and optimizes resource utilization through real-time monitoring and adjustment mechanisms, crucial for handling unpredictable application demands \cite{farhoudi2023qosaware}.  Approaches include dynamic NUMA node selection and core consolidation \cite{bensalah2024vnf}, optimizing service placement across the edge-cloud continuum by considering factors like delay and resource constraints.  Dynamic task offloading mechanisms further optimize task placement and scheduling based on latency, energy consumption, and task dependencies \cite{chen2024dynamica}.

In drone communication networks, dynamic resource allocation considers drone mobility, communication loads, and latency constraints for efficient real-time applications \cite{das2024edgecloud}.  Similarly, for cloud-edge-terminal IoT networks, collaborative policy learning using federated reinforcement learning (FRL) enables dynamic task scheduling, adapting to changing workload demands \cite{kim2024collaborative}.  Distributed serverless edge clouds utilize dynamic split computing frameworks to minimize inference latency while adhering to resource constraints by dynamically determining splitting points and maintaining container instance warm status \cite{ko2024dynamic}.
Deep reinforcement learning techniques optimize task offloading decisions in 5G edge-cloud environments, maximizing quality-of-experience (QoE) while meeting latency requirements \cite{nieto2024deep}.  Additionally, multi-target-aware dynamic resource scheduling algorithms in multi-tier computing networks consider diverse QoS requirements, including computing, storage, bandwidth, and delay, to enhance resource flexibility and efficiency \cite{zhang2024multitargetaware}.

\subsubsection{Collaborative Resource Management}\label{sec4.2.2}
Coordinated resource allocation across distributed infrastructure requires sophisticated management strategies that account for heterogeneous device capabilities and dynamic workload demands. For instance, in IoT-based distribution grids, resource optimization across cloud, edge, and end devices spans multiple timescales, with decisions at each level influencing others \cite{wang2024cloudedgeend}, necessitating sophisticated, interdependent resource management strategies.  Dynamic resource allocation is also crucial for adapting to fluctuating demands and optimizing real-time performance \cite{wang2024cloudedgeend}.

One approach leverages optimal transport (OT) and FL principles.  Specifically, an OT-based offline optimization model can initially allocate bandwidth and computation resources, followed by online optimization using federated actor-critic learning (OTFAC) \cite{gan2024optimal}.  In such dual-driven approach, edge servers can act as local aggregators within the federated learning process, improving efficiency and reducing delays \cite{moraes2019pub}.  Additionally, collaborative resource management benefits from a hybrid approach where edge servers not only collaborate with the cloud but also expand their local resources \cite{wang2024efficient}.  Such strategies allow edge servers to dynamically purchase cloud resources and expand local capacity for long-term cost optimization \cite{wang2024efficient}. Consequently, such adaptability manages dynamic workloads and varying demands.  Edge-cloud and edge-edge cooperation further enhance performance through efficient network-wide resource utilization \cite{fan2024collaborative}.

\subsubsection{Reinforcement Learning for Resource Management}\label{sec4.2.3}
Adaptive optimization mechanisms leverage reinforcement learning to optimize resource allocation by learning strategies through environmental feedback and performance-enhancing actions \cite{binh2024reinforcement}.  Specifically, these agents adapt to changing workloads and resource availability, leading to more efficient resource utilization. For example, RL optimizes task offloading decisions in vehicular edge-cloud computing (VECC) to minimize latency while maintaining quality of service \cite{binh2024reinforcement}, and addresses online dependent task offloading (ODTO) in collaborative edge computing by jointly optimizing network flow scheduling and resource allocation \cite{chen2024dynamica}.  In addition, RL dynamically adjusts model parameters and offloading decisions, improving the adaptability and performance of edge-cloud systems \cite{huang2024joint}.  Researchers have demonstrated RL's effectiveness in various scenarios, including optimizing microservice placement \cite{afachao2024efficient}, managing resources for delay-sensitive tasks \cite{binh2024reinforcement}, and handling dependent and parallel tasks \cite{chen2024realtime}.  Integrating RL with techniques like graph neural networks (GNNs) \cite{li2024offloading} and deep Q-networks (DQNs) \cite{chen2024realtime} further enhances the agents' ability to capture complex relationships and optimize offloading decisions.  Multi-agent RL extends capabilities by enabling decentralized learning and coordinated resource management across multiple servers \cite{lu2024a2cdrl}.

\subsection{Energy Efficiency Optimization}\label{sec4.3}

Sustainable power management across edge-cloud environments requires integrated optimization strategies balancing energy consumption with performance through comprehensive modeling and adaptive scheduling mechanisms~\cite{georgiou2021comprehensive,yao2023edgecloud}. Energy-aware scheduling algorithms deliver dynamic adaptation via techniques like DVFS and intelligent sleep-wake management, with DVFO achieving 40\% energy savings through co-optimization while Co-HDRL demonstrates 50-70\% power reductions in IoT scenarios~\cite{zhang2024dvfo,zhou2023cooperative}. Furthermore, joint optimization addresses fundamental trade-offs between efficiency and competing metrics through multi-objective approaches employing Pareto methods to balance conflicting goals~\cite{zhang2024manyobjective,hua2024energyefficient}. Consequently, advanced strategies demonstrate substantial impact, with green computing achieving 30-45\% reductions in data center consumption through renewable integration and intelligent load balancing~\cite{arroba2024sustainable,chen2024fast}. Comprehensive analysis is detailed in \autoref{sec:supp-energy} of the Supplementary Material.

\subsection{Lessons Learned on Resource Management}\label{sec4.4}
The review of AI-driven resource management strategies underscores a clear shift towards data-driven, adaptive control mechanisms to handle the complexity of ECCC systems. Several key lessons emerge from this analysis. Firstly, reinforcement learning-based methods, while promising for long-term performance optimization, face significant challenges in practical deployment due to high training costs, slow convergence, and a strong dependency on accurate environmental models, particularly in highly dynamic edge environments. Secondly, defining an ``optimal'' offloading decision is inherently a multi-objective optimization problem that requires balancing conflicting goals such as latency, energy consumption, and cost; the formulation and modeling of this problem are complex undertakings in themselves. Thirdly, traditional model-based optimization techniques often fall short in the face of unpredictable and rapidly changing edge conditions. Consequently, data-driven AI approaches are becoming the dominant paradigm for enabling intelligent and adaptive resource management across the edge-cloud continuum.

\section{Privacy and Security Enhancement}\label{sec5}
\input{taxo-sec5-privacy-security.tex}
To provide a structured overview, \autoref{taxo:sec5} presents a hierarchical taxonomy of privacy and security enhancement for ECCC systems, organized into three primary branches. First, data privacy protection (\autoref{sec5.1}) encompasses techniques that safeguard sensitive information. Second, model security (\autoref{sec5.2}) addresses threats to AI models themselves. Third, communication security (\autoref{sec5.3}) secures data transmission across the edge-cloud continuum. Accordingly, the following subsections are structured according to this taxonomy.

Privacy and security mechanisms in ECCC systems form an interconnected protection ecosystem where techniques often complement and build upon each other rather than operating in isolation. Our classification organizes these mechanisms according to their protection scope and technical approach: (1) \textit{Data-centric privacy protection} (\autoref{sec5.1}) encompassing techniques that protect raw data and sensitive information throughout the distributed learning process, including privacy-preserving learning, secure multi-party computation, differential privacy, and data anonymization; (2) \textit{Model-centric security protection} (\autoref{sec5.2}) addressing threats to AI models themselves, including integrity protection, adversarial attack mitigation, secure updates, and trust management; (3) \textit{Communication-centric security} (\autoref{sec5.3}) securing data transmission and inter-node communication across the edge-cloud continuum. 

It is important to note that many privacy and security techniques are inherently non-orthogonal and often complement each other in practical deployments. For example, privacy-preserving learning methods frequently incorporate multiple approaches: differential privacy for statistical privacy guarantees, secure multi-party computation for cryptographic protection, and data anonymization for additional privacy layers. Similarly, federated learning serves dual purposes as both a model adaptation technique (\autoref{sec3.2}) and a privacy protection mechanism, demonstrating the multi-faceted nature of techniques in ECCC systems. Understanding these interdependencies is crucial for designing comprehensive security frameworks that leverage synergies between different protection mechanisms while avoiding redundant overhead.
\subsection{Data Privacy Protection}\label{sec5.1}

Comprehensive protection mechanisms safeguard sensitive information throughout its lifecycle in collaborative computing paradigms, addressing critical challenges of data security within distributed environments. We begin by examining privacy-preserving learning methods, such as federated learning and secure aggregation, enabling collaborative model training without direct data sharing.  Subsequently, we delve into secure multi-party computation (SMPC), which allows joint function computation on private inputs.  Additionally, differential privacy techniques, adding noise to protect individual data points while preserving aggregate insights, are considered.  Finally, data anonymization approaches, transforming data to prevent individual identification, are explored.

\subsubsection{Privacy-preserving learning methods}\label{sec5.1.1}

Privacy enhancement in federated learning leverages techniques including differential privacy (DP) and secure aggregation to protect sensitive information \cite{gilbert2022secure}.  DP adds carefully calibrated noise to model updates before aggregation, making it difficult to infer individual data points \cite{li2019asynchronous}.  Meanwhile, secure aggregation uses cryptographic protocols like multi-party computation (MPC) to aggregate updates without revealing individual contributions to the server \cite{stevens2022efficient}.  Additionally, zero-concentrated differential privacy (zCDP) offers tighter privacy guarantees than traditional DP in FL \cite{hu2020concentrated}, and CaPC learning combines MPC, homomorphic encryption, and privately aggregated teacher models to achieve both confidentiality and privacy \cite{choquette-choo2021capc}.  Lightweight encryption and aggregation techniques further enhance security and efficiency in federated learning systems \cite{zheng2023aggregation}.

\subsubsection{Secure Multi-Party Computation}\label{sec5.1.2}
Collaborative computation frameworks enable multiple parties to jointly compute functions over private inputs without revealing sensitive data through secure multi-party computation protocols~\cite{breuer2021introducing}.  SMPC protocols ensure each party learns only the function's output, safeguarding data privacy and security, crucial for edge-cloud collaborative AI model training.  For example, multiple hospitals could train a joint model on patient data without sharing raw data due to privacy regulations~\cite{choquette-choo2021capc}.  Additionally, SMPC can decompose complex AI tasks into smaller sub-tasks distributed across edge devices and the cloud, improving training efficiency~\cite{feng2023smpc}.  Robust management frameworks, such as FlexSMC~\cite{vonmaltitz2018management}, address the dynamic edge-cloud environment by facilitating node discovery, establishing trust, and ensuring robustness against failures. Coded computation within SMPC, as in coded-MPC (CMPC)~\cite{vedadi2024efficient}, further enhances performance through resource optimization, with adaptive gap entangled (AGE) polynomial codes~\cite{vedadi2022adaptive} as a promising approach.  Ensuring security against malicious attacks during MPC-enabled privacy-preserving training is also addressed by approaches like the one in~\cite{liu2021mpcenabled}, which provides active security even with a majority of malicious parties. Systems like Helen~\cite{zheng2019helen} demonstrate practical applications of maliciously secure cooperative learning for linear models.

\subsubsection{Differential Privacy Techniques}\label{sec5.1.3}

Privacy-preserving analysis frameworks offer robust protection for individual privacy by introducing carefully calibrated noise into data or computations \cite{zhu2022more}, enabling meaningful insights extraction while safeguarding sensitive information. Specifically, DP mechanisms ensure negligible impact on analysis outputs from the presence or absence of a single individual's data, preventing identification or inference of sensitive information \cite{gupta2022privacypreserving,demelius2025recent}.  Common approaches include adding noise directly to raw data before analysis or perturbing the results of computations, such as model parameters, with the noise calibrated based on the computation's sensitivity \cite{gupta2022privacypreserving,demelius2025recent}.  For instance, in federated learning, DP protects individual device data during collaborative training by adding noise to local model updates before aggregation, preventing private data reconstruction from shared updates \cite{wang2022privacypreserving}.  Similarly, in collaborative inference, DP can protect intermediate outputs and mitigate membership information leakage \cite{duan2023privascissors}.  While noise addition inevitably impacts data utility and model accuracy \cite{naidu2021quantifying}, a key challenge is balancing privacy and utility, especially given the heterogeneous nature of edge devices and distributed edge-cloud systems \cite{tran2021differentially,chen2024edgeleakage}.

\subsubsection{Data Anonymization Approaches}\label{sec5.1.4}
Identity protection techniques transform data to prevent individual identification while enabling collaborative AI development, particularly crucial in distributed processing environments with unique privacy challenges \cite{dhinakaran2024privacypreserving}.  Several approaches achieve this goal, including StyleID, which anonymizes faces in image datasets through identity disentanglement \cite{le2022styleid}. Specifically, StyleID projects images onto a GAN latent space, manipulating specific features to protect identities while preserving other image characteristics. Another key technique, DP, adds noise to data or computations to protect individual privacy while preserving useful insights \cite{yang2024differentially}. For instance, differentially private federated tensor completion enables secure cloud-edge collaborative AIoT data prediction \cite{yang2024differentially}.  Dynamic anonymization techniques, often leveraging machine learning, adapt to specific contexts and data characteristics for more flexible and effective privacy protection \cite{dhinakaran2024privacypreserving}. Data anonymization in edge-cloud collaborative AI systems, where data may be distributed across multiple devices and locations, can be combined with other privacy-preserving techniques.  Complementary approaches include SMPC and federated learning to create a more comprehensive privacy framework \cite{yang2024differentially}.  As an example, in cloud-edge collaborative inference, anonymization complements methods like PrivaScissors, which reduces mutual information leakage between intermediate model outputs and device data \cite{duan2023privascissors}.

\subsection{Model Security}\label{sec5.2}
Comprehensive security frameworks address critical vulnerabilities throughout AI model lifecycles via multi-layered protection mechanisms against unauthorized modifications, adversarial attacks, model poisoning, and model privacy attacks such as membership inference and model extraction~\cite{li2024nextgeneration,yang2024differentially}. Robust protection encompasses integrity mechanisms utilizing cryptographic hashing and trusted execution environments, attack detection through adversarial training and anomaly detection, and secure update protocols implementing authentication and encrypted transmission~\cite{alaverdyan2023confidential,chi2024trusted,fan2024autoupdating}. Advanced approaches including FlexibleFL demonstrate effective federated learning defense through contribution-based mitigation, while blockchain-based solutions provide enhanced protection against coordinated attacks~\cite{zhao2024flexiblefl,fang2024secure,liu2024blockchainempowered}. Additionally, runtime monitoring and auto-updating mechanisms enable adaptive responses to emerging threats while maintaining system resilience against novel attack patterns~\cite{yao2024efficient,zheng2023kubeedgesedna}. Detailed technical analysis of integrity protection, attack detection, model privacy attacks, secure updates, and trust management is provided in \autoref{sec:supp-model-security} of the Supplementary Material.

\subsection{Communication Security}\label{sec5.3}

Secure data exchange across heterogeneous infrastructures requires comprehensive protection through integrated encryption, authentication, and access control mechanisms addressing distributed edge-cloud architectural challenges~\cite{alwarafy2021survey,gupta2022secure}. Adaptive security strategies balance protection strength with computational efficiency through frameworks integrating symmetric encryption for large-scale transfers, asymmetric encryption for secure key exchange, and TLS/VPN protocols establishing protected channels~\cite{zobaed2022aidriven,zhang2021steciot}. Authentication incorporates multi-factor authentication, role-based access control, and hierarchical privilege management preventing unauthorized access while maintaining operational efficiency~\cite{al-aqrabi2018scalable,xiao2024domainspecific}. Furthermore, encryption schemes leverage attribute-based approaches for fine-grained access control, while blockchain-based solutions provide decentralized trust management and immutable audit trails~\cite{yao2024efficient,ri2022blockchainbased,queralta2021blockchain}. Coverage of secure transmission protocols, authentication mechanisms, encryption schemes, and blockchain solutions is detailed in \autoref{sec:supp-communication-security} of the Supplementary Material.

\section{Applications}\label{sec6}

ECCC enables transformative capabilities across diverse application domains by leveraging the unique synergy between edge proximity and cloud resources. Unlike traditional centralized or purely edge-based approaches, ECCC provides distinctive advantages through: (1) \textit{Hierarchical intelligence deployment}, sophisticated AI models in the cloud guide lightweight edge models for real-time decision-making; (2) \textit{Dynamic workload distribution}, computational tasks are intelligently partitioned based on latency requirements, privacy constraints, and resource availability; (3) \textit{Collaborative learning}, federated and continual learning approaches enable privacy-preserving model improvement across distributed deployments; (4) \textit{Adaptive resource orchestration}, dynamic scaling between edge and cloud resources based on real-time demands and network conditions. Subsequently, the following subsections demonstrate how these ECCC-specific capabilities create unprecedented opportunities for innovation across autonomous driving, smart healthcare, industrial automation, and smart cities, highlighting applications that are either impossible or significantly limited without the collaborative edge-cloud paradigm.
\subsection{Autonomous Driving}\label{sec6.1}

Real-time processing of complex sensor data is crucial for safe and efficient autonomous driving.  Edge-cloud collaboration offers a promising solution by distributing computational workloads between resource-constrained in-vehicle edge devices and powerful cloud servers.  Specifically, edge devices can handle time-sensitive tasks like object detection and perception, potentially leveraging lightweight frameworks like Edge YOLO \cite{liang2022edge}, which combines pruned and compressed networks for efficient multi-scale prediction on resource-limited platforms.  In addition, the collaborative nature of these systems enables integration of crowdsourced data, enhancing perception capabilities by providing a more comprehensive environmental view, as demonstrated by LiveMap \cite{liu2021livemap,liu2023realtime}, which efficiently processes and integrates multi-vehicle data for real-time dynamic map generation.  Beyond perception, LLMs are emerging as powerful tools for motion planning and decision-making \cite{chen2024edgecloud}.  For example, systems like EC-Drive utilize LLMs across both cloud and edge, adapting to data drifts and optimizing communication for reduced latency.  Similarly, collaborative policy learning through DRL \cite{liu2021livemap} offers a mechanism for dynamic scheduling and resource allocation, optimizing vehicle scheduling and offloading decisions for improved system robustness and adaptability.  Adaptive edge-cloud collaboration, aided by large vision models like in LAECIPS \cite{hu2024laecips}, further enhances performance and adaptability of IoT-based perception systems through efficient hard input mining.  Concurrently, the fusion of DRL and edge computing facilitates real-time monitoring and control optimization in IoT environments \cite{xu2024fusion}, enabling predictive system state analysis and dynamic resource allocation for enhanced control.

\subsection{Smart Healthcare}\label{sec6.2}
\begin{figure}[t!]
  \centering
\includegraphics[width=0.35\textwidth]{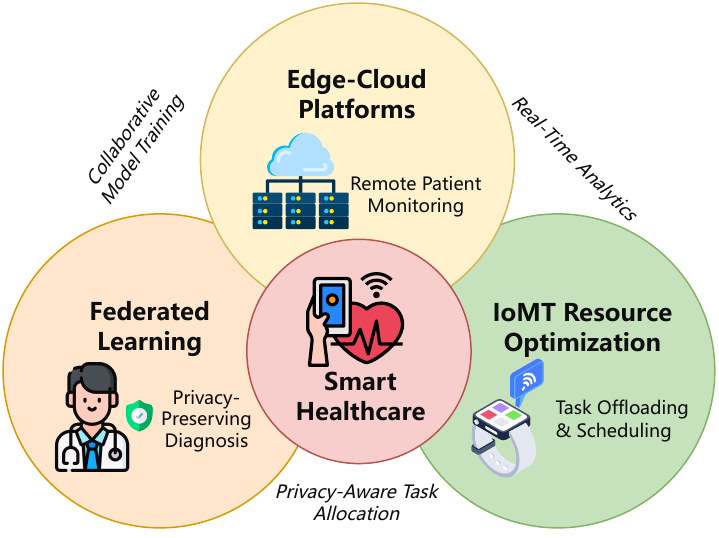}
  \caption{Integration of federated learning, edge-cloud platforms, and IoMT optimization in smart healthcare.}
  \label{fig:8}
\vspace{-15px}
\end{figure}

AI-driven collaborative computing transforms smart healthcare systems into efficient, secure, and personalized ecosystems, as shown in Fig. \ref{fig:8}. We explore how AI-driven edge-cloud collaboration revolutionizes healthcare delivery and improves patient outcomes by investigating FL for medical image analysis, edge-cloud platforms for remote patient monitoring, and resource allocation optimization for Internet of medical things (IoMT) devices.  FL offers a promising technique for training AI models on distributed medical datasets without compromising patient privacy, allowing collaborative model training across multiple healthcare providers while keeping sensitive data localized \cite{shaik2023remote}.  Specifically, FL enables robust diagnostic model development, including COVID-19 diagnosis using multimodal data like X-rays and ultrasound images \cite{qayyum2022collaborative}.  Edge-cloud platforms are crucial for enabling remote patient monitoring and personalized healthcare \cite{wu2020personalized}, leveraging edge device computation and cloud storage for real-time monitoring and personalized interventions.  For example, wearable sensors and IoMT devices collect patient data, process it at the edge for immediate feedback, and transmit it to the cloud for analysis and storage \cite{kanduri2024edgecentric}. AI algorithms deployed on edge devices analyze this data, providing personalized recommendations and improving intervention effectiveness \cite{shaik2022fedstack}.  Efficient resource allocation is also essential for optimizing IoMT device performance and ensuring timely healthcare delivery \cite{nguyen2022federated}. Consequently, AI-driven task offloading and resource management dynamically allocate tasks and resources between edge and cloud based on real-time demands and availability, minimizing latency and improving system efficiency.  Hierarchical FL approaches further enable multi-party collaboration and efficient model training for anomaly detection in IoMT data, leveraging digital twins and edge cloudlets for real-time patient health insights and timely interventions \cite{gupta2021hierarchical}.  Integrating FL within the healthcare Metaverse enhances these capabilities by improving privacy, scalability, and interoperability \cite{bashir2023survey}.

\subsection{Industrial Automation}\label{sec6.3}

Intelligent automation systems benefit significantly from collaborative computing architectures that deliver real-time fault diagnosis and predictive maintenance capabilities \cite{xu2024fusion}. Leveraging both cloud computational power and edge real-time processing capabilities, industrial automation systems achieve greater agility through optimized production processes and resource allocation \cite{zhang2024cloud}, where cloud-based platforms analyze historical data to generate optimized production schedules while edge devices adapt these schedules in real-time based on local conditions. Distributed intelligence across multiple edge devices handles complex manufacturing tasks more efficiently and enhances fault tolerance since individual device failures do not necessarily compromise the entire system \cite{zeng2021coedge}. As edge intelligence, anomaly detection, cloud-edge collaboration, and distributed intelligence converge, industrial automation systems evolve toward more sophisticated and autonomous manufacturing processes \cite{tang2024anomaly}.

\subsection{Smart Cities}\label{sec6.4}

Urban infrastructure optimization leverages collaborative computing to enhance traffic management and environmental monitoring capabilities \cite{firdaus2023joint,arroba2024sustainable}.  Edge devices collect and analyze real-time data for adaptive traffic control and congestion reduction, while distributed sensor networks monitor environmental factors like air quality and noise levels. Subsequently, cloud platforms aggregate and analyze this data to inform urban planning and environmental protection.  In addition, AI-driven resource allocation optimizes smart grid energy consumption \cite{taik2020electrical}.  Specifically, edge devices monitor energy usage, and cloud-based algorithms dynamically adjust energy distribution to minimize waste and improve grid stability.  FL is crucial for privacy-preserving data analysis and decision-making in these applications \cite{zhou2024enhancing,zhu2024flight,feng2024leveraging}.  Through FL, stakeholders such as government agencies and private companies can collaboratively train models on combined data without sharing sensitive information, leading to more accurate and robust models for urban planning and public safety.

\section{Performance Analysis}\label{sec7}
\subsection{Evaluation Metrics}\label{sec7.1}
\begin{figure}[t!]
  \centering
\includegraphics[width=0.49\textwidth]{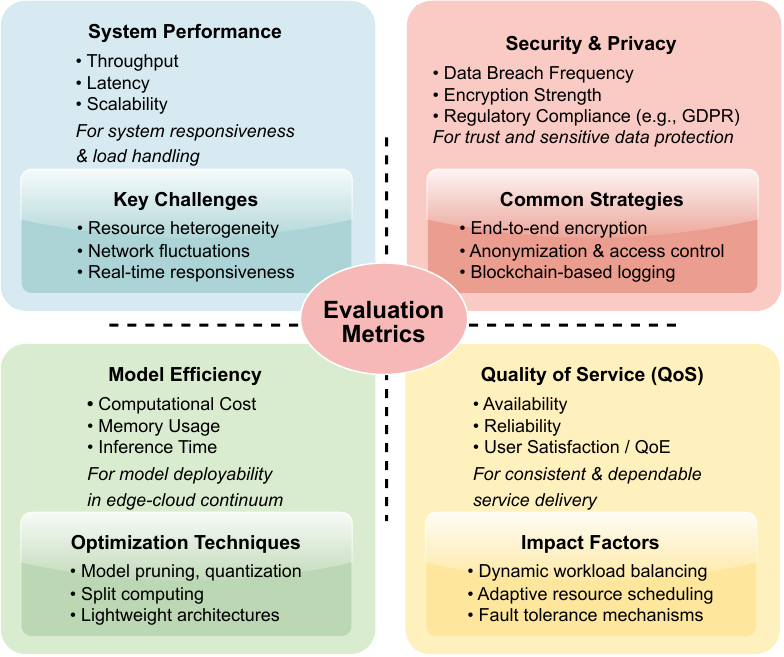}
\caption{Evaluation metrics, which are categorized into four primary groups: System performance, model efficiency, quality of service, and security and privacy. Each group contains specific evaluation dimensions (e.g., latency, inference time, availability, and data breach frequency), enabling comprehensive assessment of collaborative performance, resource usage, user experience, and trustworthiness.}
  \label{fig:10}
\vspace{-15px}
\end{figure}
Comprehensive performance assessment of collaborative computing systems requires multidimensional evaluation that addresses the unique characteristics of distributed computing paradigms. Unlike traditional centralized systems, ECCC performance evaluation must consider the complex interplay between edge proximity and cloud scalability, heterogeneous resource constraints, and dynamic network conditions. We categorize the relevant metrics into four primary groups tailored for edge-cloud environments: system performance, model efficiency, quality of service, and security and privacy. As illustrated in \autoref{fig:10}, comprehensive evaluation spans system-level, model-level, service-level, and trust-level metrics, considering real-world constraints unique to edge-cloud deployments such as network variability, compute heterogeneity, device mobility, and multi-tier resource orchestration. 

Performance evaluation in ECCC environments must address several unique challenges: (1) \textit{Cross-tier latency analysis}, measuring end-to-end response times that include network transmission delays, edge processing times, and cloud computation overhead; (2) \textit{Resource utilization efficiency}, evaluating how effectively computational, storage, and communication resources are distributed and utilized across the edge-cloud continuum; (3) \textit{Adaptation overhead}, quantifying the costs associated with dynamic model partitioning, task migration, and real-time workload redistribution; (4) \textit{Collaborative intelligence effectiveness}, assessing the performance gains achieved through edge-cloud cooperation compared to standalone edge or cloud-only deployments; (5) \textit{Fault tolerance and resilience}, measuring system robustness against edge device failures, network partitions, and cloud service interruptions. Specialized considerations necessitate specialized evaluation frameworks that capture the distributed, heterogeneous, and dynamic nature of edge-cloud systems while providing meaningful comparisons with traditional computing paradigms. 

\subsubsection{System Performance}\label{sec7.1.1}
Throughput, latency, and scalability are crucial metrics for evaluating the effectiveness and responsiveness of ECCC systems \cite{carpio2020engineering,jansen2023specrg,ali-eldin2021hidden}.  Throughput measures the volume of data processed or tasks completed within a given timeframe, with higher throughput indicating efficient handling of substantial workloads \cite{michalke2023scaling}.  Conversely, latency quantifies the delay between request initiation and completion; minimizing latency is paramount for real-time and interactive applications where responsiveness is critical \cite{ali-eldin2021hidden}.  Scalability, in turn, refers to a system's ability to adapt to increasing workloads or expanding infrastructure without performance degradation, accommodating growth in user demands or data volume while maintaining acceptable throughput and latency \cite{queralta2021blockchain}.  Given that the distributed and heterogeneous nature of edge-cloud environments amplifies the significance of these metrics, the interplay between edge and cloud resources necessitates careful optimization. For example, the study in \cite{carpio2020engineering} benchmarks a container-based edge computing system, demonstrating the impact of data synchronization on scalability across different deployment strategies.  Similarly, SPEC-RG \cite{jansen2023specrg} provides a reference framework for analyzing resource management and performance, while \cite{ali-eldin2021hidden} highlights the potential for performance inversion due to edge queuing delays.  Evaluating these metrics is essential for understanding system efficiency and informing design choices for optimized performance in resource-constrained and dynamic edge-cloud environments \cite{rublein2024improved}.

\subsubsection{Model Efficiency}\label{sec7.1.2}
Efficient AI models are essential for effective deployment and operation within the resource-constrained edge-cloud continuum.  Key metrics for quantifying model efficiency include computational cost, memory usage, and inference time, informing model selection and deployment strategies. Computational cost, typically measured in FLOPS \cite{caspart2022precise}, quantifies processing demands. Higher FLOPS values indicate greater computational requirements and potentially increased energy consumption, impacting resource-limited edge devices and overall system sustainability \cite{dehghani2022efficiency,richins2021ai}.  Memory usage reflects a model's storage footprint; large models may exceed the limited capacity of edge devices \cite{singh2025impact}, necessitating compression techniques like pruning, quantization, or knowledge distillation \cite{wu2022serving}.  Inference time, the duration for processing input and generating output, directly impacts application responsiveness \cite{singh2025impact}.  In latency-sensitive applications like autonomous driving or real-time video analytics, minimizing inference time is crucial \cite{jansen2023specrg}. Efficiency optimization often involves balancing model accuracy and computational efficiency, potentially employing strategies like model partitioning across different devices \cite{rosendo2022distributed}.

\subsubsection{Quality of Service}\label{sec7.1.3}
Evaluating the performance of edge-cloud collaborative systems necessitates an analysis of QoS metrics, which directly reflect the system's capacity to fulfill user expectations and provide consistent, dependable service \cite{zeng2024framework}.  Key QoS metrics include availability, reliability, and user satisfaction. Availability quantifies the proportion of time a system remains operational and accessible, a crucial aspect in edge-cloud systems where interruptions significantly impact user experience and application performance \cite{jansen2023specrg}.  High availability relies on robust system design, redundant infrastructure, and effective fault tolerance mechanisms, further enhanced by dynamic resource management and adaptive optimization techniques \cite{sedlak2024equilibrium}.  Reliability, in turn, assesses the system's ability to consistently deliver the expected service level over time, encompassing fault tolerance, error recovery, and performance maintenance under varying workloads \cite{simon2021parsimonious}.  Dependency-aware offloading and collaborative resource management contribute to improved reliability by optimizing task placement and resource allocation.  While often subjective, user satisfaction remains a crucial QoS indicator, reflecting the overall user experience, including performance, ease of use, and perceived value \cite{arnab2021analysis}. In edge-cloud systems, factors such as latency, throughput, and the quality of AI-driven services influence user satisfaction.  Specifically, QoE serves as a proxy for user satisfaction, directly linked to QoS parameters like latency and loss in applications like video streaming \cite{vidacs2023network}.  Similarly, in industrial automation, the relationship between QoS and QoE is critical for optimizing network resource management and ensuring customer satisfaction.

\subsubsection{Security and Privacy}\label{sec7.1.4}

In edge-cloud collaborative systems, robust security and privacy are crucial for maintaining trust and protecting sensitive information, especially given the sensitive nature of processed data \cite{xu2023blockchainbased}.  Therefore, evaluating these systems requires metrics focusing on data breaches, encryption strength, and regulatory compliance. Data breach frequency directly indicates system vulnerability by quantifying unauthorized accesses within a given timeframe \cite{zobaed2022aidriven}.  A lower frequency suggests stronger security, and analyzing breach characteristics can inform improved protocols, potentially leveraging frameworks like the transparency, accountability, and blockchain framework \cite{xu2023blockchainbased}.  Encryption strength assesses cryptographic robustness, including key length, algorithms, and resistance to attacks.  Strong encryption is paramount in distributed edge-cloud environments \cite{yao2024efficient}, and combining techniques like lightweight symmetric encryption and ciphertext-policy attribute-based encryption can enhance security \cite{yao2024efficient}.  Additionally, compliance with regulations like GDPR and CCPA is essential.  Metrics in this area assess adherence to data anonymization, user consent, and data retention policies \cite{zhao2022using}.  Evaluating privacy protectability, along with techniques like PrivaScissors, can further enhance privacy protection \cite{shi2023privacy,duan2023privascissors}.

\subsection{Benchmarking Methods}\label{sec7.2}

Performance benchmarking of edge-cloud collaborative systems necessitates standardized evaluation approaches to facilitate meaningful comparisons across diverse system designs, with a detailed analysis of tools provided in the Supplementary Material (see \autoref{t7-benchmark-comparison} in \autoref{sec:supp-benchmarking}). Essential to this process are standardized datasets and controlled testing environments that minimize variability, alongside comprehensive comparison frameworks employing simulation, emulation, and real-world testing to identify bottlenecks. For instance, selecting appropriate tools depends on application domains and evaluation objectives, Edge AIBench \cite{hao2019edge} provides realistic application modeling for healthcare and autonomous driving, while EdgeBench \cite{yang2024edgebench} offers flexibility for custom workflow configurations. Moreover, BenchFaaS \cite{carpio2023benchfaas} specializes in serverless paradigms, whereas KheOps \cite{rosendo2023kheops} facilitates collaborative research reproducibility, ultimately requiring that the optimal choice aligns workload representativeness and performance metrics with specific system requirements.

\subsubsection{Standard Datasets and Workloads}\label{sec7.2.1}

Consistent and comparable performance evaluation across different studies of ECCC systems requires the use of standard datasets and workloads \cite{mohammadi2018continuous}.  Unified standardization clarifies the strengths and weaknesses of various architectures, optimization techniques, and algorithms, enabling researchers to build upon existing work and accelerate the development of innovative solutions.  For example, Edge AIBench \cite{hao2019edge} offers benchmarks modeling typical edge scenarios like ICU patient monitoring and autonomous vehicles, focusing on data distribution and workload collaboration across client devices, the edge, and the cloud for comprehensive end-to-end performance evaluation.  Similarly, EdgeBench \cite{yang2024edgebench} provides a workflow-based benchmarking approach, allowing customization of workflow logic, data storage, and the distribution of workflow stages across computing tiers to tailor benchmarks to specific application scenarios.  Appropriate workload selection is also crucial \cite{jansen2023specrg}.  Specifically, workloads should reflect real-world application characteristics, including data intensity, computational complexity, and communication patterns, particularly in distributed intelligence scenarios where data and model distribution, as well as communication overhead for model training and inference, are critical factors. KheOps \cite{rosendo2023kheops} further emphasizes reproducibility in edge-to-cloud experiments by providing a collaborative environment with an experiment repository, a notebook environment, and a multi-platform methodology, promoting transparency and rigor in performance evaluation.  Moreover, the availability of standard datasets and workloads facilitates the development of new optimization techniques \cite{zhang2022ents}, allowing researchers to evaluate their methods on established benchmarks \cite{rosendo2022distributed}.

\subsubsection{Testing Environments}\label{sec7.2.2}

Robust benchmarking mandates controlled testing environments to accurately capture performance characteristics amidst complex interactions among edge devices, network infrastructure, and cloud servers \cite{varghese2021survey}. Achieving reliable results requires meticulous configuration of hardware, software, and network parameters to minimize external influences and ensure reproducibility across diverse platforms ranging from resource-constrained devices to powerful servers \cite{hao2019edge}. For instance, BenchFaaS \cite{carpio2023benchfaas} and KheOps \cite{rosendo2023kheops} employ heterogeneous testbeds encompassing virtual machines, Raspberry Pis, and cloud platforms to evaluate performance under varied conditions. Moreover, accurate measurement of KPIs such as latency and energy consumption demands precise control over network conditions, while standardized software configurations eliminate inconsistencies across experiments. Furthermore, performance comparison frameworks offer structured methodologies encompassing multifaceted evaluation metrics including system performance, model efficiency, and security considerations \cite{bi2023openperf,yang2019frequency}, alongside guidelines for controlled environments \cite{kahlhofer2023benchmarking} and analysis tools incorporating statistical techniques \cite{jansen2023specrg}. Additionally, evaluation methodologies spanning simulation, emulation \cite{luckow2021exploring}, and real-world testing each provide distinct advantages in capturing system complexities. Detailed frameworks and methodology analysis are provided in \autoref{sec:supp-frameworks} and \autoref{sec:supp-methodologies} of the Supplementary Material.

\subsection{System Design Evaluation}\label{sec7.3}

Performance assessment across ECCC paradigms requires multi-dimensional analysis through architectural comparisons, optimization evaluations, and systematic trade-off analysis~\cite{huang2021armada,li2021appealnet,ren2022edgematrix}. Architectural configurations reveal how cloud-centric solutions prioritize scalability versus edge-centric approaches like Armada emphasizing minimal latency, while hybrid systems including AppealNet and EdgeMatrix leverage predictive offloading and multi-agent algorithms for optimal balance. Additionally, optimization technique analysis examines load balancing effectiveness through two-timescale frameworks, collaborative edge-edge cooperation, and reinforcement learning-based adaptive management demonstrating superior performance in distributed scenarios~\cite{asghar2022survey,li2024twotimescale,fan2024collaborative}. Furthermore, trade-off analysis addresses design considerations balancing performance versus energy consumption, latency-accuracy optimization, and security-usability considerations~\cite{arroba2024sustainable,zhai2024edgecloud,rublein2024improved}. Detailed architectural comparisons, optimization analysis, and trade-off methodologies are provided in \autoref{sec:supp-system-design-eval} of the Supplementary Material.

\section{Challenges and Future Directions}\label{sec8}
\subsection{Technical Challenges}\label{sec8.1}

\begin{figure}[t!]
  \centering
\includegraphics[width=0.49\textwidth]{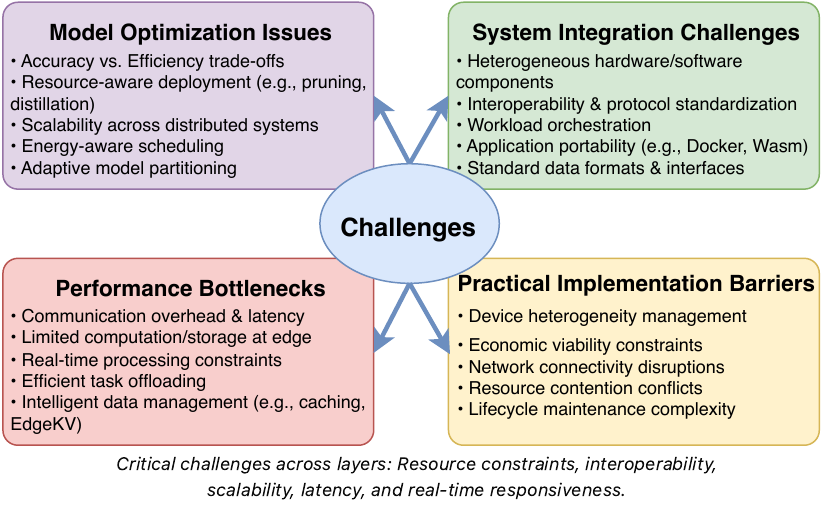}
\caption{Technical challenges, which span across model optimization, system integration, and performance bottlenecks, with domain-specific requirements shaping solution design.}
  \label{fig:9}
\vspace{-15px}
\end{figure}

As illustrated in \autoref{fig:9}, the technical challenges in AI-driven ECCC can be systematically categorized into four interconnected areas that reflect the lifecycle of system design and deployment. We begin with \textit{Model Optimization Issues} (\autoref{sec8.1.1}), which are foundational challenges related to designing and adapting AI models for resource-constrained and distributed environments. Next, we examine \textit{System Integration Challenges} (\autoref{sec8.1.2}), focusing on the difficulties of creating cohesive systems from heterogeneous hardware and software. Subsequently, we analyze \textit{Performance Bottlenecks} (\autoref{sec8.1.3}), which arise during system operation due to communication overhead and resource limitations. Finally, we address \textit{Practical Implementation Barriers} (\autoref{sec8.1.4}) that emerge during real-world deployment, encompassing economic and operational complexities. Accordingly, this structure provides a logical progression from model-level design to system-level operation and deployment.

\subsubsection{Model Optimization Issues}\label{sec8.1.1}
Model optimization for edge-cloud deployment fundamentally challenges the accuracy-efficiency trade-off, where complex models demand substantial resources unsuitable for edge deployment while smaller models sacrifice accuracy \cite{gowda2024watt}. Resource limitations including processing power, memory capacity, and energy availability, particularly for IoT devices \cite{arroba2024sustainable,ollivier2023sustainable}, compound deployment complexities alongside scaling challenges from communication overhead, resource contention, and device heterogeneity \cite{bhattacharjee2020deepedge,shi2020communicationefficient,tuli2023ai}. Effective optimization necessitates integrated approaches combining pruning, quantization, knowledge distillation \cite{tocze2018taxonomy} with resource-aware strategies like dynamic model selection, offloading \cite{rublein2024improved,wang2019edge}, energy-aware scheduling \cite{ullah2022framework}, and adaptive model partitioning \cite{zheng2023kubeedgesedna}. LLMs further intensify optimization challenges, requiring specialized inference techniques for resource-constrained environments, with solutions like Shoggoth \cite{wang2023shoggoth} demonstrating online knowledge distillation to enhance accuracy while managing data drift and computational offloading \cite{lin2024pushing}.

\subsubsection{System Integration Challenges}\label{sec8.1.2}
System integration in ECCC demands sophisticated heterogeneity management across diverse hardware capabilities, operating systems, network bandwidths, and communication protocols between resource-constrained edge nodes and powerful cloud servers \cite{rac2021edge,richins2021ai}. Interoperability challenges stem from differing data formats, communication protocols, and software interfaces, necessitating complex data transformations and sophisticated scheduling algorithms that consider device-specific capabilities \cite{menetrey2022webassembly,sowinski2023autonomous,luckow2021exploring}. Standardized protocols and well-defined interfaces become essential for seamless data exchange, simplified development, and enhanced scalability, with containerization (Docker \cite{rac2021edge}) and WebAssembly \cite{menetrey2022webassembly} providing portability solutions alongside AI-driven task placement optimization \cite{maia2024survey,sowinski2023autonomous}. Standardization particularly benefits critical applications like autonomous driving \cite{skarin2018missioncritical} and smart healthcare \cite{arroba2024sustainable} requiring reliable real-time data exchange, while promoting vendor interoperability and fostering open ecosystems \cite{ren2022edgematrix,shen2023edgematrix,sedlak2024equilibrium}.

\subsubsection{Performance Bottlenecks}\label{sec8.1.3}
ECCC systems encounter significant performance bottlenecks from communication overhead introducing latency and bandwidth consumption that impacts real-time applications \cite{carpio2020engineering,ali-eldin2021hidden,alnoman2019emerging}, while edge devices struggle with limited processing power, memory, and energy constraints when executing complex AI models and handling growing IoT data volumes \cite{vuruma2024cloud,makaya2020costeffective,baucas2020using,kokkonen2023autonomy}. Resource limitations become particularly acute during test-time adaptation scenarios where models must adjust to new data distributions while maintaining strict latency and energy budgets~\cite{ni2024fedsl}. Storage limitations exacerbated by device heterogeneity necessitate intelligent caching, data compression, distributed storage solutions like EdgeKV \cite{sonbol2020edgekv}, and comprehensive data lifecycle management \cite{tocze2018taxonomy,frascaria2021case,kolosov2023theory,joshi2024integration}. Data distribution shifts across geographically distributed edge nodes create additional performance challenges, requiring adaptive load balancing and intelligent workload redistribution to maintain system efficiency. Real-time applications including autonomous driving, industrial automation, and smart healthcare demanding ultra-low latency \cite{hu2024laecips,xing2023taskoriented,xu2024fusion} require integrated architectural innovations combining hierarchical edge computing with intelligent resource management \cite{zeng2021energyefficient}, utilizing frameworks like EdgeMatrix \cite{ren2022edgematrix} for multi-agent scaling, KubeEdge \cite{zheng2023kubeedgesedna} for automated provisioning, predictive offloading \cite{li2021appealnet}, intelligent caching \cite{wang2020survey}, and hardware acceleration technologies for optimal performance.

\subsubsection{Practical Implementation Barriers}\label{sec8.1.4}
Real-world ECCC deployment reveals substantial barriers extending beyond theoretical optimization to encompass operational complexities, economic constraints, and integration challenges, where device heterogeneity necessitates standardized middleware layers and adaptive frameworks for automatic capability detection \cite{jansen2023specrg,kanduri2024edgecentric,bensalem2023scaling,wang2024socialized,hamm2019edge}. Network connectivity issues require robust offline operation modes and intelligent synchronization mechanisms \cite{kawana2024communication}, while deployment complexity often results in 40-60\% timeline overruns due to integration challenges and vendor compatibility issues \cite{singh2025impact}. Economic viability concerns balance substantial upfront investments against long-term benefits, with edge computing reducing bandwidth costs by 30-50\% but requiring 3-5 year payback periods \cite{arroba2024sustainable,wang2023edge}, while technical pitfalls include underestimated communication overhead \cite{xu2024fusion,shi2020communicationefficient}, data consistency issues requiring eventual consistency models \cite{liu2023distributed}, and resource contention demanding sophisticated isolation algorithms \cite{lu2024a2cdrl,kokkonen2023autonomy}. Successful implementations emphasize phased rollout strategies, hybrid architectures, comprehensive security frameworks, and dedicated cross-functional teams \cite{poularakis2019joint,ren2021synergy,seifelnasr2024privacypreserving}.

\subsection{Future Research Directions}\label{sec8.2}

Future research in ECCC must address critical scalability challenges while advancing revolutionary computing paradigms~\cite{mungoli2023scalable}. \textit{Short-term objectives} include developing adaptive model partitioning algorithms~\cite{banitalebi-dehkordi2021autosplit} and standardized APIs for heterogeneous device integration~\cite{al-aqrabi2018scalable}, alongside lightweight containerization technologies optimized for resource-constrained environments. Furthermore, \textit{medium-term objectives} should focus on neuromorphic computing integration~\cite{thakur2018largescale,vogginger2024neuromorphic} for ultra-low-power edge inference and quantum-resistant security protocols~\cite{baseri2024cybersecurity,lu2024quantum} for long-term protection. Additionally, \textit{long-term research directions} encompass quantum edge computing~\cite{hossain2024quantumedge,nguyen2023iquantum} for solving intractable optimization problems and fully autonomous edge-cloud ecosystems capable of self-optimization~\cite{sowinski2023autonomous}. Consequently, implementation feasibility analysis indicates that standardization efforts represent the most immediately achievable objectives, while quantum computing integration and autonomous systems present longer-term challenges requiring significant advances in underlying technologies~\cite{paul2025quantumenhanced}.  
\subsubsection{Advanced AI Techniques}\label{sec8.2.1}
\begin{figure}[t!]
  \centering
\includegraphics[width=0.49\textwidth]{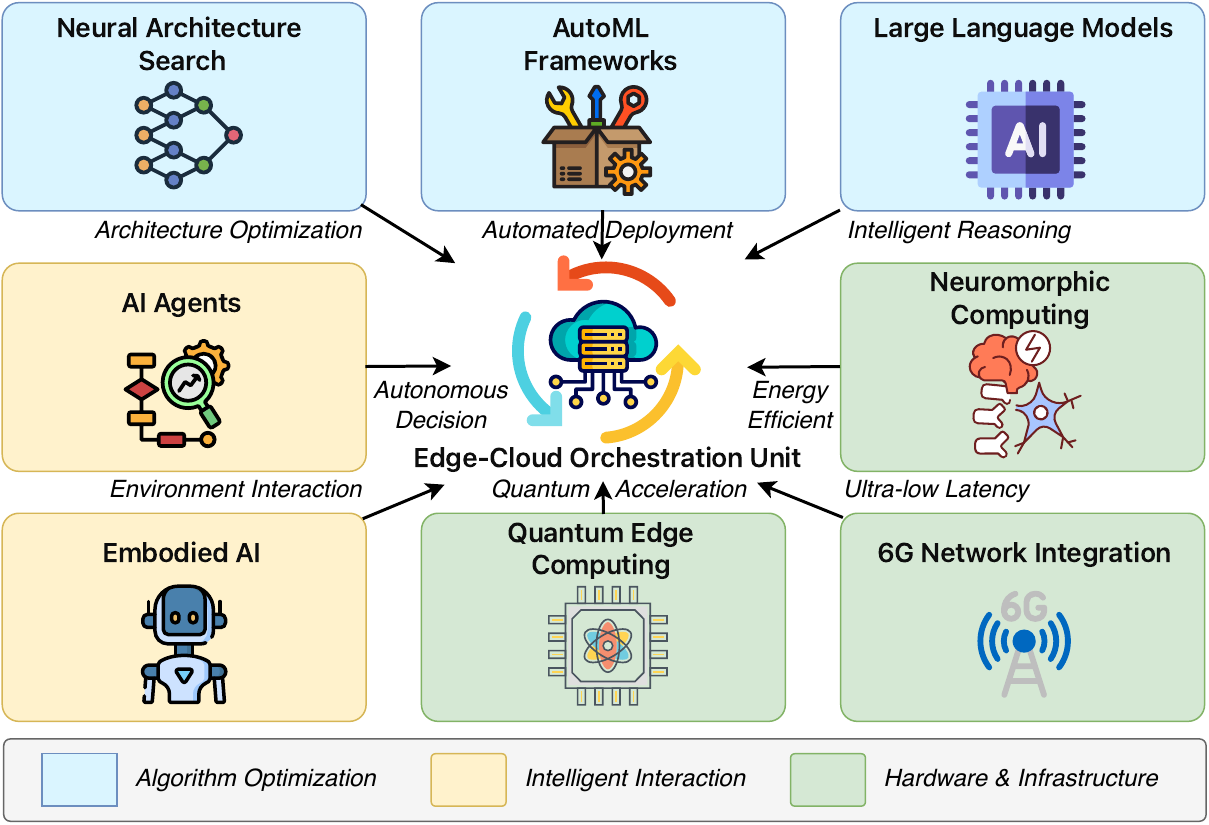}
  \caption{Overview of advanced AI techniques.}
  \label{fig:11}
\vspace{-15px}
\end{figure}

Advanced AI technologies are poised to redefine the capabilities of ECCC systems. To structure this forward-looking discussion, we categorize these emerging techniques into three distinct yet complementary areas, as illustrated in \autoref{fig:11}. First, \textit{algorithm optimization} focuses on automating and enhancing the AI models themselves through methods like neural architecture search and AutoML. Second, \textit{intelligent interaction} explores technologies such as AI agents and embodied AI, which enable more autonomous and adaptive system behaviors. Third, \textit{hardware and infrastructure} innovations, including neuromorphic computing, quantum edge computing, and 6G integration, provide the foundational advancements in compute and connectivity required to support next-generation AI. Accordingly, the following discussion details each of these promising research directions.

\noindent$\bullet$ \textbf{Neural Architecture Search}  NAS automates the design and optimization of neural network architectures, aiming to discover models that outperform manually designed ones in accuracy and efficiency for edge-cloud collaborative environments \cite{hao2021enabling}. NAS algorithms explore a vast search space of potential architectures, evaluating their performance to identify architectures tailored to the specific constraints of edge-cloud systems, such as limited resources and bandwidth \cite{jain2022latencymemory,zeng2021coedge}.  Approaches like reinforcement learning, evolutionary algorithms, and gradient-based NAS show promise in finding optimal architectures for specific tasks and resource constraints \cite{banitalebi-dehkordi2021autosplit}, and combining NAS with model optimization techniques like compression and adaptation further improves performance and efficiency \cite{mannan2020clan,mih2024ecavg}.  Ultimately, integrated optimization addresses the challenges of deploying complex AI models in heterogeneous edge-cloud environments \cite{mao2024green}.

\noindent$\bullet$ \textbf{AutoML} Automated machine learning (AutoML) streamlines AI model development and deployment in edge-cloud systems by automating processes like model selection, hyperparameter tuning, and deployment \cite{elshawi2019automated,schmitt2023automated}. Automated optimization becomes crucial in edge-cloud environments where diverse hardware and software coupled with resource constraints make manual optimization complex \cite{xue2023omniforce}.  For example, techniques like FAST-DAD \cite{fakoor2020fast} distill complex models into smaller, deployable ones for edge devices, while frameworks like H2O AutoML simplify model selection and hyperparameter optimization \cite{schmitt2023automated}.  Language-driven systems like AutoMMLab \cite{yang2024autommlab} further democratize AI development.  However, challenges remain in transparency, flexibility for complex scenarios, and consistent coverage of the entire ML workflow \cite{azevedo2024multivocal}. In large-scale environments, sampling-based AutoML and human-centered systems like OmniForce \cite{yin2024automl,xue2023omniforce} show promise, while automatically generating data-dependent configurations improves model performance \cite{kayali2022mining}.

\noindent$\bullet$ \textbf{Large Language Models} Optimizing LLM inference for edge devices involves addressing limited resources and real-time performance, necessitating research on optimized inference techniques and model compression methods like Agile-Quant \cite{yi2025edgemoe,shen2024agilequant}.  Approaches like expert-wise bitwidth adaptation in EdgeMoE \cite{yi2025edgemoe} can further enhance on-device efficiency.  Integrating LLMs for edge-cloud orchestration, including dynamic resource allocation and task offloading, promises more efficient systems, as demonstrated by VELO and EC-Drive \cite{yao2024velo,chen2024edgecloud}.  Developing specialized LLM architectures for edge-cloud environments, including low-rank adaptation (NetGPT \cite{chen2024netgpt}), model partitioning, and split learning/inference \cite{lin2024pushing}, is a critical research direction.

\noindent$\bullet$ \textbf{AI Agents} Intelligent agents capable of autonomous decision-making and learning within edge-cloud systems enhance adaptability and efficiency \cite{tuli2023ai,wang2020convergence}.  AI agents facilitate continuous learning on edge devices, addressing bandwidth limitations and connectivity issues through approaches like NeuroEvolutionary learning \cite{mannan2020clan}. Collaborative learning improves model accuracy, particularly in heterogeneous networks \cite{yao2023edgecloud,liu2023distributed}, and these agents optimize system performance through dynamic task offloading and resource allocation \cite{shi2020communicationefficient,mao2024green}.  Integration with lifelong learning frameworks like KubeEdge-Sedna further enhances automation \cite{zheng2023kubeedgesedna}.

\noindent$\bullet$ \textbf{Embodied AI} Embodied AI, focusing on agents that interact with their environment, requires new algorithms and frameworks tailored for resource-constrained edge devices \cite{mu2023embodiedgpt}.  Algorithms for robot control and coordination must consider latency and bandwidth limitations \cite{tian2019fog}, while efficient task offloading and resource allocation optimize performance \cite{ren2022edgematrix}. Furthermore, exploring embodied AI applications for multi-robot collaboration and dynamic visual servoing presents significant research opportunities~\cite{tahir2023consensusbased,tian2019fog}. Ultimately, developing robust and adaptive algorithms that effectively leverage distributed computational resources represents a critical research direction for advancing embodied AI in edge-cloud environments~\cite{wan2022circuit}.

\noindent$\bullet$ \textbf{Neuromorphic Computing} Inspired by the biological brain, neuromorphic computing promises energy-efficient AI models for edge-cloud environments \cite{thakur2018largescale,date2022neuromorphic}. Neuromorphic hardware, like SNNs, offers potential energy savings compared to traditional hardware \cite{vogginger2024neuromorphic}, while inherent fault tolerance improves robustness \cite{rastogi2021selfrepair}.  Integrating neuromorphic hardware into data centers presents challenges in algorithms, software, and benchmarks \cite{dang2022neucasl}.  However, its potential benefits for edge robotics drive research towards ultra-low-power designs and 3D integration technologies \cite{wan2022circuit,kurshan2020case}.

\noindent$\bullet$ \textbf{Quantum Edge Computing} Quantum computing offers revolutionary potential for edge-cloud applications by tackling classically intractable problems \cite{hossain2024quantumedge}. Consequently, developing edge-tailored quantum algorithms \cite{nguyen2023iquantum} and robust quantum communication protocols \cite{nguyen2024quantum}, potentially in hybrid approaches, are key research avenues. However, significant security \cite{baseri2024cybersecurity} and privacy concerns \cite{lu2024quantum} necessitate quantum-resistant techniques. Investigating the associated societal impacts \cite{paul2025quantumenhanced} is also crucial for responsible deployment.

\noindent$\bullet$ \textbf{6G Network Integration} 6G networks, with ultra-low latency and high bandwidth, promise to enhance edge-cloud systems \cite{letaief2022edge}.  Low latency enables real-time processing, while high bandwidth facilitates seamless data transfer, supporting resource-intensive AI tasks \cite{han2021timeenergyconstrained,you20236g}.  Advanced features like network slicing and cell-free MIMO optimize performance \cite{you20236g}, and integration with digital twins and pervasive AI unlocks new possibilities \cite{tariq2022experiencedriven,baccour2023zero}.  Addressing challenges in network architecture, resource allocation, and security through AI-driven solutions is crucial \cite{adem2021how,li2021slicingbased}.

\subsubsection{System Improvements}\label{sec8.2.2}
Efficient and dependable AI-driven ECCC necessitates crucial system improvements, primarily focusing on dynamic resource management, adaptive optimization, and fault tolerance.

\noindent$\bullet$ \textbf{Dynamic Resource Management} Dynamic resource management optimizes performance and resource utilization in the fluctuating demands of edge-cloud environments through efficient, real-time resource allocation and reallocation.  Key aspects include algorithms that dynamically adjust resource allocation based on real-time workload characteristics and available resources~\cite{fan2024collaborative}, such as the energy-aware and trust-collaboration cross-domain resource allocation (ETCRA) algorithm, which minimizes system functions while adhering to latency and trust constraints~\cite{li2024energyaware}.  Collaborative resource management coordinates resource allocation across multiple edge nodes and the cloud for enhanced system efficiency~\cite{tuli2023ai}.  Additionally, integrating RL allows systems to learn optimal resource allocation strategies based on environmental feedback~\cite{xu2024fusion}, while dynamic algorithms with initial allocation and real-time updating based on trust assessment improve performance~\cite{cao2020edge}. Multi-timescale algorithms address the varying optimization periods of different sub-problems like service placement and task scheduling within edge-cloud cooperation~\cite{fan2024collaborative}.

\noindent$\bullet$ \textbf{Adaptive Optimization}  Adaptive optimization dynamically adjusts system parameters and optimization strategies in response to real-time data and evolving conditions to maximize efficiency. For instance, the dynamic resource allocation mechanism in~\cite{xu2024fusion} ensures efficient scheduling of edge resources for global optimization.  Adaptive training with small batches, as proposed in Shoggoth~\cite{wang2023shoggoth}, allows edge model adaptation under limited computing power.  Similarly, incremental learning approaches periodically retrain edge AI models using real-time data~\cite{joshi2024integration}, optimizing prediction accuracy and reducing cloud dependency.  Dynamic model selection based on resources and application requirements~\cite{tuli2023ai} and decentralized adaptive scheduling using A3C learning and R2N2~\cite{tuli2022dynamic} are other crucial aspects.  Additionally, the two-stage iterated greedy optimization (TIGO) algorithm addresses changing user requirements for dynamic microservice placement~\cite{he2024efficient}, while adaptive switching between fog and cloud instances based on response time enhances performance~\cite{athanasopoulos2022ai}.

\noindent$\bullet$ \textbf{Fault Tolerance}  Fault tolerance ensures reliable operation by implementing fault detection, isolation, and recovery strategies.  DeepFT~\cite{tuli2023deepft}, for example, uses a deep surrogate model for proactive fault prediction and optimized task scheduling.  Ensuring service resilience for critical nodes is addressed by confidence-aware resilience models like CAROL~\cite{tuli2022carol}, which leverages generative neural networks for QoS prediction and system recovery.  Programmable asset orchestration at the edge contributes to operational resilience, with systems like RECS~\cite{moura2022resilience} dynamically managing services and resources during incidents.  Finally, distributed data auditing schemes, often incorporating techniques like FSS~\cite{li2024nextgeneration}, ensure data integrity for reliable AI applications.

\section{Conclusion}\label{sec9}

In this survey, we presented a holistic analysis of AI-driven edge-cloud collaborative computing, emphasizing the foundational principle that ``design-time'' model optimization serves as a critical enabler for effective ``run-time'' resource management. Our work synthesizes key developments in model optimization, resource management, and security to construct a unified framework for the field, highlighting the synergistic relationship between creating efficient AI models and orchestrating their deployment across distributed infrastructure. By systematically connecting these domains, this review provides a comprehensive understanding of how to architect intelligent systems that are both powerful and resource-aware. By systematically structuring the co-design challenges and opportunities in model optimization and resource management, this survey extends beyond the scope of recent works and charts a clear path for future research in edge intelligence.

Despite substantial advancements, several open challenges persist that require focused research. Key issues include guaranteeing robust security and privacy in highly decentralized environments, achieving scalable and reliable performance with heterogeneous hardware, establishing standardized benchmarking protocols to ensure fair and reproducible evaluations, and promoting integrated hardware-software co-design for optimized performance. To address these gaps, future research should prioritize several high-impact directions, including: (1) Creating end-to-end confidential computing frameworks to protect data throughout the ECCC lifecycle; (2) Engineering unified orchestration platforms that streamline management and improve resource utilization across the computing continuum; (3) Establishing community-driven benchmarking suites to standardize evaluation and foster collaborative innovation; and (4) Investigating cross-layer optimizations that jointly consider AI models and emerging technologies like 6G, neuromorphic computing, and quantum edge computing to unlock new frontiers in distributed intelligence.

\ifCLASSOPTIONcaptionsoff
  \newpage
\fi

{\small
\bibliographystyle{IEEEtran}
\begin{spacing}{0.95} 
  \bibliography{refs.bib}
  \end{spacing}
}
\includepdf[pages=-,pagecommand={},fitpaper=true]{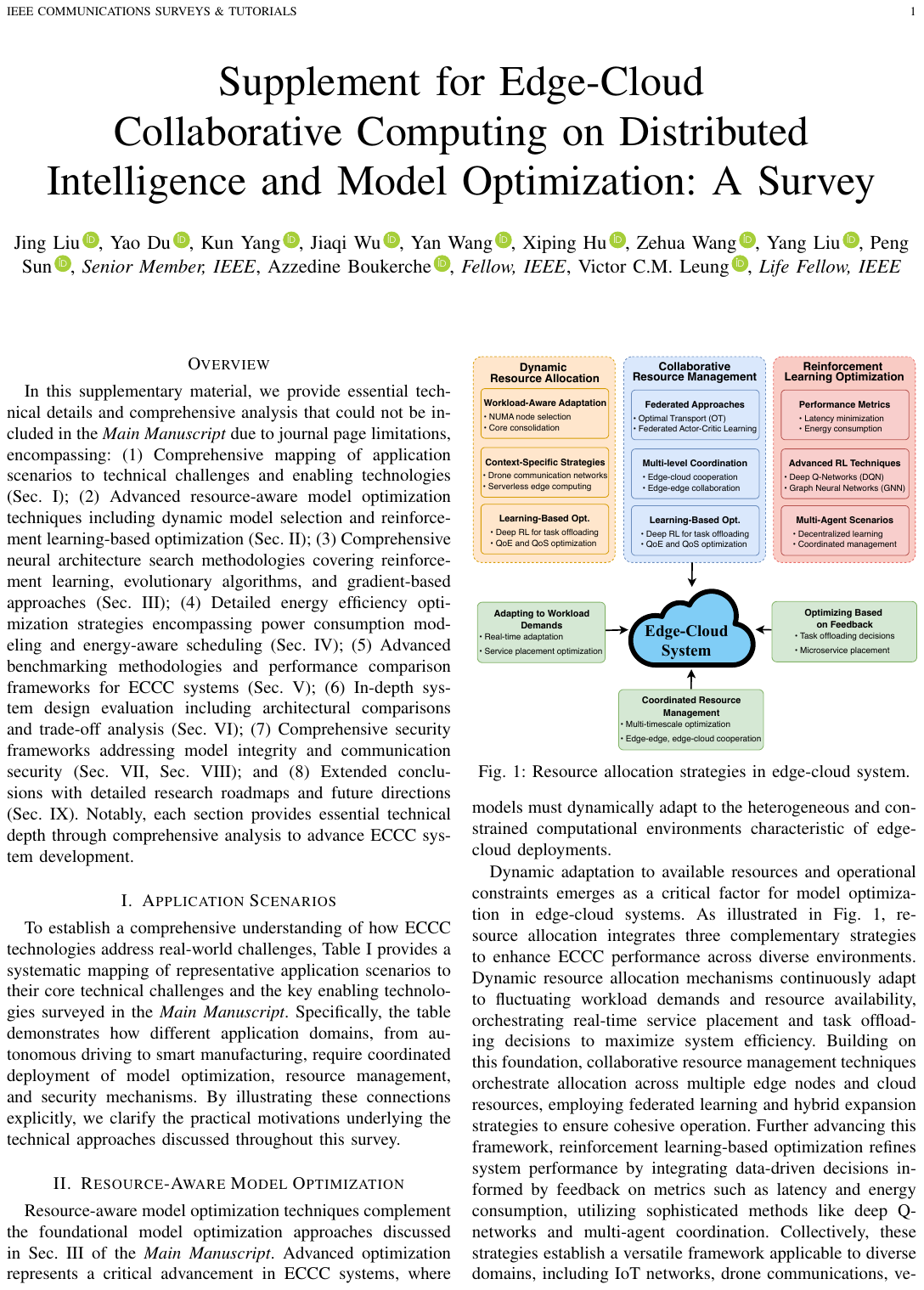}
\end{document}

%% file: t1_survey_1.tex
\begin{table*}[tb!]
  \centering
  \caption{Comparative analysis of existing surveys related to AI-driven edge-cloud collaborative computing. Our work uniquely provides comprehensive coverage across the distributed computing continuum, technical focuses, and applications.  
  }
  \label{tab:1}
  \begin{center}
  \setlength{\tabcolsep}{0.5mm}{
  \resizebox{0.98\textwidth}{!}{
    \begin{tabular}{@{}ccc|cccc|cccccc|ccc@{}}
      \toprule
      \multirow{2}{*}{\textbf{Ref.}} & \multirow{2}{*}{\textbf{Year}} & \multirow{2}{*}{\textbf{Topic}} & \multicolumn{4}{c|}{\textbf{Computing Continuum}} & \multicolumn{6}{c|}{\textbf{Technical Focuses}} & \multicolumn{3}{c}{\textbf{Applications}} \\ \cmidrule(l){4-7} \cmidrule(l){8-13} \cmidrule(l){14-16} 
       &  &  & \textbf{Edge} & \textbf{Cloud} & \textbf{ECCo} & \textbf{ECC} & \textbf{Model Opt.} & \textbf{Task Off.} & \textbf{Res. Mgmt.} & \textbf{Privacy} & \textbf{Security} & \textbf{LLMs} & \textbf{Manuf.} & \textbf{Auto. Drive} & \textbf{Health} \\  \midrule
      \cite{abbas2018mobile} & 2018 & Mobile Edge Comp., Network Architecture & \CIRCLE & \LEFTcircle & \LEFTcircle & \LEFTcircle & \Circle & \LEFTcircle & \LEFTcircle & \LEFTcircle & \LEFTcircle & \Circle & \Circle & \LEFTcircle & \LEFTcircle \\
      \cite{ferrer2019decentralised} & 2019 & Decentralized Cloud, Ad Hoc Computing & \CIRCLE & \CIRCLE & \LEFTcircle & \LEFTcircle & \Circle & \LEFTcircle & \LEFTcircle & \LEFTcircle & \Circle & \Circle & \Circle & \LEFTcircle & \Circle \\
      \cite{liu2019survey} & 2019 & Edge Computing, Systems, Tools & \CIRCLE & \LEFTcircle & \LEFTcircle & \Circle & \LEFTcircle & \LEFTcircle & \CIRCLE & \LEFTcircle & \LEFTcircle & \Circle & \LEFTcircle & \LEFTcircle & \LEFTcircle \\
      \cite{wang2019edge} & 2019 & Edge Cloud, Offloading Algorithms & \CIRCLE & \LEFTcircle & \CIRCLE & \CIRCLE & \LEFTcircle & \CIRCLE & \CIRCLE & \LEFTcircle & \LEFTcircle & \Circle & \Circle & \LEFTcircle & \LEFTcircle \\
      \cite{cao2021survey} & 2021 & Edge Computing, ECC, CPS & \CIRCLE & \CIRCLE & \CIRCLE & \CIRCLE & \Circle & \LEFTcircle & \CIRCLE & \CIRCLE & \CIRCLE & \Circle & \CIRCLE & \LEFTcircle & \LEFTcircle \\
      \cite{arthurs2022taxonomy} & 2022 & EC Computing, ITS, Connected Vehicles & \CIRCLE & \CIRCLE & \CIRCLE & \LEFTcircle & \Circle & \LEFTcircle & \LEFTcircle & \Circle & \Circle & \Circle & \Circle & \CIRCLE & \Circle \\
      \cite{vanhuynh2022edge}           & 2022                              & Edge Intel., URLLC, Digital Twin          & \CIRCLE                                     & \LEFTcircle                                  & \LEFTcircle                                       & \CIRCLE                                               & \Circle                                        & \CIRCLE                                      & \CIRCLE                                          & \LEFTcircle                                     & \Circle                                     & \Circle                          & \Circle                                    & \LEFTcircle                                 & \Circle                                 \\
      \cite{kar2023offloading} & 2023 & Offloading, Federated Cloud-Edge-Fog & \CIRCLE & \CIRCLE & \CIRCLE & \CIRCLE & \Circle & \CIRCLE & \CIRCLE & \LEFTcircle & \LEFTcircle & \Circle & \LEFTcircle & \LEFTcircle & \Circle \\
      \cite{wang2023edge} & 2023 & Edge Computing, Sensor-Cloud & \CIRCLE & \CIRCLE & \CIRCLE & \LEFTcircle & \Circle & \LEFTcircle & \CIRCLE & \CIRCLE & \CIRCLE & \Circle & \LEFTcircle & \LEFTcircle & \LEFTcircle \\
      \cite{yao2023edgecloud} & 2023 & EC Polarization, Collab. AI, Edge Comp. & \CIRCLE & \CIRCLE & \CIRCLE & \CIRCLE & \CIRCLE & \CIRCLE & \CIRCLE & \LEFTcircle & \LEFTcircle & \LEFTcircle & \LEFTcircle & \LEFTcircle & \LEFTcircle \\
      \cite{xu2024unleashing} & 2024 & AIGC, Mobile Networks, EC Generative AI & \CIRCLE & \CIRCLE & \LEFTcircle & \Circle & \LEFTcircle & \LEFTcircle & \LEFTcircle & \CIRCLE & \CIRCLE & \CIRCLE & \Circle & \Circle & \Circle \\
      \cite{gu2024aienhanced} & 2024 & AI, CET Network, Collaborative Computing & \CIRCLE & \CIRCLE & \CIRCLE & \CIRCLE & \LEFTcircle & \CIRCLE & \CIRCLE & \LEFTcircle & \CIRCLE & \LEFTcircle & \CIRCLE & \CIRCLE & \CIRCLE \\
      \cite{wang2024endedgecloud} & 2024 & DL, Distributed Computing, Model Opt. & \CIRCLE & \CIRCLE & \CIRCLE & \CIRCLE & \CIRCLE & \CIRCLE & \CIRCLE & \LEFTcircle & \LEFTcircle & \LEFTcircle & \LEFTcircle & \LEFTcircle & \LEFTcircle \\
\cite{yang2025edge}               & 2025                              & RL, Mobile Edge Comp.,  Res. Mgmt.   & \CIRCLE                                     & \LEFTcircle                                  & \LEFTcircle                                       & \LEFTcircle                                           & \LEFTcircle                                    & \CIRCLE                                      & \CIRCLE                                          & \LEFTcircle                                     & \CIRCLE                                     & \LEFTcircle                      & \LEFTcircle                                & \CIRCLE                                     & \LEFTcircle                             \\ 
\cite{qu2025mobile}               & 2025                              & Mobile Edge Intel., LLMs, Res. Mgmt. & \CIRCLE                                     & \CIRCLE                                      & \CIRCLE                                           & \LEFTcircle                                           & \CIRCLE                                        & \CIRCLE                                      & \CIRCLE                                          & \CIRCLE                                         & \LEFTcircle                                 & \CIRCLE                          & \Circle                                    & \Circle                                     & \CIRCLE                                 \\ \midrule
\textit{Ours} & 2025 & ECCC, AI, Distributed Intel., Model Opt. & \CIRCLE & \CIRCLE & \CIRCLE & \CIRCLE & \CIRCLE & \CIRCLE & \CIRCLE & \CIRCLE & \CIRCLE & \CIRCLE & \CIRCLE & \CIRCLE & \CIRCLE \\ \bottomrule
  \end{tabular}}}
  \end{center}
  \fontsize{7.5pxpt}{2pt}\selectfont {
  \begin{minipage}{0.98\textwidth}
    \vspace{1px}
\textit{Notes:} $\CIRCLE$: ``Fully Covered''; $\LEFTcircle$: ``Partially Covered''; $\Circle$: ``Not Covered''. Column abbreviations: Edge: ``Edge Computing''; Cloud: ``Cloud Computing''; ECCo: ``Edge-Cloud Computing''; ECC: ``Edge-Cloud Collaboration''; Model Opt.: ``Model Optimization''; Task Off.: ``Task Offloading''; Res. Mgmt.: ``Resource Management''; LLMs: ``Large Language Models''; Manuf.: ``Manufacturing''; Auto. Drive: ``Autonomous Driving''; Health: ``Healthcare''.
  \end{minipage}
  }
\vspace{-15px}
\end{table*}

%% file: t1.1_tech_comparison.tex
\begin{table*}[!htbp]
   \centering
   \caption{Comprehensive comparison of edge-cloud collaborative computing technical categories.}
   \label{t-tech-comparison}
   \renewcommand{\arraystretch}{1.2}
   \setlength{\tabcolsep}{5pt}
   \resizebox{\textwidth}{!}{%
   \begin{tabular}{>{\centering\arraybackslash}p{1.5cm}
   >{\centering\arraybackslash}p{4cm}
   >{\centering\arraybackslash}p{5cm}
   >{\centering\arraybackslash}p{5cm}
   >{\centering\arraybackslash}p{5cm}}
   \toprule
   \textbf{Category}   & \textbf{Core Techniques}       & \textbf{Advantages}& \textbf{Limitations}         & \textbf{Applicable Scenarios}   \\ \midrule
   
   \rotatebox{270}{\hspace{25px}\parbox{3cm}{\centering Model Optimization for EC Systems 
   }}  &  Advanced model compression techniques including pruning, quantization, and knowledge distillation, combined with adaptive learning approaches like transfer learning, federated learning, and continual learning for optimal deployment across resource-constrained edge-cloud infrastructures. & 
   \vspace{-8px}
   \begin{itemize}
     \item Enables deployment of complex AI models on resource-constrained edge devices
     \item Reduces computational overhead and memory requirements significantly
     \item Maintains acceptable performance while optimizing for efficiency
     \item Supports adaptive learning and dynamic model selection
     \item Facilitates collaborative training without centralized data sharing   \vspace{-15px}
   \end{itemize}

   & 
   \vspace{-8px}
   \begin{itemize}
     \item Model compression may lead to accuracy degradation
     \item Federated learning faces challenges with non-IID data distributions
     \item Continual learning suffers from catastrophic forgetting issues
     \item NAS requires significant computational resources for architecture search
     \item Trade-offs between model accuracy, efficiency, and deployment constraints
   \vspace{-15px}
   \end{itemize}
   & 
   \vspace{-8px}
   \begin{itemize}
     \item Deployment of deep learning models on IoT devices and edge nodes
     \item Privacy-sensitive applications requiring distributed training
     \item Dynamic environments with evolving data distributions
     \item Resource-constrained scenarios requiring efficient inference
     \item Real-time applications with strict latency requirements
   \vspace{-15px}
   \end{itemize} \\ \midrule
   
   \rotatebox{270}{\hspace{35px}\parbox{3cm}{\centering AI-Driven Resource Management 
   }}  &  Intelligent task offloading mechanisms employing latency-aware, energy-efficient, and dependency-aware strategies, coupled with dynamic resource allocation and collaborative management approaches to optimize system performance and energy efficiency across distributed heterogeneous infrastructure. & 
   \vspace{-8px}
   \begin{itemize}
     \item Optimizes resource utilization across heterogeneous infrastructure
     \item Enables intelligent task placement and load balancing
     \item Reduces overall system energy consumption and operational costs
     \item Supports dynamic adaptation to changing workload demands
     \item Facilitates collaborative resource sharing across edge-cloud continuum
   \vspace{-15px}
   \end{itemize}
   & 
   \vspace{-8px}
   \begin{itemize}
     \item Complex optimization problems with multiple conflicting objectives
     \item Requires accurate prediction of system state and workload patterns
     \item Communication overhead between edge and cloud components
     \item Scalability challenges in large-scale distributed environments
     \item Difficulty in handling unpredictable workload fluctuations
   \vspace{-15px}
   \end{itemize} 
   & 
   \vspace{-8px}
   \begin{itemize}
     \item Large-scale IoT deployments with heterogeneous devices
     \item Time-critical applications requiring intelligent task scheduling
     \item Energy-constrained environments prioritizing sustainability
     \item Multi-tenant edge-cloud systems requiring fair resource sharing
     \item Dynamic workloads with varying computational demands
   \vspace{-15px}
   \end{itemize}
   \\ \midrule
   
   \rotatebox{270}{\hspace{35px}\parbox{3cm}{\centering Privacy and Security \\ Enhancement 
   }}    &  Comprehensive data privacy protection through privacy-preserving learning, secure multi-party computation, and differential privacy, integrated with model security mechanisms for integrity protection and attack detection, plus robust communication security using encryption and blockchain technologies. & 
   \vspace{-8px}
   \begin{itemize}
     \item Protects sensitive data throughout the edge-cloud computing pipeline
     \item Enables secure collaborative learning and inference
     \item Provides robust defense mechanisms against various attack vectors
     \item Ensures compliance with privacy regulations and standards
     \item Maintains data integrity and system trustworthiness
   \vspace{-15px}
   \end{itemize}
   & 
   \vspace{-8px}
   \begin{itemize}
     \item Privacy-preserving techniques often introduce computational overhead
     \item Trade-offs between security level and system performance
     \item Complex key management and authentication in distributed systems
     \item Potential vulnerabilities in heterogeneous edge devices
     \item Evolving threat landscape requiring continuous security updates
   \vspace{-15px}
   \end{itemize}
   & 
   \vspace{-8px}
   \begin{itemize}
     \item Healthcare and financial applications with strict privacy requirements
     \item Multi-party collaborative learning scenarios
     \item Industrial IoT systems handling sensitive operational data
     \item Government and defense applications requiring high security
     \item Cross-domain data sharing with regulatory compliance needs
   \vspace{-15px}
   \end{itemize} \\ \bottomrule
   \end{tabular}
   }
\end{table*}

%% file: taxo-sec3-model-opt.tex
\begin{figure}[t]
  \centering
  \makebox[\columnwidth][c]{%
  \resizebox{\columnwidth}{!}{
\begin{tikzpicture}[node distance=0.6cm, >=latex,  
  Root/.style={
    fill=red!10,              
    rounded corners=1pt,       
    inner sep=2pt,  
    font=\fontsize{12pt}{2pt}\selectfont, 
    text width=5.3cm,          
    align=center               
  },
  L1/.style={
    fill=blue!10,              
    rounded corners=1pt,       
    inner sep=2pt,             
    font=\fontsize{12pt}{2pt}\selectfont,          
    text width=5.5cm,          
    align=center               
  },
  L2/.style={
    fill=green!10,             
    rounded corners=1pt,       
    inner sep=2pt,             
    font=\fontsize{12pt}{2pt}\selectfont,          
    text width=4.5cm,          
    align=center               
  },
]
  \node[Root] (root) [rotate=90] at (0,0) {Model Optimization (\S\ref{sec3})};
  
  \node[L1] (compress) at (4, 2.3) {Model Compression (\S\ref{sec3.1})};   
  \node[L1] (adapt) at (4, 0.1) {Model Adaptation (\S\ref{sec3.2})};       
  \node[L1] (resource) at (4, -1.6) {Resource-Aware Opt. (\S\ref{sec3.3})}; 
  \node[L1] (nas) at (4, -3.2) {Neural Arch. Search (\S\ref{sec3.4})};     
  
  \node[L2] (pruning) at (10, 2.9) {Pruning};
  \node[L2] (quant) at (10, 2.3) {Quantization};
  \node[L2] (kd) at (10, 1.7) {Knowledge Distillation};
  
  \node[L2] (tl) at (10, 1.0) {Transfer Learning};
  \node[L2] (fl) at (10, 0.40) {Federated Learning};
  \node[L2] (cl) at (10, -0.2) {Continual Learning};
  \node[L2] (sl) at (10, -0.7) {Split Learning};
  
  \node[L2] (dms) at (10, -1.3) {Dynamic Model Sel.};
  \node[L2] (mso) at (10, -1.9) {Model Splitting};
  
  \node[L2] (rlnas) at (10, -2.6) {RL-based NAS};
  \node[L2] (eanas) at (10, -3.2) {Evolutionary NAS};
  \node[L2] (gnas) at (10, -3.8) {Gradient-based NAS};
  
  \node[font=\fontsize{12pt}{2pt}\selectfont, text width=5.8cm, align=left, anchor=west] (p-ex) at (13.5, 2.9) {\cite{ko2024twophase,yan2024hybrid,zheng2024semantic}};
  \node[font=\fontsize{12pt}{2pt}\selectfont, text width=5.8cm, align=left, anchor=west] (q-ex) at (13.5, 2.3) {\cite{dong2020cdc,liu2024quasyncfl,chen2024implementing}};
  \node[font=\fontsize{12pt}{2pt}\selectfont, text width=5.8cm, align=left, anchor=west] (k-ex) at (13.5, 1.7) {\cite{yao2024gkt,wu2024agglomerative,wang2024clouddevice}};
  
  \node[font=\fontsize{12pt}{2pt}\selectfont, text width=5.8cm, align=left, anchor=west] (tl-ex) at (13.5, 1.0) {\cite{li2024transfer,wang2024cloudedgea,sharma2024deep}};
  \node[font=\fontsize{12pt}{2pt}\selectfont, text width=5.8cm, align=left, anchor=west] (fl-ex) at (13.5, 0.40) {\cite{yu2024fedcae,le2024sfetec,zhao2024flexiblefl}};
  \node[font=\fontsize{12pt}{2pt}\selectfont, text width=5.8cm, align=left, anchor=west] (cl-ex) at (13.5, -0.2) {\cite{li2024digital,lin2024online,zhang2024crossfcl}};
  \node[font=\fontsize{12pt}{2pt}\selectfont, text width=5.8cm, align=left, anchor=west] (sl-ex) at (13.5, -0.7) {\cite{lin2024efficient,ahmed2024heterogeneous,kim2021splitbridge}};
  
  \node[font=\fontsize{12pt}{2pt}\selectfont, text width=5.8cm, align=left, anchor=west] (dms-ex) at (13.5, -1.3) {\cite{carpio2020engineering,rublein2024improved,wang2024efficient}};
  \node[font=\fontsize{12pt}{2pt}\selectfont, text width=5.8cm, align=left, anchor=west] (mso-ex) at (13.5, -1.9) {\cite{guo2024madrlom,hua2024energyefficient,wang2024cooperative}};
  
  \node[font=\fontsize{12pt}{2pt}\selectfont, text width=5.8cm, align=left, anchor=west] (rlnas-ex) at (13.5, -2.6) {\cite{goudarzi2023distributed,bensalem2023scaling,wen2023adaptivenet}};
  \node[font=\fontsize{12pt}{2pt}\selectfont, text width=5.8cm, align=left, anchor=west] (eanas-ex) at (13.5, -3.2) {\cite{he2024efficient,chen2024fast,materwala2022performance}};
  \node[font=\fontsize{12pt}{2pt}\selectfont, text width=5.8cm, align=left, anchor=west] (gnas-ex) at (13.5, -3.8) {\cite{yao2020efficient,dutta2023searchtime,li2021appealnet}};
  
  \draw[->] (root.south) -- (compress.west);
  \draw[->] (root.south) -- (adapt.west);
  \draw[->] (root.south) -- (resource.west);
  \draw[->] (root.south) -- (nas.west);
  
  \draw[->] (compress.east) -- (pruning.west);
  \draw[->] (compress.east) -- (quant.west);
  \draw[->] (compress.east) -- (kd.west);
  
  \draw[->] (adapt.east) -- (tl.west);
  \draw[->] (adapt.east) -- (fl.west);
  \draw[->] (adapt.east) -- (cl.west);
  \draw[->] (adapt.east) -- (sl.west);
  
  \draw[->] (resource.east) -- (dms.west);
  \draw[->] (resource.east) -- (mso.west);
  
  \draw[->] (nas.east) -- (rlnas.west);
  \draw[->] (nas.east) -- (eanas.west);
  \draw[->] (nas.east) -- (gnas.west);
  
  \draw[->] (pruning.east) -- (p-ex.west);
  \draw[->] (quant.east) -- (q-ex.west);
  \draw[->] (kd.east) -- (k-ex.west);
  \draw[->] (tl.east) -- (tl-ex.west);
  \draw[->] (fl.east) -- (fl-ex.west);
  \draw[->] (cl.east) -- (cl-ex.west);
  \draw[->] (sl.east) -- (sl-ex.west);
  \draw[->] (dms.east) -- (dms-ex.west);
  \draw[->] (mso.east) -- (mso-ex.west);
  \draw[->] (rlnas.east) -- (rlnas-ex.west);
  \draw[->] (eanas.east) -- (eanas-ex.west);
  \draw[->] (gnas.east) -- (gnas-ex.west);
  
\end{tikzpicture}
}
}
  \caption{Taxonomy of model optimization.}
  \label{taxo:sec3}
\end{figure} 

%% file: t2_mod_compre_3.1.tex
\begin{table}[t!]
  \centering
  \caption{Summary of model compression techniques.}
  \label{tab:2}
  \vspace{-0.1cm}
  \begin{center}
  \setlength{\tabcolsep}{0.5mm}{
  \resizebox{0.49\textwidth}{!}{
    \begin{tabular}{@{}cccc>{\centering\arraybackslash}p{2.5cm}>{\centering\arraybackslash}p{3.5cm}cc@{}}
      \toprule
      \textbf{Method} & \textbf{Year} & \textbf{Venue} & \textbf{Type} & \textbf{Domain} & \textbf{Performance Metrics} & \textbf{Edge} & \textbf{Cloud} \\ \midrule
      TSCF \cite{ko2024twophase} & 2024 & IEEE IoTJ & P & Edge-Cloud Systems & Inference Latency & \greencheck & \greencheck \\
      Hybrid SD \cite{yan2024hybrid} & 2024 & ArXiv & P & Image Synthesis & Parameter Efficiency, Image Quality & \greencheck & \greencheck \\ 
      PSFed \cite{zheng2024semantic} & 2024 & IEEE JSAC & P & Satellite Edge Cloud & Latency, Energy, Privacy & \greencheck & \greencheck \\
      3CE2C \cite{chen2024implementing} & 2024 & ASC & Q & Healthcare/ECG & Accuracy, Energy Efficiency & \greencheck & \greencheck \\
      CDC \cite{dong2020cdc} & 2020 & IJCAI & P & Edge-Cloud Systems & Bandwidth Efficiency & \greencheck & \greencheck \\
      QuAsyncFL \cite{liu2024quasyncfl} & 2024 & IEEE IoTJ & Q & AIoT & Communication Efficiency & \greencheck & \greencheck \\
      Zhang et al. \cite{zhang2025large} & 2025 & IEEE JSAC & Q & Aerial Edge Computing & Mean Average Precision (mAP) & \greencheck & \greencheck \\
      DTS-KD \cite{tang2024anomaly} & 2024 & IEEE IoTJ & K & Industrial IoT & F1 Score, Accuracy, Precision, Recall & \greencheck & \greencheck \\
      Wang et al. \cite{wang2024clouddevice} & 2024 & CVPR & K & Multimodal LLMs & Generalization Performance & \greencheck & \greencheck \\
      Wu et al. \cite{wu2024systematic} & 2024 & IEEE TII & P & Aviation Manufacturing & Edge Detection Accuracy & \redcross & \redcross \\
      FedAgg \cite{wu2024agglomerative} & 2024 & INFOCOM & K & End-Edge-Cloud Computing & Model Accuracy, Convergence Rate & \greencheck & \greencheck \\
      Yang et al. \cite{yang2025growthadaptive} & 2025 & ADHOC & K & IoT Traffic Identification & Classification Accuracy, Resource Efficiency & \greencheck & \greencheck \\
      GKT \cite{yao2024gkt} & 2024 & ACL & K & LLM Inference & Inference Speed, Accuracy, Cost & \greencheck & \greencheck \\
      OTO \cite{zeng2024offlinetransferonline} & 2024 & IEEE TPDS & K & Reinforcement Learning & Convergence Speed, Reward & \greencheck & \greencheck \\
      Zhang et al. \cite{zhang2024cloud} & 2024 & ESWA & K & Manufacturing Processes & Fault Detection Accuracy & \greencheck & \greencheck \\ \bottomrule     
  \end{tabular}}}
  \end{center}
  \fontsize{7.5pt}{2pt}\selectfont {
  \begin{minipage}{0.48\textwidth}
    \vspace{1px}
    \* \textit{Note for 'Type'}: P = Pruning, Q = Quantization, K = Knowledge Distillation.
  \end{minipage}
  }
\vspace{-15px}
\end{table}

%% file: t3_mod_ada_3.2.tex
\begin{table}[t!]
  \centering
  \caption{Summary of TL, FL, and CL techniques for enhancing AI model adaptability in ECCC systems.}
  \label{tab:3}
  \vspace{-0.1cm}
  \begin{center}
  \setlength{\tabcolsep}{0.5mm}{
  \resizebox{0.49\textwidth}{!}{
    \begin{tabular}{@{}cccc>{\centering\arraybackslash}p{3.5cm}cccc@{}}
      \toprule
      {\textbf{Method}} & {\textbf{Year}} & {\textbf{Venue}} & {\textbf{Type}} & {\textbf{Domain}} & {\textbf{Edge}} & {\textbf{Cloud}} & {\textbf{ECC}} & {\textbf{MC}} \\ \midrule
      {CoLLaRS \cite{hu2024collars}} & {2024} & {FGCS} & {CL} & {AIoT Applications} & {\greencheck} & {\greencheck} & {\greencheck} & {\redcross} \\
      {Li et al. \cite{li2024transfer}} & {2024} & {IEEE TNSM} & {TL} & {Manufacturing} & {\greencheck} & {\greencheck} & {\greencheck} & {\redcross} \\
      {Pan et al. \cite{pan2024cloudedge}} & {2024} & {IEEE Netw.} & {FL} & {Large Model Services} & {\greencheck} & {\greencheck} & {\greencheck} & {\greencheck} \\
      {MADDPG \cite{she2024efficient}} & {2024} & {IEEE IoTJ} & {TL} & {6G Networks} & {\greencheck} & {\greencheck} & {\greencheck} & {\redcross} \\
      {DTS-KD \cite{tang2024anomaly}} & {2024} & {IEEE IoTJ} & {TL} & {Industrial IoT} & {\greencheck} & {\greencheck} & {\greencheck} & {\greencheck} \\
      {Wang et al. \cite{wang2024clouddevice}} & {2024} & {CVPR} & {CL} & {Multimodal Large Language Models} & {\greencheck} & {\greencheck} & {\greencheck} & {\greencheck} \\
      {Wang et al. \cite{wang2024cloudedgea}} & {2024} & {ESWA} & {FL} & {Rotating Machinery Fault Diagnosis} & {\greencheck} & {\greencheck} & {\greencheck} & {\redcross} \\
      {FedAgg \cite{wu2024agglomerative}} & {2024} & {INFOCOM} & {FL} & {End-Edge-Cloud Systems} & {\greencheck} & {\greencheck} & {\greencheck} & {\redcross} \\
      {OTO \cite{zeng2024offlinetransferonline}} & {2024} & {IEEE TPDS} & {TL} & {Cloud-Edge Collaborative DRL} & {\greencheck} & {\greencheck} & {\greencheck} & {\redcross} \\
      {U-Federator \cite{ali2024universal}} & {2024} & {JNCA} & {FL} & {Cloud-Edge-Fog Computing} & {\greencheck} & {\greencheck} & {\greencheck} & {\redcross} \\
      {OTFAC \cite{gan2024optimal}} & {2024} & {IEEE IoTJ} & {FL} & {Cloud-Edge Collaborative IoT} & {\greencheck} & {\greencheck} & {\greencheck} & {\redcross} \\
      {FLAT \cite{garofalo2024webcentric}} & {2024} & {IEEE ISCC} & {FL} & {Cloud-Edge-Client Continuum} & {\greencheck} & {\greencheck} & {\greencheck} & {\redcross} \\
      {FedITA \cite{he2024fedita}} & {2024} & {AEI} & {FL} & {Industrial Motor Fault Diagnosis} & {\greencheck} & {\greencheck} & {\greencheck} & {\redcross} \\
      {Khuen et al. \cite{khuencheng2024integration}} & {2024} & {IEEE Access} & {FL} & {Precision Aquaculture} & {\greencheck} & {\greencheck} & {\greencheck} & {\redcross} \\
      {Kim et al. \cite{kim2024collaborative}} & {2024} & {IEEE IoTJ} & {FL} & {Cloud-Edge-Terminal IoT Networks} & {\greencheck} & {\greencheck} & {\greencheck} & {\redcross} \\
      {QuAsyncFL \cite{liu2024quasyncfl}} & {2024} & {IEEE IoTJ} & {FL} & {Cloud-Edge-Terminal AIoT} & {\greencheck} & {\greencheck} & {\greencheck} & {\greencheck} \\
      {Loconte et al. \cite{loconte2024expanding}} & {2024} & {FGCS} & {FL} & {Cloud-to-Edge-IoT Continuum} & {\greencheck} & {\greencheck} & {\greencheck} & {\redcross} \\
      {BAFL-DT \cite{qu2024blockchained}} & {2024} & {IEEE TSC} & {FL} & {Digital Twin Edge-Cloud Continuum} & {\greencheck} & {\greencheck} & {\greencheck} & {\redcross} \\
      {FDRL \cite{wang2024federated}} & {2024} & {IJPR} & {FL} & {Cloud-Edge Manufacturing Systems} & {\greencheck} & {\greencheck} & {\greencheck} & {\redcross} \\
      {Wei et al. \cite{wei2024joint}} & {2024} & {IEEE TPDS} & {FL} & {Edge Clouds} & {\greencheck} & {\greencheck} & {\greencheck} & {\redcross} \\
      {Yang et al. \cite{yang2024differentially}} & {2024} & {IEEE IoTJ} & {FL} & {AIoT Data Prediction} & {\greencheck} & {\greencheck} & {\greencheck} & {\redcross} \\
      {FedCAE \cite{yu2024fedcae}} & {2024} & {IEEE TIE} & {FL} & {Machine Fault Diagnosis} & {\greencheck} & {\greencheck} & {\greencheck} & {\greencheck} \\
      {FlexibleFL \cite{zhao2024flexiblefl}} & {2024} & {INS} & {FL} & {Cloud-Edge Federated Learning} & {\greencheck} & {\greencheck} & {\greencheck} & {\redcross} \\
      {PSPFL \cite{zhou2024accelerating}} & {2024} & {IEEE TMC} & {FL} & {Mobile Edge-Cloud Networks} & {\greencheck} & {\greencheck} & {\greencheck} & {\greencheck} \\
      {CCSFLF \cite{zhou2024ccsflf}} & {2024} & {CCPE} & {FL} & {Federated Learning Systems} & {\greencheck} & {\greencheck} & {\greencheck} & {\redcross} \\
      {Zhou et al. \cite{zhou2024enhancing}} & {2024} & {ADHOC} & {FL} & {AIoT Quality of Service} & {\greencheck} & {\greencheck} & {\greencheck} & {\redcross} \\
      {CADCL-DTM \cite{li2024conditionadaptive}} & {2024} & {CAC} & {CL} & {Rolling Bearings Monitoring} & {\greencheck} & {\greencheck} & {\redcross} & {\redcross} \\
      {Li et al. \cite{li2024digital}} & {2024} & {IEEE TMC} & {CL} & {Edge Computing Service Provisioning} & {\greencheck} & {\greencheck} & {\greencheck} & {\redcross} \\
      {Lian et al. \cite{lian2024cloudedge}} & {2024} & {SDI} & {CL} & {ITS Object Detection} & {\greencheck} & {\greencheck} & {\greencheck} & {\greencheck} \\
      {EdgeC3 \cite{lin2024online}} & {2024} & {IEEE TMC} & {CL} & {Distributed Deep Learning} & {\greencheck} & {\greencheck} & {\greencheck} & {\redcross} \\ \bottomrule
           
  \end{tabular}}}
  \end{center}
  \fontsize{7.5pt}{2pt}\selectfont {
  \begin{minipage}{0.48\textwidth}
    \vspace{1px}
    \textit{Note for 'Type'}: TL = Transfer Learning, FL = Federated Learning, CL = Continual Learning, MC = Model Compression.
  \end{minipage}
  }
\vspace{-15px}
\end{table}

%% file: taxo-sec4-resource-mgmt.tex
\begin{figure}[t]
  \centering
  \makebox[\columnwidth][c]{%
  \resizebox{\columnwidth}{!}{
\begin{tikzpicture}[node distance=0.6cm, >=latex,
  Root/.style={fill=blue!10, rounded corners=1pt, inner sep=2pt, font=\fontsize{12pt}{2pt}\selectfont, text width=5cm, align=center},
  L1/.style={fill=orange!10, rounded corners=1pt, inner sep=2pt, font=\fontsize{12pt}{2pt}\selectfont, text width=5cm, align=center},
  L2/.style={fill=violet!10, rounded corners=1pt, inner sep=2pt, font=\fontsize{12pt}{2pt}\selectfont, text width=5.5cm, align=center}
]
  \node[Root] (root) [rotate=90] at (0,0) {Resource Management (\S\ref{sec4})};
  
  \node[L1] (offload) at (4, 2.0) {Task Offloading (\S\ref{sec4.1})};
  \node[L1] (alloc) at (4, -0.2) {Resource Allocation (\S\ref{sec4.2})};
  \node[L1] (energy) at (4, -2.2) {Energy Efficiency (\S\ref{sec4.3})};
  
  \node[L2] (latency) at (10, 2.9) {Latency-Aware};
  \node[L2] (ener-off) at (10, 2.3) {Energy-Efficient};
  \node[L2] (depend) at (10, 1.7) {Dependency-Aware};
  \node[L2] (multi) at (10, 1.1) {Multi-Objective};
  
  \node[L2] (dynamic) at (10, 0.4) {Dynamic Allocation};
  \node[L2] (collab) at (10, -0.2) {Collaborative Management};
  \node[L2] (rl) at (10, -0.8) {RL-based Allocation};
  
  \node[L2] (power) at (10, -1.6) {Power Management};
  \node[L2] (sched) at (10, -2.2) {Energy-Aware Scheduling};
  \node[L2] (green) at (10, -2.8) {Green Computing};
  
  \node[font=\fontsize{12pt}{2pt}\selectfont, text width=5.8cm, align=left, anchor=west] (lat-ex) at (13.8, 2.9) {\cite{sun2025latencyaware,wang2024deep,zhai2024edgecloud}};
  \node[font=\fontsize{12pt}{2pt}\selectfont, text width=5.8cm, align=left, anchor=west] (ene-ex) at (13.8, 2.3) {\cite{su2024primaldualbased,hua2024energyefficient,li2024energyaware}};
  \node[font=\fontsize{12pt}{2pt}\selectfont, text width=5.8cm, align=left, anchor=west] (dep-ex) at (13.8, 1.7) {\cite{feng2024dependencyaware,pang2024multimobile,chen2024realtime}};
  \node[font=\fontsize{12pt}{2pt}\selectfont, text width=5.8cm, align=left, anchor=west] (mul-ex) at (13.8, 1.1) {\cite{zhang2024manyobjective,jain2022latencymemory,guo2023edge}};
  
  \node[font=\fontsize{12pt}{2pt}\selectfont, text width=5.8cm, align=left, anchor=west] (dyn-ex) at (13.8, 0.4) {\cite{farhoudi2023qosaware,kim2024collaborative,chen2024dynamica}};
  \node[font=\fontsize{12pt}{2pt}\selectfont, text width=5.8cm, align=left, anchor=west] (col-ex) at (13.8, -0.2) {\cite{gan2024optimal,wang2024cloudedgeend,moraes2019pub}};
  \node[font=\fontsize{12pt}{2pt}\selectfont, text width=5.8cm, align=left, anchor=west] (rl-ex) at (13.8, -0.8) {\cite{binh2024reinforcement,huang2024joint,afachao2024efficient}};
  
  \node[font=\fontsize{12pt}{2pt}\selectfont, text width=5.8cm, align=left, anchor=west] (pow-ex) at (13.8, -1.6) {\cite{georgiou2021comprehensive,yao2023edgecloud,zhang2024dvfo}};
  \node[font=\fontsize{12pt}{2pt}\selectfont, text width=5.8cm, align=left, anchor=west] (sch-ex) at (13.8, -2.2) {\cite{nieto2024deep,zhang2024multitargetaware,bensalah2024vnf}};
  \node[font=\fontsize{12pt}{2pt}\selectfont, text width=5.8cm, align=left, anchor=west] (gre-ex) at (13.8, -2.8) {\cite{arroba2024sustainable,chen2024fast,abouaomar2021resource}};
  
  \draw[->] (root.south) -- (offload.west);
  \draw[->] (root.south) -- (alloc.west);
  \draw[->] (root.south) -- (energy.west);
  
  \draw[->] (offload.east) -- (latency.west);
  \draw[->] (offload.east) -- (ener-off.west);
  \draw[->] (offload.east) -- (depend.west);
  \draw[->] (offload.east) -- (multi.west);
  
  \draw[->] (alloc.east) -- (dynamic.west);
  \draw[->] (alloc.east) -- (collab.west);
  \draw[->] (alloc.east) -- (rl.west);
  
  \draw[->] (energy.east) -- (power.west);
  \draw[->] (energy.east) -- (sched.west);
  \draw[->] (energy.east) -- (green.west);
  
  \draw[->] (latency.east) -- (lat-ex.west);
  \draw[->] (ener-off.east) -- (ene-ex.west);
  \draw[->] (depend.east) -- (dep-ex.west);
  \draw[->] (multi.east) -- (mul-ex.west);
  \draw[->] (dynamic.east) -- (dyn-ex.west);
  \draw[->] (collab.east) -- (col-ex.west);
  \draw[->] (rl.east) -- (rl-ex.west);
  \draw[->] (power.east) -- (pow-ex.west);
  \draw[->] (sched.east) -- (sch-ex.west);
  \draw[->] (green.east) -- (gre-ex.west);
  
\end{tikzpicture}
}
}
  \caption{Taxonomy of AI-driven resource management.}
  \label{taxo:sec4}
\end{figure}

%% file: t4_task_4.1.tex
\begin{table}[t!]
  \centering
  \caption{Summary of task offloading mechanisms, categorized by strategy type and performance optimization targets.}
  \label{tab:4}
  \vspace{-0.1cm}
  \begin{center}
  \setlength{\tabcolsep}{0.4mm}
  \resizebox{0.49\textwidth}{!}{
    \begin{tabular}{@{}l>{\centering\arraybackslash}p{1.8cm}>{\centering\arraybackslash}p{2.5cm}>{\centering\arraybackslash}p{3.0cm}c>{\centering\arraybackslash}p{0.8cm}@{}}
      \toprule
      {\textbf{Type}} & {\textbf{Method}} & {\textbf{Metrics}} & {\textbf{Achievements}} & {\textbf{Gain}} & {\textbf{ECC}} \\ \midrule
      \multirow{4}{*}{\rotatebox[origin=c]{270}{\hspace{35px}Latency-Aware}} 
      & iGATS \cite{sun2025latencyaware} & Response latency, Load balancing & Outperforms baseline techniques for latency reduction & High & \greencheck \\
      & Fan et al. \cite{fan2024collaborative} & Task processing delay, Stability & Superior performance vs. 4 comparison schemes & Medium & \greencheck \\
      & Zhai et al. \cite{zhai2024edgecloud} & Response delay, Carbon emissions & 33.42× delay reduction for sensitive workloads & Very High & \greencheck \\
      & Luckow et al. \cite{luckow2021exploring} & Performance impact, Task placement & 5-65\% performance improvement across scenarios & High & \greencheck \\
      \midrule
      \multirow{4}{*}{\rotatebox[origin=c]{270}{\hspace{35px}Energy-Efficient}} 
      & PDCO \cite{su2024primaldualbased} & Energy consumption, Resource util. & Approaches optimal performance in energy metrics & High & \greencheck \\
      & Hua et al. \cite{hua2024energyefficient} & Energy consumption of mobile devices & Significantly reduced energy consumption & High & \greencheck \\
      & ETCRA \cite{li2024energyaware} & Energy, Time, Reliability & Outperforms baselines on four key measurements & High & \greencheck \\
      & Hu et al. \cite{hu2022energy} & Energy efficiency, Service delay & Achieves [O(1/V), O(V)] EE-delay tradeoff & Medium & \redcross \\
      \midrule
      \multirow{2}{*}{\rotatebox{270}{\hspace{-15px}\parbox{3cm}{\centering Dependency-\\Aware}}}
      & PRAISE \cite{feng2024dependencyaware} & System benefits, Resource costs & Higher benefits and lower costs vs. baselines & Medium & \redcross \\
      & Peng et al. \cite{peng2024redundancy} & Computing time, Blocking probability & Optimal replication with varying parameters & Medium & \redcross \\
      \midrule
      \multirow{5}{*}{\rotatebox[origin=c]{270}{\hspace{55px}Multi-Objective}} 
      & Zhang et al. \cite{zhang2024manyobjective} & Time, Cost, Load balance, Trust & Found $>$50\% of best solutions on benchmarks & High & \redcross \\
      & DRLIS \cite{wang2024deep} & Load balancing, Response time, Cost & Up to 55\% cost reduction across metrics & Very High & \greencheck \\
      & DT-MADDPG \cite{wang2024cooperative} & Success rate, Latency, Energy & Improved success rate with reduced resource usage & High & \greencheck \\
      & Zhang et al. \cite{zhang2024hybrid} & Time, Cost, Energy, Reliability & Outperforms existing design approaches & Medium & \greencheck \\
      & EdgeOptimizer \cite{qiao2024edgeoptimizer} & Success rate, Overall overload & Improved success rate for time-critical tasks & Medium & \greencheck \\ \bottomrule
  \end{tabular}}
  \end{center}
  \scriptsize {
  \begin{minipage}{0.48\textwidth}
    \vspace{1px}
    \textit{Notes}: Performance Gain: Very High ($>$50\%), High (20-50\%), Medium (5-20\%).
  \end{minipage}
  }
\vspace{-15px}
\end{table}

%% file: t5_res_alloc_4.2.tex
\begin{table}[t!]
  \centering
    \caption{Summary of resource allocation methods.}
  \label{tab:5}
  \vspace{-0.1cm}
  \begin{center}
  \setlength{\tabcolsep}{0.4mm}{
  \resizebox{0.49\textwidth}{!}{
    \begin{tabular}{@{}>{\centering\arraybackslash}m{1.5cm}>{\centering\arraybackslash}p{2.2cm}>{\centering\arraybackslash}p{2.8cm}>{\centering\arraybackslash}p{2.8cm}>{\centering\arraybackslash}p{0.8cm}>{\centering\arraybackslash}p{0.8cm}>{\centering\arraybackslash}p{0.8cm}@{}}
      \toprule
      {\textbf{Type}} & {\textbf{Method}} & {\textbf{Domain}} & {\textbf{Metrics}} & {\textbf{MC}} & {\textbf{RA}} & {\textbf{ECC}} \\ \midrule
      
      \multirow{5}{*}{\rotatebox[origin=c]{270}{\hspace{-2px}Deep Reinforcement Learning}} & A2C-DRL \cite{lu2024a2cdrl} & Edge-Cloud Environments & Reward Value, Load-balancing & \redcross & \greencheck & \greencheck \\
      & AC-DRL \cite{nieto2024deep} & 5G Edge-Cloud Continuum & QoE, Energy Consumption & \redcross & \greencheck & \greencheck \\
      & DA-HADRL \cite{huang2024joint} & SAGINs with Cloud-Edge & Energy, Latency & \redcross & \greencheck & \greencheck \\
      & GRLDO \cite{li2024offloading} & Client-Edge-Cloud Environment & User Satisfaction & \greencheck & \greencheck & \greencheck \\
      & MADDPG \cite{she2024efficient} & End-Edge-Cloud in 6G Networks & Delay, Energy Consumption & \redcross & \greencheck & \greencheck \\
      \midrule
      \multirow{4}{*}{\rotatebox[origin=c]{270}{\hspace{9px}Federated Learning}} & FRL \cite{kim2024collaborative} & Cloud-Edge-Terminal IoT & Learning Speed, Adaptation & \redcross & \greencheck & \greencheck \\
      & FDRL \cite{wang2024federated} & Cloud-Edge Manufacturing & Makespan, Energy Consumption & \redcross & \greencheck & \greencheck \\
      & MARL-SF \cite{lu2024dynamic} & IoV Edge Cloud & Quality of Service & \redcross & \greencheck & \greencheck \\
      & MARL-CA \cite{yang2024multiagent} & Industrial Edge-Cloud Web3 & Communication Rate, Stability & \redcross & \greencheck & \greencheck \\
      \midrule
      \multirow{4}{*}{\rotatebox[origin=c]{270}{\hspace{5px}Optimization-Based}} & DySCo \cite{bensalah2024vnf} & Edge-Cloud Infrastructure & Deployment Cost, Delay & \redcross & \greencheck & \greencheck \\
      & OTFAC \cite{gan2024optimal} & Cloud-Edge Collaborative IoT & Delay, Energy Consumption & \redcross & \greencheck & \greencheck \\
      & HDMO \cite{zhang2024hybrid} & Edge-Cloud Collaboration & Time, Cost, Energy, Reliability & \redcross & \greencheck & \greencheck \\
      & DaCM \cite{wang2024efficient} & Edge-Cloud Computing & Long-term Profit, Delay & \redcross & \greencheck & \greencheck \\
      \midrule
      \multirow{3}{*}{\rotatebox[origin=c]{270}{\hspace{10px}Model-Aware}} & DSCF \cite{ko2024dynamic} & Distributed Serverless Edge & Inference Latency, Res. Consump. & \greencheck & \greencheck & \greencheck \\
      & DRL-LOA \cite{saranya2024enhanced} & Healthcare Cloud-Edge & Power Usage, Latency & \greencheck & \greencheck & \greencheck \\
      & Wang et al. \cite{wang2024cloudedgeend} & IoT-Based Distribution Grid & Convergence Rate, Stability & \greencheck & \greencheck & \greencheck \\
      \midrule
      \multirow{4}{*}{\rotatebox[origin=c]{270}{\hspace{17px}Domain-Specific}} & Das et al. \cite{das2024edgecloud} & Drone Communication Networks & Latency, Scalability, Resource Util. & \redcross & \greencheck & \greencheck \\
      & DA-DDQN \cite{pang2024multimobile} & Vehicle-Edge-Cloud & Latency, Energy Consumption & \redcross & \greencheck & \greencheck \\
      & QRSRL \cite{zhu2024collaborative} & Train-Edge-Cloud Smart Systems & Task Processing Delay & \redcross & \greencheck & \greencheck \\
      & Informer-A3C \cite{zhu2024vehicular} & Vehicular Edge Cloud Computing & Content Access Cost & \redcross & \greencheck & \greencheck \\ \bottomrule
  \end{tabular}}}
  \end{center}
  \fontsize{7.5pt}{2pt}\selectfont {
  \begin{minipage}{0.48\textwidth}
   \vspace{1px}
    \textit{Notes}: RA = Resource Allocation.
  \end{minipage}
  }
\vspace{-15px}
\end{table}

%% file: taxo-sec5-privacy-security.tex
\begin{figure}[t]
  \centering
  \makebox[\columnwidth][c]{%
  \resizebox{\columnwidth}{!}{
\begin{tikzpicture}[node distance=0.6cm, >=latex,
  Root/.style={fill=yellow!20, rounded corners=1pt, inner sep=2pt, font=\fontsize{12pt}{2pt}\selectfont, text width=5cm, align=center},
  L1/.style={fill=cyan!15, rounded corners=1pt, inner sep=2pt, font=\fontsize{12pt}{2pt}\selectfont, text width=5cm, align=center},
  L2/.style={fill=red!10, rounded corners=1pt, inner sep=2pt, font=\fontsize{12pt}{2pt}\selectfont, text width=5cm, align=center}
]
  \node[Root] (root) [rotate=90] at (0,0) {Privacy \& Security (\S\ref{sec5})};
  
  \node[L1] (privacy) at (4, 2.1) {Data Privacy Protect. (\S\ref{sec5.1})};
  \node[L1] (model) at (4, -0.1) {Model Security (\S\ref{sec5.2})};
  \node[L1] (comm) at (4, -2.1) {Comm. Security (\S\ref{sec5.3})};
  
  \node[L2] (ppl) at (10.5, 3.0) {Privacy-preserving};
  \node[L2] (smpc) at (10.5, 2.4) {SMPC};
  \node[L2] (dp) at (10.5, 1.8) {Differential Privacy};
  \node[L2] (anon) at (10.5, 1.2) {Data Anonymization};
  
  \node[L2] (integrity) at (10.5, 0.5) {Model Integrity};
  \node[L2] (attack) at (10.5, -0.1) {Attack Detection};
  \node[L2] (update) at (10.5, -0.7) {Secure Updates};
  
  \node[L2] (transmit) at (10.5, -1.5) {Secure Transmission};
  \node[L2] (auth) at (10.5, -2.1) {Auth. \& Access};
  \node[L2] (blockchain) at (10.5, -2.7) {Blockchain Security};
  
  \node[font=\fontsize{12pt}{2pt}\selectfont, text width=5.8cm, align=left, anchor=west] (ppl-ex) at (14.5, 3.0) {\cite{stevens2022efficient,hu2020concentrated,choquette-choo2021capc}};
  \node[font=\fontsize{12pt}{2pt}\selectfont, text width=5.8cm, align=left, anchor=west] (smpc-ex) at (14.5, 2.4) {\cite{vonmaltitz2018management,vedadi2024efficient,feng2023smpc}};
  \node[font=\fontsize{12pt}{2pt}\selectfont, text width=5.8cm, align=left, anchor=west] (dp-ex) at (14.5, 1.8) {\cite{wang2022privacypreserving,duan2023privascissors,gupta2022privacypreserving}};
  \node[font=\fontsize{12pt}{2pt}\selectfont, text width=5.8cm, align=left, anchor=west] (anon-ex) at (14.5, 1.2) {\cite{le2022styleid,yang2024differentially,dhinakaran2024privacypreserving}};
  
  \node[font=\fontsize{12pt}{2pt}\selectfont, text width=5.8cm, align=left, anchor=west] (int-ex) at (14.5, 0.5) {\cite{alaverdyan2023confidential,chi2024trusted,li2024nextgeneration}};
  \node[font=\fontsize{12pt}{2pt}\selectfont, text width=5.8cm, align=left, anchor=west] (att-ex) at (14.5, -0.1) {\cite{fan2024autoupdating,zhao2024flexiblefl,fang2024secure}};
  \node[font=\fontsize{12pt}{2pt}\selectfont, text width=5.8cm, align=left, anchor=west] (upd-ex) at (14.5, -0.7) {\cite{liu2021communicationefficient,seifelnasr2024privacypreserving,liu2024blockchainempowered}};
  
  \node[font=\fontsize{12pt}{2pt}\selectfont, text width=5.8cm, align=left, anchor=west] (tra-ex) at (14.5, -1.5) {\cite{alwarafy2021survey,gupta2022secure,zobaed2022aidriven}};
  \node[font=\fontsize{12pt}{2pt}\selectfont, text width=5.8cm, align=left, anchor=west] (aut-ex) at (14.5, -2.1) {\cite{al-aqrabi2018scalable,xiao2024domainspecific,zaghloul2020pmod}};
  \node[font=\fontsize{12pt}{2pt}\selectfont, text width=5.8cm, align=left, anchor=west] (blo-ex) at (14.5, -2.7) {\cite{queralta2021blockchain,al-rakhami2021decentralized,liu2024enhancing}};
  
  \draw[->] (root.south) -- (privacy.west);
  \draw[->] (root.south) -- (model.west);
  \draw[->] (root.south) -- (comm.west);
  
  \draw[->] (privacy.east) -- (ppl.west);
  \draw[->] (privacy.east) -- (smpc.west);
  \draw[->] (privacy.east) -- (dp.west);
  \draw[->] (privacy.east) -- (anon.west);
  
  \draw[->] (model.east) -- (integrity.west);
  \draw[->] (model.east) -- (attack.west);
  \draw[->] (model.east) -- (update.west);
  
  \draw[->] (comm.east) -- (transmit.west);
  \draw[->] (comm.east) -- (auth.west);
  \draw[->] (comm.east) -- (blockchain.west);
  
  \draw[->] (ppl.east) -- (ppl-ex.west);
  \draw[->] (smpc.east) -- (smpc-ex.west);
  \draw[->] (dp.east) -- (dp-ex.west);
  \draw[->] (anon.east) -- (anon-ex.west);
  \draw[->] (integrity.east) -- (int-ex.west);
  \draw[->] (attack.east) -- (att-ex.west);
  \draw[->] (update.east) -- (upd-ex.west);
  \draw[->] (transmit.east) -- (tra-ex.west);
  \draw[->] (auth.east) -- (aut-ex.west);
  \draw[->] (blockchain.east) -- (blo-ex.west);
  
\end{tikzpicture}
}
}
  \caption{Taxonomy of privacy and security enhancement.}
  \label{taxo:sec5}
\end{figure}

%% file: main_all.bbl
\begin{thebibliography}{100}
\providecommand{\url}[1]{#1}
\csname url@samestyle\endcsname
\providecommand{\newblock}{\relax}
\providecommand{\bibinfo}[2]{#2}
\providecommand{\BIBentrySTDinterwordspacing}{\spaceskip=0pt\relax}
\providecommand{\BIBentryALTinterwordstretchfactor}{4}
\providecommand{\BIBentryALTinterwordspacing}{\spaceskip=\fontdimen2\font plus
\BIBentryALTinterwordstretchfactor\fontdimen3\font minus \fontdimen4\font\relax}
\providecommand{\BIBforeignlanguage}[2]{{%
\expandafter\ifx\csname l@#1\endcsname\relax
\typeout{** WARNING: IEEEtran.bst: No hyphenation pattern has been}%
\typeout{** loaded for the language `#1'. Using the pattern for}%
\typeout{** the default language instead.}%
\else
\language=\csname l@#1\endcsname
\fi
#2}}
\providecommand{\BIBdecl}{\relax}
\BIBdecl

\bibitem{poularakis2019joint}
K.~Poularakis, J.~Llorca, A.~M. Tulino, I.~Taylor, and L.~Tassiulas, ``Joint service placement and request routing in multi-cell mobile edge computing networks,'' in \emph{IEEE INFOCOM}, 2019, pp. 10--18.

\bibitem{hu2024laecips}
S.~Hu, R.~Deng, X.~Du, Z.~Lu, Q.~Duan, Y.~He, S.-C. Huang, and J.~Wu, ``Laecips: Large vision model assisted adaptive edge-cloud collaboration for iot-based perception system,'' \emph{arXiv:2404.10498}, 2024.

\bibitem{liu2021livemap}
Q.~Liu, T.~Han, J.~L. Xie, and B.~Kim, ``Livemap: Real-time dynamic map in automotive edge computing,'' in \emph{IEEE INFOCOM}, 2021, pp. 1--10.

\bibitem{chen2024edgecloud}
J.~Chen, S.~Dai, F.~Chen, Z.~Lv, and J.~Tang, ``Edge-cloud collaborative motion planning for autonomous driving with large language models,'' \emph{arXiv:2408.09972}, 2024.

\bibitem{vanhuynh2022edge}
D.~Van~Huynh, S.~R. Khosravirad, A.~Masaracchia, O.~A. Dobre, and T.~Q. Duong, ``Edge intelligence-based ultra-reliable and low-latency communications for digital twin-enabled metaverse,'' \emph{IEEE Wirel. Commun.}, vol.~11, no.~8, pp. 1733--1737, 2022.

\bibitem{hua2024energyefficient}
W.~Hua, P.~Liu, and L.~Huang, ``Energy-efficient resource allocation for heterogeneous edge--cloud computing,'' \emph{IEEE Internet Things J.}, vol.~11, no.~2, pp. 2808--2818, 2024.

\bibitem{arroba2024sustainable}
P.~Arroba, R.~Buyya, R.~C{\'a}rdenas, J.~L. {Risco-Mart{\'i}n}, and J.~M. Moya, ``Sustainable edge computing: Challenges and future directions,'' \emph{Softw. Pract. Exp.}, vol.~54, no.~11, pp. 2272--2296, 2024.

\bibitem{ren2021synergy}
H.~Ren, D.~Anicic, and T.~A. Runkler, ``The synergy of complex event processing and tiny machine learning in industrial iot,'' in \emph{DEBS}, 2021, pp. 126--135.

\bibitem{yu2024fedcae}
Y.~Yu, L.~Guo, H.~Gao, Y.~He, Z.~You, and A.~Duan, ``Fedcae: A new federated learning framework for edge-cloud collaboration based machine fault diagnosis,'' \emph{IEEE Trans. Ind. Electron.}, vol.~71, no.~4, pp. 4108--4119, 2024.

\bibitem{zhang2024cloud}
X.~Zhang, L.~Ma, K.~Peng, C.~Zhang, and M.~A. Shahid, ``A cloud--edge collaboration based quality-related hierarchical fault detection framework for large-scale manufacturing processes,'' \emph{Expert Syst. Appl.}, vol. 256, p. 124909, 2024.

\bibitem{tang2024anomaly}
L.~Tang, C.~Xue, Y.~Zhao, and Q.~Chen, ``Anomaly detection of service function chain based on distributed knowledge distillation framework in cloud--edge industrial internet of things scenarios,'' \emph{IEEE Internet Things J.}, vol.~11, no.~6, pp. 10\,843--10\,855, 2024.

\bibitem{carpio2020engineering}
F.~Carpio, M.~Delgado, and A.~Jukan, ``Engineering and experimentally benchmarking a container-based edge computing system,'' in \emph{IEEE ICC}, 2020, pp. 1--6.

\bibitem{rublein2024improved}
C.~Rublein, F.~Mehmeti, M.~Mahon, and T.~F.~L. Porta, ``Improved methods of task assignment and resource allocation with preemption in edge computing systems,'' \emph{arXiv:2403.15665}, 2024.

\bibitem{qi2024service}
Y.~Qi, L.~Pan, and S.~Liu, ``Service provisioning based on edge-cloud collaboration: A two-timescale online scheduling algorithm,'' \emph{IEEE Internet Things J.}, vol.~11, no.~19, pp. 31\,999--32\,011, 2024.

\bibitem{zhang2024manyobjective}
J.~Zhang, Z.~Ning, R.~H. Ali, M.~Waqas, S.~Tu, and I.~Ahmad, ``A many-objective ensemble optimization algorithm for the edge cloud resource scheduling problem,'' \emph{IEEE Trans. Mobile Comput.}, vol.~23, no.~2, pp. 1330--1346, 2024.

\bibitem{wang2024efficient}
C.~Wang, Y.~Yang, M.~Qi, and H.~Ma, ``Efficient cloud-edge collaborative inference for object re-identification,'' \emph{arXiv:2401.02041}, 2024.

\bibitem{li2024energyaware}
J.~Li, Z.~Qin, W.~Liu, and X.~Yu, ``Energy-aware and trust-collaboration cross-domain resource allocation algorithm for edge-cloud workflows,'' \emph{IEEE Internet Things J.}, vol.~11, no.~4, pp. 7249--7264, 2024.

\bibitem{guo2024madrlom}
Y.~Guo, X.~Xu, and F.~Xiao, ``Madrlom: A computation offloading mechanism for software-defined cloud-edge computing power network,'' \emph{Comput. Netw.}, vol. 245, p. 110352, 2024.

\bibitem{wang2024cooperative}
Y.~Wang, J.~Fang, Y.~Cheng, H.~She, Y.~Guo, and G.~Zheng, ``Cooperative end-edge-cloud computing and resource allocation for digital twin enabled 6g industrial iot,'' \emph{IEEE J. Sel. Topics Signal Process.}, vol.~18, no.~1, pp. 124--137, 2024.

\bibitem{jain2022latencymemory}
T.~Jain, {Avaneesh}, R.~Verma, and R.~Shorey, ``Latency-memory optimized splitting of convolution neural networks for resource constrained edge devices,'' in \emph{COMSNETS}, 2022, pp. 531--539.

\bibitem{zeng2021energyefficient}
Q.~Zeng, Y.~Du, K.~Huang, and K.~K. Leung, ``Energy-efficient resource management for federated edge learning with cpu-gpu heterogeneous computing,'' \emph{IEEE Trans. Wireless Commun.}, vol.~20, no.~12, pp. 7947--7962, 2021.

\bibitem{wen2020joint}
D.~Wen, M.~Bennis, and K.~Huang, ``Joint parameter-and-bandwidth allocation for improving the efficiency of partitioned edge learning,'' \emph{IEEE Trans. Wireless Commun.}, vol.~19, no.~12, pp. 8272--8286, 2020.

\bibitem{nieto2024deep}
G.~Nieto, I.~{de~la~Iglesia}, U.~{Lopez-Novoa}, and C.~Perfecto, ``Deep reinforcement learning techniques for dynamic task offloading in the 5g edge-cloud continuum,'' \emph{J. Cloud Comput.}, vol.~13, no.~1, p.~94, 2024.

\bibitem{sharma2024deep}
N.~Sharma, A.~Ghosh, R.~Misra, and S.~K. Das, ``Deep meta q-learning based multi-task offloading in edge-cloud systems,'' \emph{IEEE Trans. Mobile Comput.}, vol.~23, no.~4, pp. 2583--2598, 2024.

\bibitem{goudarzi2024mmddrl}
M.~Goudarzi, M.~A. Rodriguez, M.~Sarvi, and R.~Buyya, ``{$M\mu$}-ddrl: A qos-aware distributed deep reinforcement learning technique for service offloading in fog computing environments,'' \emph{IEEE Trans. Serv. Comput.}, vol.~17, no.~1, pp. 47--59, 2024.

\bibitem{xu2024fusion}
J.~Xu, W.~Wan, L.~Pan, W.~Sun, and Y.~Liu, ``The fusion of deep reinforcement learning and edge computing for real-time monitoring and control optimization in iot environments,'' in \emph{EPECE}, 2024, pp. 193--196.

\bibitem{binh2024reinforcement}
T.~H. Binh, D.~B. Son, H.~Vo, B.~M. Nguyen, and H.~T.~T. Binh, ``Reinforcement learning for optimizing delay-sensitive task offloading in vehicular edge--cloud computing,'' \emph{IEEE Internet Things J.}, vol.~11, no.~2, pp. 2058--2069, 2024.

\bibitem{afachao2024efficient}
K.~Afachao, A.~M. {Abu-Mahfouz}, and G.~P. Hanke, ``Efficient microservice deployment in the edge-cloud networks with policy-gradient reinforcement learning,'' \emph{IEEE Access}, vol.~12, pp. 133\,110--133\,124, 2024.

\bibitem{goudarzi2023distributed}
M.~Goudarzi, M.~Palaniswami, and R.~Buyya, ``A distributed deep reinforcement learning technique for application placement in edge and fog computing environments,'' \emph{IEEE Trans. Mobile Comput.}, vol.~22, no.~5, pp. 2491--2505, 2023.

\bibitem{bensalem2023scaling}
M.~Bensalem, E.~Ipek, and A.~Jukan, ``Scaling serverless functions in edge networks: A reinforcement learning approach,'' in \emph{IEEE GLOBECOM}, 2023, pp. 1777--1782.

\bibitem{wen2023adaptivenet}
H.~Wen, Y.~Li, Z.~Zhang, S.~Jiang, X.~Ye, Y.~Ouyang, Y.~Zhang, and Y.~Liu, ``Adaptivenet: Post-deployment neural architecture adaptation for diverse edge environments,'' in \emph{MobiCom}, 2023, pp. 1--17.

\bibitem{shahhosseini2022online}
S.~Shahhosseini, D.~Seo, A.~Kanduri, T.~Hu, S.-S. Lim, B.~Donyanavard, A.~M. Rahmani, and N.~Dutt, ``Online learning for orchestration of inference in multi-user end-edge-cloud networks,'' \emph{ACM Trans. Embed. Comput. Syst.}, vol.~21, no.~6, pp. 73:1--73:25, 2022.

\bibitem{she2024efficient}
H.~She, L.~Yan, and Y.~Guo, ``Efficient end--edge--cloud task offloading in 6g networks based on multiagent deep reinforcement learning,'' \emph{IEEE Internet Things J.}, vol.~11, no.~11, pp. 20\,260--20\,270, 2024.

\bibitem{he2024efficient}
X.~He, H.~Xu, X.~Xu, Y.~Chen, and Z.~Wang, ``An efficient algorithm for microservice placement in cloud-edge collaborative computing environment,'' \emph{IEEE Trans. Serv. Comput.}, vol.~17, no.~5, pp. 1983--1997, 2024.

\bibitem{chen2024fast}
Y.~Chen, S.~Ye, J.~Wu, B.~Wang, H.~Wang, and W.~Li, ``Fast multi-type resource allocation in local-edge-cloud computing for energy-efficient service provision,'' \emph{Inf. Sci.}, vol. 668, p. 120502, 2024.

\bibitem{materwala2022performance}
H.~Materwala and L.~Ismail, ``Performance and energy-aware bi-objective tasks scheduling for cloud data centers,'' \emph{Procedia Comput. Sci.}, vol. 197, pp. 238--246, 2022.

\bibitem{cortellessa2024exploring}
V.~Cortellessa, D.~Di~Pompeo, and M.~Tucci, ``Exploring sustainable alternatives for the deployment of microservices architectures in the cloud,'' in \emph{IEEE ICSA}, 2024, pp. 34--45.

\bibitem{lin2019timedriven}
B.~Lin, F.~Zhu, J.~Zhang, J.~Chen, X.~Chen, N.~N. Xiong, and J.~Lloret~Mauri, ``A time-driven data placement strategy for a scientific workflow combining edge computing and cloud computing,'' \emph{IEEE Trans. Ind. Informat.}, vol.~15, no.~7, pp. 4254--4265, 2019.

\bibitem{li2021appealnet}
M.~Li, Y.~Li, Y.~Tian, L.~Jiang, and Q.~Xu, ``Appealnet: An efficient and highly-accurate edge/cloud collaborative architecture for dnn inference,'' in \emph{DAC}, 2021, pp. 409--414.

\bibitem{yao2020efficient}
Q.~Yao, J.~Xu, W.-W. Tu, and Z.~Zhu, ``Efficient neural architecture search via proximal iterations,'' \emph{AAAI}, vol.~34, no.~04, pp. 6664--6671, 2020.

\bibitem{dutta2023searchtime}
O.~Dutta, T.~Kanvar, and S.~Agarwal, ``Search-time efficient device constraints-aware neural architecture search,'' in \emph{PReMI}, 2023, pp. 38--48.

\bibitem{yao2023edgecloud}
J.~Yao, S.~Zhang, Y.~Yao, F.~Wang, J.~Ma, J.~Zhang, Y.~Chu, L.~Ji, K.~Jia, T.~Shen, A.~Wu, F.~Zhang, Z.~Tan, K.~Kuang, C.~Wu, F.~Wu, J.~Zhou, and H.~Yang, ``Edge-cloud polarization and collaboration: A comprehensive survey for ai,'' \emph{IEEE Trans. Knowl. Data Eng.}, vol.~35, no.~7, pp. 6866--6886, 2023.

\bibitem{georgiou2021comprehensive}
K.~Georgiou, Z.~Chamski, K.~Nikov, and K.~Eder, ``A comprehensive and accurate energy model for arm's cortex-m0 processor,'' \emph{arXiv:2104.01055}, 2021.

\bibitem{zhang2022more}
L.~Zhang, Z.~Fu, B.~Shi, X.~Li, R.~Lai, C.~Yang, A.~Zhou, X.~Ma, S.~Wang, and M.~Xu, ``More is different: Prototyping and analyzing a new form of edge server with massive mobile socs,'' 2022.

\bibitem{rocha2019abeona}
I.~Rocha, G.~Vinha, A.~Brito, P.~Felber, M.~Pasin, and V.~Schiavoni, ``Abeona: An architecture for energy-aware task migrations from the edge to the cloud,'' in \emph{SRDS}, 2019, pp. 378--3782.

\bibitem{zhai2024edgecloud}
X.~Zhai, Y.~Peng, and X.~Guo, ``Edge-cloud collaboration for low-latency, low-carbon, and cost-efficient operations,'' \emph{Comput. Electr. Eng.}, vol. 120, p. 109758, 2024.

\bibitem{zhang2024dvfo}
Z.~Zhang, Y.~Zhao, H.~Li, C.~Lin, and J.~Liu, ``Dvfo: Learning-based dvfs for energy-efficient edge-cloud collaborative inference,'' \emph{IEEE Trans. Mobile Comput.}, vol.~23, no.~10, pp. 9042--9059, 2024.

\bibitem{zhou2023cooperative}
H.~Zhou, M.~Elsayed, M.~Bavand, R.~Gaigalas, S.~Furr, and M.~{Erol-Kantarci}, ``Cooperative hierarchical deep reinforcement learning based joint sleep and power control in ris-aided energy-efficient ran,'' 2023.

\bibitem{fan2025madrlbased}
W.~Fan, P.~Chen, X.~Chun, and Y.~Liu, ``Madrl-based model partitioning, aggregation control, and resource allocation for cloud-edge-device collaborative split federated learning,'' \emph{IEEE Trans. Mobile Comput.}, vol.~24, no.~6, pp. 5324--5341, 2025.

\bibitem{guo2023edge}
Y.~Guo, Y.~Zhang, L.~Wu, M.~Li, X.~Cai, and J.~Chen, ``Edge computing service deployment and task offloading based on multi-task high-dimensional multi-objective optimization,'' \emph{arXiv:2312.04101}, 2023.

\bibitem{khaleel2024enhancing}
M.~I. Khaleel, ``Enhancing the resilience of error-prone computing environments using a hybrid multi-objective optimization algorithm for edge-centric cloud computing systems,'' \emph{Neural Comput. Appl.}, vol.~36, no.~18, pp. 10\,733--10\,760, 2024.

\bibitem{hao2019edge}
T.~Hao, Y.~Huang, X.~Wen, W.~Gao, F.~Zhang, C.~Zheng, L.~Wang, H.~Ye, K.~Hwang, Z.~Ren, and J.~Zhan, ``Edge aibench: Towards comprehensive end-to-end edge computing benchmarking,'' in \emph{Bench}, 2019, pp. 23--30.

\bibitem{yang2024edgebench}
Q.~Yang, R.~Jin, N.~Gandhi, X.~Ge, H.~A. Khouzani, and M.~Zhao, ``Edgebench: A workflow-based benchmark for~edge computing,'' in \emph{Adv. Inf. Commun.}, 2024, pp. 150--170.

\bibitem{carpio2023benchfaas}
F.~Carpio, M.~Michalke, and A.~Jukan, ``Benchfaas: Benchmarking serverless functions in an edge computing network testbed,'' \emph{IEEE Netw.}, vol.~37, no.~5, pp. 81--88, 2023.

\bibitem{rosendo2023kheops}
D.~Rosendo, K.~Keahey, A.~Costan, M.~Simonin, P.~Valduriez, and G.~Antoniu, ``Kheops: Cost-effective repeatability, reproducibility, and replicability of edge-to-cloud experiments,'' in \emph{ACM REP}, 2023, pp. 62--73.

\bibitem{bi2023openperf}
F.~Bi, F.~Han, S.~Zhao, J.~Li, Y.~Zhang, and W.~Wang, ``Openperf: A benchmarking framework for the sustainable development of the open-source ecosystem,'' \emph{arXiv:2311.15212}, 2023.

\bibitem{mohammadi2018continuous}
M.~Mohammadi and T.~Bazhirov, ``Continuous evaluation of the performance of cloud infrastructure for scientific applications,'' \emph{arXiv:1812.05257}, 2018.

\bibitem{yang2019frequency}
C.~Yang, Z.~Du, X.~Meng, Y.~Du, and Z.~Duan, ``A frequency scaling based performance indicator framework for big data systems,'' in \emph{DASFAA}, 2019, pp. 19--35.

\bibitem{varghese2021survey}
B.~Varghese, N.~Wang, D.~Bermbach, C.-H. Hong, E.~D. Lara, W.~Shi, and C.~Stewart, ``A survey on edge performance benchmarking,'' \emph{ACM Comput. Surv.}, vol.~54, no.~3, pp. 66:1--66:33, 2021.

\bibitem{kahlhofer2023benchmarking}
M.~Kahlhofer, P.~Kern, S.~Henning, and S.~Rass, ``Benchmarking function hook latency in cloud-native environments,'' \emph{arXiv:2310.12702}, 2023.

\bibitem{jansen2023specrg}
M.~Jansen, A.~{Al-Dulaimy}, A.~V. Papadopoulos, A.~Trivedi, and A.~Iosup, ``The spec-rg reference architecture for the compute continuum,'' in \emph{CCGRID}, 2023, pp. 469--484.

\bibitem{luckow2021exploring}
A.~Luckow, K.~Rattan, and S.~Jha, ``Exploring task placement for edge-to-cloud applications using emulation,'' in \emph{ICFEC}, 2021, pp. 79--83.

\bibitem{becker2022network}
S.~Becker, T.~Pfandzelter, N.~Japke, D.~Bermbach, and O.~Kao, ``Network emulation in large-scale virtual edge testbeds: A note of caution and the way forward,'' in \emph{IEEE IC2E}, 2022, pp. 1--7.

\bibitem{huang2021armada}
L.~Huang, ``Armada: A robust latency-sensitive edge cloud in heterogeneous edge-dense environments,'' Master's thesis, University of Minnesota, 2021.

\bibitem{ali-eldin2021hidden}
A.~{Ali-Eldin}, B.~Wang, and P.~Shenoy, ``The hidden cost of the edge: A performance comparison of edge and cloud latencies,'' in \emph{SC}, 2021, pp. 1--12.

\bibitem{ren2022edgematrix}
Y.~Ren, S.~Shen, Y.~Ju, X.~Wang, W.~Wang, and V.~C. Leung, ``Edgematrix: A resources redefined edge-cloud system for prioritized services,'' in \emph{IEEE INFOCOM}, 2022, pp. 610--619.

\bibitem{wang2022ace}
L.~Wang, C.~Zhao, S.~Yang, X.~Yang, and J.~Mccann, ``Ace: Toward application-centric, edge-cloud, collaborative intelligence,'' \emph{Commun. ACM}, vol.~66, no.~1, pp. 62--73, 2022.

\bibitem{asghar2022survey}
H.~Asghar and E.-S. Jung, ``A survey on scheduling techniques in the edge cloud: Issues, challenges and future directions,'' \emph{arXiv:2202.07799}, 2022.

\bibitem{li2024twotimescale}
Y.~Li, H.~Wang, J.~Sun, H.~Lv, W.~Zheng, and G.~Feng, ``Two-timescale joint service caching and resource allocation for task offloading with edge--cloud cooperation,'' \emph{Comput. Netw.}, vol. 254, p. 110771, 2024.

\bibitem{fan2024collaborative}
W.~Fan, L.~Zhao, X.~Liu, Y.~Su, S.~Li, F.~Wu, and Y.~Liu, ``Collaborative service placement, task scheduling, and resource allocation for task offloading with edge-cloud cooperation,'' \emph{IEEE Trans. Mobile Comput.}, vol.~23, no.~1, pp. 238--256, 2024.

\bibitem{caiazza2022edge}
C.~Caiazza, S.~Giordano, V.~Luconi, and A.~Vecchio, ``Edge computing vs centralized cloud: Impact of communication latency on the energy consumption of lte terminal nodes,'' \emph{Comput. Commun.}, vol. 194, pp. 213--225, 2022.

\bibitem{su2024primaldualbased}
Q.~Su, Q.~Zhang, W.~Li, and X.~Zhang, ``Primal-dual-based computation offloading method for energy-aware cloud-edge collaboration,'' \emph{IEEE Trans. Mobile Comput.}, vol.~23, no.~2, pp. 1534--1549, 2024.

\bibitem{li2024nextgeneration}
Y.~Li, J.~Shen, P.~Vijayakumar, C.-F. Lai, A.~Sivaraman, and P.~K. Sharma, ``Next-generation consumer electronics data auditing scheme toward cloud--edge distributed and resilient machine learning,'' \emph{IEEE Trans. Consum. Electron.}, vol.~70, no.~1, pp. 2244--2256, 2024.

\bibitem{alaverdyan2023confidential}
Y.~Alaverdyan, S.~Poghosyan, and V.~Poghosyan, ``Confidential computing in edge-cloud hierarchy,'' \emph{IJCSIT}, vol.~15, no.~3, pp. 21--30, 2023.

\bibitem{chi2024trusted}
C.~Chi, Z.~Yin, Y.~Liu, and S.~Chai, ``A trusted cloud--edge decision architecture based on blockchain and mlp for aiot,'' \emph{IEEE Internet Things J.}, vol.~11, no.~1, pp. 201--216, 2024.

\bibitem{yang2024differentially}
Z.~Yang, B.~Xiong, K.~Chen, L.~T. Yang, X.~Deng, C.~Zhu, and Y.~He, ``Differentially private federated tensor completion for cloud--edge collaborative aiot data prediction,'' \emph{IEEE Internet Things J.}, vol.~11, no.~1, pp. 256--267, 2024.

\bibitem{ali2024universal}
A.~Ali, Y.-D. Lin, J.~Liu, and C.-T. Huang, ``The universal federator: A third-party authentication solution to federated cloud, edge, and fog,'' \emph{J. Netw. Comput. Appl.}, vol. 229, p. 103922, 2024.

\bibitem{zhao2024flexiblefl}
Y.~Zhao, Y.~Cao, J.~Zhang, H.~Huang, and Y.~Liu, ``Flexiblefl: Mitigating poisoning attacks with contributions in cloud-edge federated learning systems,'' \emph{Inf. Sci.}, vol. 664, p. 120350, 2024.

\bibitem{al-hawawreh2024digital}
M.~{Al-Hawawreh} and M.~S. Hossain, ``Digital twin-driven secured edge-private cloud industrial internet of things (iiot) framework,'' \emph{J. Netw. Comput. Appl.}, vol. 226, p. 103888, 2024.

\bibitem{chen2024intelligent}
X.~Chen, C.~Lin, and B.~Lin, ``An intelligent workflow scheduling scheme for complex network robustness in fuzzy edge-cloud environments,'' \emph{IEEE Trans. Netw. Sci. Eng.}, vol.~11, no.~1, pp. 1106--1123, 2024.

\bibitem{fan2024autoupdating}
C.~Fan, J.~Cui, H.~Jin, H.~Zhong, I.~Bolodurina, and D.~He, ``Auto-updating intrusion detection system for vehicular network: A deep learning approach based on cloud-edge-vehicle collaboration,'' \emph{IEEE Trans. Veh. Technol.}, vol.~73, no.~10, pp. 15\,372--15\,384, 2024.

\bibitem{li2024endedgecloud}
H.~Li, C.~Dou, D.~Yue, G.~P. Hancke, Z.~Zeng, W.~Guo, and L.~Xu, ``End-edge-cloud collaboration-based false data injection attack detection in distribution networks,'' \emph{IEEE Trans. Ind. Informat.}, vol.~20, no.~2, pp. 1786--1797, 2024.

\bibitem{yu2024game}
K.~Yu, S.~Wang, and X.~Tao, ``Game theory for 5g cloud- edge-terminal distributed networks under dos attacks,'' in \emph{IEEE WCNC}, 2024, pp. 1--6.

\bibitem{yao2024efficient}
Y.~Yao, J.~Chang, and A.~Zhang, ``Efficient data sharing scheme with fine-grained access control and integrity auditing in terminal--edge--cloud network,'' \emph{IEEE Internet Things J.}, vol.~11, no.~16, pp. 26\,944--26\,954, 2024.

\bibitem{fang2024secure}
W.~Fang, C.~Zhu, and W.~Zhang, ``Toward secure and lightweight data transmission for cloud--edge--terminal collaboration in artificial intelligence of things,'' \emph{IEEE Internet Things J.}, vol.~11, no.~1, pp. 105--113, 2024.

\bibitem{su2024robust}
Y.~Su, J.~Li, J.~Li, Z.~Su, W.~Meng, H.~Yin, and R.~Lu, ``Robust and lightweight data aggregation with histogram estimation in edge-cloud systems,'' \emph{IEEE Trans. Netw. Sci. Eng.}, vol.~11, no.~3, pp. 2864--2875, 2024.

\bibitem{gu2020confidential}
Z.~Gu, H.~Huang, J.~Zhang, D.~Su, H.~Jamjoom, A.~Lamba, D.~Pendarakis, and I.~Molloy, ``Confidential inference via ternary model partitioning,'' \emph{arXiv:1807.00969}, 2020.

\bibitem{xiao2024domainspecific}
M.~Xiao, Q.~Huang, W.~Chen, C.~Lyu, and W.~Susilo, ``Domain-specific fine-grained access control for cloud-edge collaborative iot,'' \emph{IEEE Trans. Inf. Forensics Security}, vol.~19, pp. 6499--6513, 2024.

\bibitem{khuencheng2024integration}
W.~Khuen~Cheng, J.~Cheng~Khor, W.~Zheng~Liew, K.~Thye~Bea, and Y.-L. Chen, ``Integration of federated learning and edge-cloud platform for precision aquaculture,'' \emph{IEEE Access}, vol.~12, pp. 124\,974--124\,989, 2024.

\bibitem{zheng2023kubeedgesedna}
Z.~Zheng, ``Kubeedge-sedna v0.3: Towards next-generation automatically customized ai engineering scheme,'' \emph{arXiv:2304.05985}, 2023.

\bibitem{abdelmoniem2023leveraging}
A.~M. Abdelmoniem, ``Leveraging the edge-to-cloud continuum for scalable machine learning on decentralized data,'' \emph{arXiv:2306.10848}, 2023.

\bibitem{liu2024blockchainempowered}
Y.~Liu, H.~Du, D.~Niyato, J.~Kang, Z.~Xiong, C.~Miao, X.~Shen, and A.~Jamalipour, ``Blockchain-empowered lifecycle management for ai-generated content products in edge networks,'' \emph{IEEE Wirel. Commun.}, vol.~31, no.~3, pp. 286--294, 2024.

\bibitem{ke2024groupvehicles}
C.~Ke, F.~Xiao, Y.~Cao, and Z.~Huang, ``A group-vehicles oriented reputation assessment scheme for edge vanets,'' \emph{IEEE Trans. Cloud Comput.}, vol.~12, no.~3, pp. 859--875, 2024.

\bibitem{liu2024eca}
J.~Liu, X.~Wang, G.~Xue, T.~Liu, and M.~Li, ``An eca regret learning game for cross-tier computation offloading against swarm attacks in sensor edge cloud,'' \emph{IEEE Internet Things J.}, vol.~11, no.~1, pp. 1201--1216, 2024.

\bibitem{mcgrath2024collaborative}
M.~J. McGrath, A.~Duenser, J.~Lacey, and C.~Paris, ``Collaborative human-ai trust (chai-t): A process framework for active management of trust in human-ai collaboration,'' \emph{arXiv:2404.01615}, 2024.

\bibitem{chen2024edgeleakage}
K.~Chen, Y.~Lin, H.~Luo, B.~Mi, Y.~Xiao, C.~Ma, and J.~S. Silva, ``Edgeleakage: Membership information leakage in distributed edge intelligence systems,'' \emph{arXiv:2404.16851}, 2024.

\bibitem{duan2023privascissors}
L.~Duan, J.~Sun, Y.~Chen, and M.~Gorlatova, ``Privascissors: Enhance the privacy of collaborative inference through the lens of mutual information,'' \emph{arXiv:2306.07973}, 2023.

\bibitem{seifelnasr2024privacypreserving}
M.~Seifelnasr, R.~AlTawy, A.~Youssef, and E.~Ghadafi, ``Privacy-preserving mutual authentication protocol with forward secrecy for iot--edge--cloud,'' \emph{IEEE Internet Things J.}, vol.~11, no.~5, pp. 8105--8117, 2024.

\bibitem{liu2021communicationefficient}
Y.~Liu, R.~Zhao, J.~Kang, A.~Yassine, D.~Niyato, and J.~Peng, ``Towards communication-efficient and attack-resistant federated edge learning for industrial internet of things,'' \emph{ACM Trans. Internet Technol.}, vol.~22, no.~3, pp. 59:1--59:22, 2021.

\bibitem{li2023trustworthy}
B.~Li, P.~Qi, B.~Liu, S.~Di, J.~Liu, J.~Pei, J.~Yi, and B.~Zhou, ``Trustworthy ai: From principles to practices,'' \emph{ACM Comput. Surv.}, vol.~55, no.~9, pp. 1--46, 2023.

\bibitem{tidjon2022never}
L.~N. Tidjon and F.~Khomh, ``Never trust, always verify : A roadmap for trustworthy ai?'' \emph{arXiv:2206.11981}, 2022.

\bibitem{wang2024surveyb}
X.~Wang, B.~Wang, Y.~Wu, Z.~Ning, S.~Guo, and F.~R. Yu, ``A survey on trustworthy edge intelligence: From security and reliability to transparency and sustainability,'' \emph{IEEE Commun. Surveys Tuts.}, pp. 1--1, 2024.

\bibitem{sanchez2022edge}
J.~M.~G. S{\'a}nchez, N.~J{\"o}rgensen, M.~T{\"o}rngren, R.~Inam, A.~Berezovskyi, L.~Feng, E.~Fersman, M.~R. Ramli, and K.~Tan, ``Edge computing for cyber-physical systems: A systematic mapping study emphasizing trustworthiness,'' \emph{ACM Trans. Cyber-Phys. Syst.}, vol.~6, no.~3, pp. 26:1--26:28, 2022.

\bibitem{hutiri2022trustworthy}
W.~T. Hutiri and A.~Yi~Ding, ``Towards trustworthy edge intelligence: Insights from voice-activated services,'' in \emph{IEEE SCC}, 2022, pp. 239--248.

\bibitem{alwarafy2021survey}
A.~Alwarafy, K.~A. {Al-Thelaya}, M.~Abdallah, J.~Schneider, and M.~Hamdi, ``A survey on security and privacy issues in edge-computing-assisted internet of things,'' \emph{IEEE Internet Things J.}, vol.~8, no.~6, pp. 4004--4022, 2021.

\bibitem{gupta2022secure}
M.~Gupta, J.~Benson, F.~Patwa, and R.~Sandhu, ``Secure v2v and v2i communication in intelligent transportation using cloudlets,'' \emph{IEEE Trans. Serv. Comput.}, vol.~15, no.~4, pp. 1912--1925, 2022.

\bibitem{zobaed2022aidriven}
S.~Zobaed, \emph{AI-Driven Confidential Computing across Edge-to-Cloud Continuum}, 2022.

\bibitem{bhardwaj2018spx}
K.~Bhardwaj, M.-W. Shih, A.~Gavrilovska, T.~Kim, and C.~Song, ``Spx: Preserving end-to-end security for edge computing,'' \emph{arXiv:1809.09038}, 2018.

\bibitem{zhang2021steciot}
P.~Zhang, C.~Jiang, X.~Pang, and Y.~Qian, ``Stec-iot: A security tactic by virtualizing edge computing on iot,'' \emph{IEEE Internet Things J.}, vol.~8, no.~4, pp. 2459--2467, 2021.

\bibitem{al-aqrabi2018scalable}
H.~{Al-Aqrabi} and R.~Hill, ``A scalable model for secure multiparty authentication,'' in \emph{IEEE SmartWorld}, 2018, pp. 17--22.

\bibitem{zaghloul2020pmod}
E.~Zaghloul, K.~Zhou, and J.~Ren, ``P-mod: Secure privilege-based multilevel organizational data-sharing in cloud computing,'' \emph{IEEE Transactions on Big Data}, vol.~6, no.~4, pp. 804--815, 2020.

\bibitem{ri2022blockchainbased}
O.-C. Ri, Y.-J. Kim, and Y.-J. Jong, ``Blockchain-based rbac model with separation of duties constraint in cloud environment,'' \emph{arXiv:2203.00351}, 2022.

\bibitem{queralta2021blockchain}
J.~P. Queralta and T.~Westerlund, ``Blockchain for mobile edge computing: Consensus mechanisms and scalability,'' in \emph{Mobile Edge Comput.}, 2021, pp. 333--357.

\bibitem{al-rakhami2021decentralized}
M.~S. {Al-Rakhami}, A.~Gumaei, S.~M.~M. Rahman, and A.~{Al-Amri}, ``Decentralized blockchain-based model for edge computing,'' \emph{arXiv:2106.15050}, 2021.

\bibitem{liu2024enhancing}
J.~Liu, C.~Chen, Y.~Li, L.~Sun, Y.~Song, J.~Zhou, B.~Jing, and D.~Dou, ``Enhancing trust and privacy in distributed networks: A comprehensive survey on blockchain-based federated learning,'' \emph{Knowl. Inf. Syst.}, vol.~66, no.~8, pp. 4377--4403, 2024.

\bibitem{patil2024mitigating}
P.~Patil, P.~Tulsiani, and D.~S. Mane, ``Mitigating data sharing in public cloud using blockchain,'' vol.~12, no.~2, 2024.

\bibitem{shahrukh2024aid}
R.~H. Shahrukh, T.~Rahman, and N.~Mansoor, ``Aid nexus : A blockchain-based financial distribution system,'' in \emph{ISS}, 2024, pp. 131--144.

\bibitem{zhou2024enhancing}
J.~Zhou, S.~Pal, C.~Dong, and K.~Wang, ``Enhancing quality of service through federated learning in edge-cloud architecture,'' \emph{Ad Hoc Netw.}, vol. 156, p. 103430, 2024.

\bibitem{yang2025growthadaptive}
Y.~Yang, C.~Fan, S.~Chen, Z.~Gao, and L.~Rui, ``Growth-adaptive distillation compressed fusion model for network traffic identification based on iot cloud--edge collaboration,'' \emph{Ad Hoc Netw.}, vol. 167, p. 103676, 2025.

\bibitem{liu2023distributed}
X.~Liu, J.~Yu, Y.~Liu, Y.~Gao, T.~Mahmoodi, S.~Lambotharan, and D.~H.-K. Tsang, ``Distributed intelligence in wireless networks,'' \emph{IEEE Open J. Commun. Soc.}, vol.~4, pp. 1001--1039, 2023.

\bibitem{yang2025edge}
N.~Yang, S.~Chen, H.~Zhang, and R.~Berry, ``Beyond the edge: An advanced exploration of reinforcement learning for mobile edge computing, its applications, and future research trajectories,'' \emph{IEEE Commun. Surveys Tuts.}, vol.~27, no.~1, pp. 546--594, 2025.

\bibitem{lu2024a2cdrl}
J.~Lu, J.~Yang, S.~Li, Y.~Li, W.~Jiang, J.~Dai, and J.~Hu, ``A2c-drl: Dynamic scheduling for stochastic edge--cloud environments using a2c and deep reinforcement learning,'' \emph{IEEE Internet Things J.}, vol.~11, no.~9, pp. 16\,915--16\,927, 2024.

\bibitem{paul2025quantumenhanced}
A.~Paul, K.~Singh, A.~Kaushik, C.-P. Li, O.~A. Dobre, M.~Di~Renzo, and T.~Q. Duong, ``Quantum-enhanced drl optimization for doa estimation and task offloading in isac systems,'' \emph{IEEE J. Sel. Areas Commun.}, vol.~43, no.~1, pp. 364--381, 2025.

\bibitem{baseri2024cybersecurity}
Y.~Baseri, V.~Chouhan, and A.~Ghorbani, ``Cybersecurity in the quantum era: Assessing the impact of quantum computing on infrastructure,'' \emph{arXiv:2404.10659}, 2024.

\bibitem{lu2024quantum}
C.~Lu, E.~Telang, A.~Aysu, and K.~Basu, ``Quantum leak: Timing side-channel attacks on cloud-based quantum services,'' \emph{arXiv:2401.01521}, 2024.

\bibitem{qu2025mobile}
G.~Qu, Q.~Chen, W.~Wei, Z.~Lin, X.~Chen, and K.~Huang, ``Mobile edge intelligence for large language models: A contemporary survey,'' \emph{IEEE Commun. Surveys Tuts.}, pp. 1--1, 2025.

\bibitem{kokkonen2023autonomy}
H.~Kokkonen, L.~Lov{\'e}n, N.~H. Motlagh, A.~Kumar, J.~Partala, T.~Nguyen, V.~C. Pujol, P.~Kostakos, T.~Lepp{\"a}nen, A.~{Gonz{\'a}lez-Gil}, E.~Sola, I.~Angulo, M.~Liyanage, M.~Bennis, S.~Tarkoma, S.~Dustdar, S.~Pirttikangas, and J.~Riekki, ``Autonomy and intelligence in the computing continuum: Challenges, enablers, and future directions for orchestration,'' \emph{arXiv:2205.01423}, 2023.

\bibitem{thakur2018largescale}
C.~S. Thakur, J.~L. Molin, G.~Cauwenberghs, G.~Indiveri, K.~Kumar, N.~Qiao, J.~Schemmel, R.~Wang, E.~Chicca, J.~Olson~Hasler, J.-s. Seo, S.~Yu, Y.~Cao, A.~{van Schaik}, and R.~{Etienne-Cummings}, ``Large-scale neuromorphic spiking array processors: A quest to mimic the brain,'' \emph{Front. Neurosci.}, vol.~12, 2018.

\bibitem{hossain2024quantumedge}
M.~I. Hossain, S.~A. Sumon, H.~M. Hasan, F.~Akter, M.~B. Badhon, and M.~N.~U. Islam, ``Quantum-edge cloud computing: A future paradigm for iot applications,'' \emph{arXiv:2405.04824}, 2024.

\bibitem{li2021slicingbased}
M.~Li, J.~Gao, C.~Zhou, Xuemin, Shen, and W.~Zhuang, ``Slicing-based ai service provisioning on network edge,'' \emph{arXiv:2105.07052}, 2021.

\bibitem{kawana2024communication}
T.~Kawana, R.~Nakagawa, and N.~Yamai, ``Communication multiplexing of server with quic and sdn in multihomed networks,'' in \emph{ICOIN}, 2024, pp. 345--350.

\bibitem{vogginger2024neuromorphic}
B.~Vogginger, A.~Rostami, V.~Jain, S.~Arfa, A.~Hantsch, D.~Kappel, M.~Sch{\"a}fer, U.~Faltings, H.~A. Gonzalez, C.~Liu, C.~Mayr, and W.~Maa{\ss}, ``Neuromorphic hardware for sustainable ai data centers,'' \emph{arXiv:2402.02521}, 2024.

\bibitem{letaief2022edge}
K.~B. Letaief, Y.~Shi, J.~Lu, and J.~Lu, ``Edge artificial intelligence for 6g: Vision, enabling technologies, and applications,'' \emph{IEEE J. Sel. Areas Commun.}, vol.~40, no.~1, pp. 5--36, 2022.

\bibitem{you20236g}
X.~You, Y.~Huang, S.~Liu, D.~Wang, J.~Ma, C.~Zhang, H.~Zhan, C.~Zhang, J.~Zhang, Z.~Liu, J.~Li, M.~Zhu, J.~You, D.~Liu, Y.~Cao, S.~He, G.~He, F.~Yang, Y.~Liu, J.~Wu, J.~Lu, G.~Li, X.~Chen, W.~Chen, and W.~Gao, ``Toward 6g tku extreme connectivity: Architecture, key technologies and experiments,'' \emph{IEEE Wirel. Commun.}, vol.~30, no.~3, pp. 86--95, 2023.

\bibitem{nguyen2023iquantum}
H.~T. Nguyen, M.~Usman, and R.~Buyya, ``iquantum: A case for modeling and simulation of quantum computing environments,'' in \emph{IEEE QSW}, 2023, pp. 21--30.

\bibitem{tuli2023ai}
S.~Tuli, F.~Mirhakimi, S.~Pallewatta, S.~Zawad, G.~Casale, B.~Javadi, F.~Yan, R.~Buyya, and N.~R. Jennings, ``Ai augmented edge and fog computing: Trends and challenges,'' \emph{J. Netw. Comput. Appl.}, vol. 216, p. 103648, 2023.

\bibitem{masaracchia20256genabled}
A.~Masaracchia, V.-L. Nguyen, D.~B. {da Costa}, E.~Ak, B.~Canberk, V.~Sharma, and T.~Q. Duong, ``Toward 6g-enabled urllcs: Digital twin, open ran, and semantic communications,'' \emph{IEEE Commun. Stand. Mag.}, vol.~9, no.~1, pp. 13--20, 2025.

\bibitem{wang2024endedgecloud}
Y.~Wang, C.~Yang, S.~Lan, L.~Zhu, and Y.~Zhang, ``End-edge-cloud collaborative computing for deep learning: A comprehensive survey,'' \emph{IEEE Commun. Surveys Tuts.}, vol.~26, no.~4, pp. 1--1, 2024.

\bibitem{moreschini2024edge}
S.~Moreschini, E.~Younesian, D.~H{\"a}stbacka, M.~Albano, J.~Ho{\v s}ek, and D.~Taibi, ``Edge to cloud tools: A multivocal literature review,'' \emph{J. Syst. Softw., journal}, vol. 210, p. 111942, 2024.

\bibitem{kanduri2024edgecentric}
A.~Kanduri, S.~Shahhosseini, E.~K. Naeini, H.~Alikhani, P.~Liljeberg, N.~Dutt, and A.~M. Rahmani, ``Edge-centric optimization of multi-modal ml-driven ehealth applications,'' in \emph{Embed. Mach. Learn. Cyber-Phys. IoT Edge Comput.}, 2024, pp. 95--125.

\end{thebibliography}
